\documentclass[]{aa}  
\usepackage[utf8]{inputenx}

\usepackage{graphicx}
\usepackage{fixltx2e}
\usepackage{url}

\usepackage{todonotes}

\usepackage{blindtext}

\usepackage{tikz}
\usetikzlibrary{patterns}
\usepackage{grffile}

\newcommand{\calaraltothanks}{Based on observations
    collected at the Centro Astronómico Hispano Alemán (CAHA) at Calar
    Alto, operated jointly by the Max-Planck Institut für Astronomie
    and the Instituto de Astrofísica de Andalucía (CSIC)}

\usepackage{txfonts}
\bibpunct{(}{)}{;}{a}{}{,} % to follow the A&A style
\begin{document} 

\title{The Lyman alpha reference sample. \\ VII. Spatially resolved
  H$\alpha$ kinematics\thanks{\calaraltothanks}} 

\titlerunning{LARS VII. Spatially resolved H$\alpha$ kinematics}
\authorrunning{E.~C.~Herenz et al.}

\author{Edmund~Christian~Herenz\inst{\ref{inst1}}\and 
        Pieter~Gruyters\inst{\ref{inst2}}\and
        Ivana~Orlitova\inst{\ref{inst3}}\and
        Matthew~Hayes\inst{\ref{stockh}}\and
        G\"{o}ran~\"{O}stlin\inst{\ref{stockh}}\and
        John~M.~Cannon\inst{\ref{inst5}}\and
        Martin~M.~Roth\inst{\ref{inst1}}\and
        Arjan~Bik\inst{\ref{stockh}}\and
        Stephen~Pardy\inst{\ref{inst6}}\and
        Héctor   Ot\'{i}-Floranes\inst{\ref{hector_inst_1},}\inst{\ref{hector_inst_2}}\and
        J. Miguel Mas-Hesse\inst{\ref{miguel_inst}}\and
        Angela~Adamo\inst{\ref{stockh}}\and
        Hakim~Atek\inst{\ref{inst7}}\and
        Florent~Duval\inst{\ref{stockh},}\inst{\ref{duval_inst}}\and
        Lucia Guaita\inst{\ref{stockh},}\inst{\ref{inst10}}\and
        Daniel Kunth\inst{\ref{inst12}}\and
        Peter Laursen\inst{\ref{dark}}\and
        Jens Melinder\inst{\ref{stockh}}\and
        Johannes Puschnig\inst{\ref{stockh}}\and
        Th{\o}ger E. Rivera-Thorsen\inst{\ref{stockh}}\and
        Daniel Schaerer\inst{\ref{inst4},}\inst{\ref{inst13}}\and
        Anne Verhamme\inst{\ref{inst4}}
       }
   \institute{Leibniz-Institut für Astrophysik Potsdam (AIP), An der
     Sternware 16, 14482 Potsdam, Germany, \email{cherenz@aip.de}\label{inst1}
   \and
   Lund Observatory, Box 43, SE-221 00 Lund, Sweden\label{inst2}
   \and
   Astronomical Institute, Academy of Sciences of the Czech Republic,
   Bo\v{c}n\'{i} II 1401, 141 00 Prague, Czech Republic\label{inst3}
   \and 
   Department of Astronomy, Oskar Klein Centre for Cosmoparticle Physics, Stockholm University,
   AlbaNova University Centre, SE-106 91 Stockholm,
   Sweden\label{stockh}
   \and
   Department of Physics \& Astronomy, Macalester College, 1600 Grand
   Avenue, Saint Paul, MN 55105, USA\label{inst5}
   \and
   Department of Astronomy, University of Wisconsin, 475 North Charter
   Street, Madison, WI 53706, USA\label{inst6}
   \and
   Consejo Nacional de Ciencia y Tecnolog\'ia, Av. Insurgentes Sur
   1582, Col. Cr\'edito Constructor, Del. Benito Ju\'arez, C.P 03940,
   Mexico D.F., Mexico\label{hector_inst_1}
   \and
   Instituto de Radioastronom\'ia y Astrof\'isica, UNAM, Campus Morelia,
   Michoac\'an, C.P. 58089, Mexico\label{hector_inst_2}
   \and
   Centro de Astrobiolog\'{i}a (CSIC–INTA), Departamento de
   Astrof\'{i}sica, P.O. Box 78, E-28691 Villanueva de la Ca\~{n}ada,
   Spain\label{miguel_inst}
   \and
   Laboratoire d'Astrophysique, \'{E}cole Polytechnique
   F\'{e}d\'{e}rale de Lausanne (EPFL), Observatoire, CH-1290
   Sauverny, Switzerland\label{inst7}
   \and
   Institute for Cosmic Ray Research, The University of Tokyo, 5-1-5
   Kashiwanoha, Kashiwa, Chiba 277-8582, Japan\label{duval_inst}
   \and
   INAF--Osservatorio Astronomico di Roma, Via Frascati 33, I-00040,
   Monteporzio, Italy\label{inst10}
   \and
   Institut d'Astrophysique de Paris, UMR 7095 CNRS \& UPMC, 98 bis Bd
   Arago, F-75014 Paris, France \label{inst12}
   \and
   Dark Cosmology Centre, Niels Bohr Institute, University of
   Copenhagen, Juliane Maries Vej 30, DK-2100 Copenhagen,
   Denmark\label{dark}
   \and
   Observatoire de Gen\`{e}ve, Universit\'{e} de Gen\`{e}ve, 51
   Ch. des Maillettes, 1290 Versoix, Switzerland\label{inst4}
   \and
   CNRS, IRAP, 14 Avenue E. Belin, 31400 Toulouse, France\label{inst13}
            }
   \date{Received \dots; accepted \dots}

% \abstract{}{}{}{}{} 
% 5 {} token are mandatory

   \abstract{We present integral field spectroscopic observations with
     the Potsdam Multi Aperture Spectrophotometer of all 14 galaxies
     in the $z\sim 0.1$ Lyman Alpha Reference Sample (LARS). We
     produce 2D line of sight velocity maps and velocity dispersion
     maps from the Balmer $\alpha$ (H$\alpha$) emission in our data
     cubes. These maps trace the spectral and spatial properties of
     the LARS galaxies' intrinsic Ly$\alpha$ radiation field. We show
     our kinematic maps spatially registered onto the Hubble Space
     Telescope H$\alpha$ and Lyman $\alpha$ (Ly$\alpha$) images. Only
     for individual galaxies a causal connection between spatially
     resolved H$\alpha$ kinematics and Ly$\alpha$ photometry can be
     conjectured. However, no general trend can be established for the
     whole sample.  Furthermore, we compute non-parametric global
     kinematical statistics -- intrinsic velocity dispersion
     $\sigma_0$, shearing velocity $v_\mathrm{shear}$, and the
     $v_\mathrm{shear}/\sigma_0$ ratio -- from our kinematic maps.
     In general LARS galaxies are characterised by high intrinsic
     velocity dispersions (54\,km\,s$^{-1}$ median) and low shearing
     velocities (65\,km\,s$^{-1}$ median). $v_\mathrm{shear}/\sigma_0$
     values range from 0.5 to 3.2 with an average of 1.5. Noteworthy,
     five galaxies of the sample are dispersion dominated systems with
     $v_\mathrm{shear}/\sigma_0 <1$ and are thus kinematically similar
     to turbulent star forming galaxies seen at high redshift.  When
     linking our kinematical statistics to the global LARS Ly$\alpha$
     properties, we find that dispersion dominated systems show higher
     Ly$\alpha$ equivalent widths and higher Ly$\alpha$ escape
     fractions than systems with $v_\mathrm{shear}/\sigma_0 > 1$. Our
     result indicates that turbulence in actively star-forming systems
     is causally connected to interstellar medium conditions that
     favour an escape of Ly$\alpha$ radiation. }


   \maketitle
%
%________________________________________________________________

\section{Introduction}
\label{sec:introduction}

As already envisioned by \cite{Partridge1967} the hydrogen Lyman
$\alpha$ (Ly$\alpha$, $\lambda_\mathrm{Ly\alpha}=1215.67$\AA{}) line
has become a prominent target in succsesful systematic searches for
galaxies in the early Universe. Redshifted into the optical, this
narrow high equivalent width line provides enough contrast to be
effectively singled out in specifically designed observational
campaigns.  As of yet mostly narrow-band selection techniques have
been employed
\citep[e.g.][]{Hu1998,Taniguchi2003,Malhotra2004,Shimasaku2006,Tapken2006,Gronwall2007,Ouchi2008,Grove2009,Shioya2009,Hayes2010,Ciardullo2012,Sandberg2015},
but multi-object and integral-field spectroscopic techniques are now
also frequently used to deliver large Lyman $\alpha$ emitter (LAE)
samples
\citep{Cassata2011,Adams2011,Mallery2012,Cassata2015,Bacon2015}. Moreover,
all galaxy redshift record holders in the last decade were
spectroscopically confirmed by virtue of their Ly$\alpha$ line
\citep{Iye2011a,Ono2012,Finkelstein2013,Oesch2015,Zitrin2015} and a
bright LAE at $z=6.6$ is even believed to contain a significant amount
of stars made exclusively from primordial material \citep[so called
Pop-III stars, ][]{Sobral2015}.

As important as Ly$\alpha$ radiation is in successfully unveiling
star formation processes in the early Universe, as notoriously
complicated appears its correct interpretation.  Resonant scatterings
in the interstellar and circum-galactic medium diffuse the intrinsic
Ly$\alpha$ radiation field in real and frequency space. These
scatterings also increase the path length of Ly$\alpha$ photons within
a galaxy and consequently Ly$\alpha$ is more susceptible for being
destroyed by dust \citep[see][for a comprehensive review covering the
Ly$\alpha$ radiative transfer fundamentals]{Dijkstra2014}.
Consequently, a galaxy's Ly$\alpha$ observables are influenced by a
large number of its physical properties. Understanding these
influences is crucial to correctly interpret high-$z$ LAE samples. In
other words, we have to answer the question: What differentiates an
LAE from other star-forming galaxies that do not show Ly$\alpha$ in
emission?

Regarding Ly$\alpha$ radiation transport in individual galaxies, a
large body of theoretical work investigated analytically and
numerically how within simplified geometries certain parameters
(e.g. density, temperature, dust content, kinematics and clumpiness)
affect the observed Ly$\alpha$ radiation field
\citep[e.g.][]{Neufeld1990,Ahn2003,Dijkstra2006,Verhamme2006,Laursen2013,Gronke2014,Duval2014}. Commonly
a spherical shell model is adopted. In this model a thin shell of
expanding, contracting or static neutral gas represents the medium
responsible for scattering Ly$\alpha$ photons.  As a result from
scattering in the shell complex Ly$\alpha$ line morphologies arise and
the expansion velocity and the neutral hydrogen column density of the
shell are of pivotal importance in shaping the observable Ly$\alpha$
line.  By introducing deviations from pure spherical symmetry
\cite{Zheng2014a} and \cite{Behrens2014a} show that the observed
Ly$\alpha$ properties also depend on the viewing angle under which a
system is observed. The latter result is also found  in large-scale
cosmological simulations that were post-processed
with Ly$\alpha$ radiative transport simulations
\citep[e.g.][]{Laursen2007,Laursen2009a,Barnes2011}.  Recently more
realistic hydro dynamic simulations of isolated galaxies have been
paired with Ly$\alpha$ radiation transport simulations
\citep{Verhamme2012,Behrens2014}. These studies underline again the
viewing angle dependence of the Ly$\alpha$ observables. In particular
they show, that disks observed face-on are expected to exhibit higher
Ly$\alpha$ equivalent widths and Ly$\alpha$ escape fractions than if
they were observed edge on.  More importantly, these state of the art
simulations also emphasise the importance of small-scale interstellar
medium structure that was previously not included in simple models. For
example \cite{Behrens2014} demonstrate how supernova-blown cavities
are able to produce favoured escape channels for Ly$\alpha$ photons.

Observationally, when Ly$\alpha$ is seen in emission, the spectral
line profiles can be typified by their distinctive shapes. In a large
number of LAE spectra the Ly$\alpha$ line often appears asymmetric,
with a relatively sharp drop on the blue and a more extended wing on
the red side. A significant fraction of spectra also shows
characteristic double peaks, with the red peak often being stronger
than the blue
\citep[e.g.][]{Tapken2004,Tapken2007,Yamada2012,Hong2014,Henry2015,Yang2015}.
Interestingly, a large fraction of double peaked and asymmetric
Ly$\alpha$ profiles appear well explained by the spherical symmetric
scenarios mentioned above.  Double-peaked profiles are successfully
reproduced by slowly-expanding shells
($v_\mathrm{exp} \lesssim 100$\,km\,s$^{-1}$) or low neutral hydrogen
column densities $N_\mathrm{\ion{H}{i}} \lesssim 10^{19}$cm$^{-2}$, while high
expansion velocities
($v_\mathrm{exp} \sim 150\,$--$\,300$\,km\,s$^{-1}$) with neutral
hydrogen column densities of $N_\mathrm{\ion{H}{i}} \gtrsim 10^{20}$cm$^{-2}$
produce the characteristic asymmetric profile with an extended red
wing
\citep[][]{Tapken2006,Tapken2007,Verhamme2008,Schaerer2011a,Gronke2014}.

Further observational constraints on the kinematics of the scattering
medium can be obtained by measuring the offset from non-resonant
rest-frame optical emission lines (e.g. H$\alpha$ or [\ion{O}{III}])
to interstellar low-ionisation state metal absorption lines
(e.g. \ion{O}{i} $\lambda1302$ or \ion{Si}{ii} $\lambda1304$). While
the emission lines provide the systemic redshift, some of the
low-ionisation state metal absorption lines trace the kinematics of
cold neutral gas phase. Both observables are challenging to obtain for
high-$z$ LAEs and require long integration times on 8--10\,m class
telescopes \citep[e.g.][]{Shapley2003,Tapken2004} or even additional
help from gravitational lenses
\citep[e.g.][]{Schaerer2008,Christensen2012}.  As the continuum
absorption often remains undetected, just the offset between the
rest-frame optical lines and the Ly$\alpha$ peaks are measured
\citep[e.g.][]{McLinden2011,Guaita2013,Erb2014}.  Curiously, the
observed Ly$\alpha$ profiles also agree well with those predicted by
the simple shell model, when the measured offsets (typically
$\Delta v \sim 200$\,km\,s$^{-1}$) are associated with shell expansion
velocities in the simple shell model
\citep[][]{Verhamme2008,Hashimoto2013,Song2014,Hashimoto2015}.  Only
few profiles appear incompatible with the expanding shells
\citep[][]{Chonis2013}. These profiles are characterised by extended
wings or bumps in the blue side of the profile
\citep{Martin2014,Henry2015}.  However, given aforementioned viewing
angle dependencies in more complex scenarios, this overall success of
the simple shell model appears surprising and is therefore currently
under scrutiny \citep{Gronke2015}.  Nevertheless, at least
qualitatively the observations demonstrate  in concert with theoretical
predictions that LAEs predominantly have outflow kinematics and that
such outflows promote the Ly$\alpha$ escape \citep[see
also][]{Kunth1998,Mas-Hesse2003}.

Kinematic information is moreover encoded in the rest-frame optical
line emission alone. These emission lines trace the ionised gas
kinematics and especially the hydrogen recombination lines such as
H$\alpha$ relate directly to the spatial and spectral properties of a
galaxy's intrinsic Ly$\alpha$ radiation field.  Therefore spatially
resolved spectroscopy of a galaxy's H$\alpha$ radiation field
constrains the initial conditions for the subsequent Ly$\alpha$
radiative transfer through the interstellar, circum-galactic, and
intergalactic medium to the observer.  But at high $z$ most LAEs are
so compact that they cannot be spatially resolved from the ground and
hence in the analysis of the integrated spectra all spatial
information is lost.  Resolving the intrinsic Ly$\alpha$ radiation of
typical high-$z$ LAEs spatially and spectrally at such small scales
would require integral field spectroscopy, preferably with adaptive
optics, in the near infrared with long integration times. Although
large samples of continuum-selected $z\sim2-3$ star-forming galaxies
have already been observed with this method \citep[see][for a
comprehensive review]{Glazebrook2013}, little is known about the
Ly$\alpha$ properties of the galaxies in those samples.

In the present manuscript we present first results obtained from our
integral-field spectroscopic observations with the aim to relate
spatially and spectrally resolved intrinsic Ly$\alpha$ radiation field
to its observed Ly$\alpha$ properties.  Therefore we targeted the
H$\alpha$ line in all galaxies of the $z\sim0.03$\,--\,$0.18$ Lyman
Alpha Reference Sample (hereafter LARS).  LARS consists of 14 nearby
laboratory galaxies, that have far-UV (FUV, $\lambda \sim$ 1500\AA{})
luminosities similar to those of high-$z$ star-forming
galaxies. Moreover, to ensure a strong intrinsic Ly$\alpha$ radiation
field, galaxies with large H$\alpha$ equivalent widths were selected
($\mathrm{EW}_\mathrm{H\alpha} \geq 100$\AA{}).  The backbone of LARS
is a substantial program with the Hubble Space Telescope.  In this
program each galaxy was observed with a combination of ultra-violet
long pass filters, optical broad band filters, as well as H$\alpha$
and H$\beta$ narrow-band filters. These images were used to accurately
reconstruct Ly$\alpha$ images of those 14 galaxies.  Moreover,
UV-spectroscopy with HSTs Cosmic Origins Spectrograph (COS) is
available for the whole sample.

\defcitealias{Ostlin2014}{Paper I} 
\defcitealias{Hayes2014}{Paper II}
\defcitealias{Pardy2014}{Paper III}
\defcitealias{Guaita2015}{Paper IV}
\defcitealias{Rivera-Thorsen2015}{Paper V}

This paper is the seventh in a series presenting results of the LARS
project. In \cite{Ostlin2014} (hereafter \citetalias{Ostlin2014}) we
detailed the sample selection, the observations with HST and the
process to reconstruct Ly$\alpha$ images from the HST data. In
\cite{Hayes2014} (hereafter \citetalias{Hayes2014}) we presented a
detailed analysis of the imaging results. We found that 6 of the 14
galaxies are indeed analogous to high-$z$ LAEs, i.e. they would be
selected by the conventional narrow-band survey selection
requirement\footnote{We adopt the convention
  established in high-$z$ narrow-band surveys of calling galaxies with
  $\mathrm{EW}_\mathrm{Ly\alpha}\geq20$\AA{} LAEs, and galaxies with
  $\mathrm{EW}_\mathrm{Ly\alpha}<20$\AA{} non-LAEs.}:
$\mathrm{EW}_\mathrm{Ly\alpha} > 20$\,\AA{}. The main result of
\citetalias{Hayes2014} is that a galaxy's morphology seen in
Ly$\alpha$ is usually very different compared to its morphological
appearance in H$\alpha$ and the FUV; especially Ly$\alpha$ is often
less centrally concentrated and so these galaxies are embedded in a
faint low surface-brightness Ly$\alpha$ halo. The results in
\citetalias{Hayes2014} \citep[see also][]{Hayes2013} therefore provide
clear observational evidence for resonant scattering of Ly$\alpha$
photons in the neutral interstellar medium. 21cm observations tracing
the neutral hydrogen content of the LARS galaxies were subsequently
presented in \cite{Pardy2014} (hereafter \citetalias{Pardy2014}) and
the results supported the complex coupling between Ly$\alpha$
radiative transfer and the properties of the neutral medium. In
\cite{Guaita2015} (hereafter \citetalias{Guaita2015}) the morphology
of the LARS galaxies was thoroughly re-examined. By artificially
redshifting the LARS imaging data-products, we were able to confirm
that morphologically LARS galaxies indeed resemble $z\sim2-3$ star
forming galaxies.  \cite{Rivera-Thorsen2015} (herafter
\citetalias{Rivera-Thorsen2015}) then presented high-resolution far-UV
COS spectroscopy of the full sample. Analysing the
neutral-interstellar medium kinematics as traced by the low-ionisation
state metal absorption lines, we showed that all galaxies with global
Ly$\alpha$ escape fractions $>$5\% appear to have outflowing
winds. Finally in Duval et al. (2015, A\&A submitted - hereafter Paper
VI) we performed a detailed radiative transfer study of one LARS
galaxy (Mrk\,1468 - LARS 5) using all the observational constraints
assembled within the LARS project, also including the data that is
presented in this paper. In particular we show that this galaxy's
spatial and spectral Ly$\alpha$ emission properties are consistent
with scattering of Ly$\alpha$ photons by outflowing cool material
along the minor axis of the disk.

In the here presented first analysis of our LARS integral-field
spectroscopic data we focus on a comparison of results obtainable from
spatially resolved H$\alpha$ kinematics to results from the LARS HST
Ly$\alpha$ imaging and 21cm \ion{H}{i} observations. In a subsequent
publication (Orlitova, in prep.) we will combine information from our
3D H$\alpha$ spectroscopy with our COS UV spectra to
constrain the parameters of outflowing winds.

We note, that this paper is not the first in relating observed
spatially resolved H$\alpha$ observations to a local galaxy's Ly$\alpha$
radiation field. Recently in a pioneering study exploiting MUSE
\citep{Bacon2014} science-verification data, \cite{Bik2015} showed,
that an asymmetric Ly$\alpha$ halo around the main star-forming knot
of \object{ESO338-IG04} \citep{Hayes2005,Ostlin2009,Sandberg2013} can
be linked to outflows seen in the H$\alpha$ radial velocity
field. Moreover, the kinematic constraints provided by our
observations were already used for modelling the Ly$\alpha$ scattering
in one LARS galaxy (Paper VI), and also here galactic scale outflows
were required to explain the galaxy's Ly$\alpha$ radiation.  With the
full data set presented here, we now study whether such theoretical
expected effects are indeed common among Ly$\alpha$-emitting galaxies.

The outline of this manuscript is as follows: In
Sect.~\ref{sec:observ} our PMAS observations of the LARS sample are
detailed.  The reduction of our PMAS data is explained in
Sect.~\ref{sec:data-reduction}.  Ancillary LARS data products used in
this manuscript are described briefly in
Sect.~\ref{sec:ancill-lars-datapr}. The derivation and analysis of the
H$\alpha$ velocity and dispersion maps is presented in
Sect.~\ref{sec:analysis}. In Sect.~\ref{sec:discussion} the results
are discussed and finally we summarise and conclude in
Sect.~\ref{sec:summary-conclusions}. Notes on individual objects are
given in  Appendix~\ref{sec:notes-indiv-objects}.

%\newpage

\section{PMAS Observations}
\label{sec:observ}

\begin{table*}
  \caption{Log of PMAS lens array observations of the LARS sample.}
  \centering 
  \begin{tabular}{cccccccccc} \hline \hline
    LARS & 
    Observing- &
    $t_\mathrm{exp.}$ & 
    FoV &
    $\lambda$-Range &
    Resolving Power &
    Seeing & 
    Observing- \\
    ID & 
    Date & 
    [s] &
    [arcsec$^2$] & 
    $\lambda_\mathrm{start} - \lambda_\mathrm{end}$ [\AA] &
    $\overline{R}_\mathrm{FWHM}$ &
    FWHM [\arcsec] &
    Conditions \\ \hline
1 & 2012-03-16 & 2$\times$1800 s & 16$\times$16 & 5948 - 7773 & 5306 & 1.8 & phot. \\ 
2 & 2012-03-15 & 2$\times$1800 s & 16$\times$16 & 5752 - 7581 & 5205 & 1.1 & non-phot. \\
3 & 2012-03-14 & 3$\times$1800 s & 16$\times$16 & 5948 - 7773 & 5362 & 1.1 & phot. \\
4 & 2012-03-13 & 3$\times$1800 s & 16$\times$16 & 5752 - 7581 & 5883& 1.0 & phot. \\
5 & 2012-03-14 & 3$\times$1800 s & 8$\times$8   & 5874 - 7700 & 5886& 1.3 & phot. \\
6 & 2012-03-13 & 2$\times$1800 s & 16$\times$16 & 5752 - 7581 & 5865 & 1.0 & phot. \\
7 & 2012-03-16 & 3$\times$1800 s & 16$\times$16 & 5948 - 7773 & 3925 & 1.3 & non-phot. \\
8 & 2012-03-15 & 3$\times$1800 s & 16$\times$16 & 5948 - 7773 & 4415  & 0.9 & non-phot. \\
9 & 2012-03-12 & 3$\times$1800 s & 16$\times$16 & 5752 - 7581 & 5561 & 0.9 & phot. \\ 
{} & 2012-03-14 & 3$\times$1800 s & 16$\times$16 & 5948 - 7773 & 4772 & 0.9 & non-phot. \\
10 & 2012-03-14 & 2$\times$1800 s & 8$\times$8   & 5874 - 7700 & 5438 & 1.5 & phot. \\
11 & 2012-03-15 & 3$\times$1800 s & 16$\times$16 & 5948 - 7773 & 4593 & 1.0 & non-phot. \\
12 & 2012-03-14 & 3$\times$1800 s & 8$\times$8   & 6553 - 8363 & 6756 & 0.9 & phot. \\
13 & 2011-10-02 & 4$\times$1800 s & 8$\times$8   & 6792 - 8507 & 7766 & 1.2 & phot. \\
{} & 2011-10-02 & 4$\times$1800 s & 8$\times$8   & 6792 - 8507 & 7771 &1.2 & phot. \\
14 & 2012-03-13 & 2$\times$1800 s & 8$\times$8   & 6553 - 8363 & 7718 &0.9 & phot. \\
\hline %\hline
  \end{tabular}
  \tablefoot{See \citetalias{Ostlin2014} for coordinates and common names of
    the galaxies in the LARS sample. LARS 09 and LARS 13 were covered
    with two pointings each, since their extent was larger than the used FoV.}
  \label{tab:log}
\end{table*}

We observed all LARS galaxies using the Potsdam Multi-Aperture
Spectrophotometer \citep[PMAS;][]{Roth2005} at the Calar Alto 3.5\,m
telescope during 4 nights from March 12 to March 15, 2012 (PMAS
run212), except for LARS 13, which was observed on October 10 2011
(PMAS run197).  PMAS was used in the lens array configuration, where
256 fibers are coupled to a $16\times16$ lens array that contiguously
samples the sky.  Depending on the extent of the targeted galaxy we
used either the standard magnification mode, that provides an
8\arcsec{}$\times$8\arcsec{} field of view (FoV) or the double
magnification mode\footnote{The naming of the mode refers to the
  instruments internal magnification of the telescopes focal plane;
  doubling the magnification of the focal plane doubles the
  extent of the FoV.}, where the FoV is
16\arcsec{}$\times$16\arcsec{}. The backward-blazed R1200 grating was
mounted on the spectrograph.
 To ensure
proper sampling of the line spread function the 4k$\times$4k CCD
\citep{Roth2010} was read out unbinned along the dispersion axis.
This setup delivers a nominal resolving power
from $R\sim5000$ to $R\sim8000$ within the targeted wavelength
ranges\footnote{Values taken from the PMAS online grating tables,
  available at
  \url{http://www.caha.es/pmas/PMAS_COOKBOOK/TABLES/pmas_gratings.html#4K_1200_1BW}
\label{fn:online_grating_footnote}}. In Sect. \ref{sec:determ-spectr-resol} we will show that while
the nominal resolving powers are met on average, the instrumental
broadening varies at small amplitudes from fibre to fibre. 

On-target exposures were usually flanked by 400\,s exposures of empty
sky near the target.  These sky frames serve as a reference for
removing the telluric background emission.  Due to an error in our
observing schedule, no sky frames were taken for LARS\,4, LARS\,7 and
LARS\,9 (2012-03-12 pointing). Fortunately, this did not render the
observations unusable, since there are enough blank-sky spectral
pixels (so called spaxels) within those on-target frames to provide us with a
reference sky (see paragraph on sky subtraction in
Sect.~\ref{sec:data-reduction}).  

Observing blocks of one hour were usually flanked by continuum and
HgNe arc lamp exposures used for wavelength and photometric
calibration.  We obtained several bias frames throughout each night
when the detector was idle during target acquisition.
Spectrophotometric standard stars  were observed at the
beginning and at the end of each astronomical night \citep[BD+75d325 \& Feige 67
from][$t_\mathrm{exp.} = 600$\,s]{Oke1990}. Twilight flat
exposures were taken during dawn and dusk.

In Table~\ref{tab:log} we provide a log of our observations. By
changing the rotation of the grating, we adjusted the wavelength
ranges covered by the detector, such that the galaxies H$\alpha$ -
[\ion{N}{ii}] complex is located near the center of the CCD. Note,
that $\pm350$\AA{} at the upper/lower end of the quoted spectral
ranges are affected by vignetting \citep[a known ``feature'' of the
PMAS detector - see ][]{Roth2010}.  We also quote the average spectral
resolution of our final datasets near H$\alpha$. The determination of
this quantity is described 
in Sect.~\ref{sec:determ-spectr-resol}.  The tabulated seeing values
refer always to the average FWHM of the guide star PSF measured during
the exposures with the acquisition and guiding camera of PMAS. This
value is on average $\approx 0.2$\arcsec{} higher than the DIMM
measurements \citep[see also][]{Husemann2013}.

\begin{figure}
  \centering
  \includegraphics[width=0.5\textwidth,trim=8 24 0 30,clip]{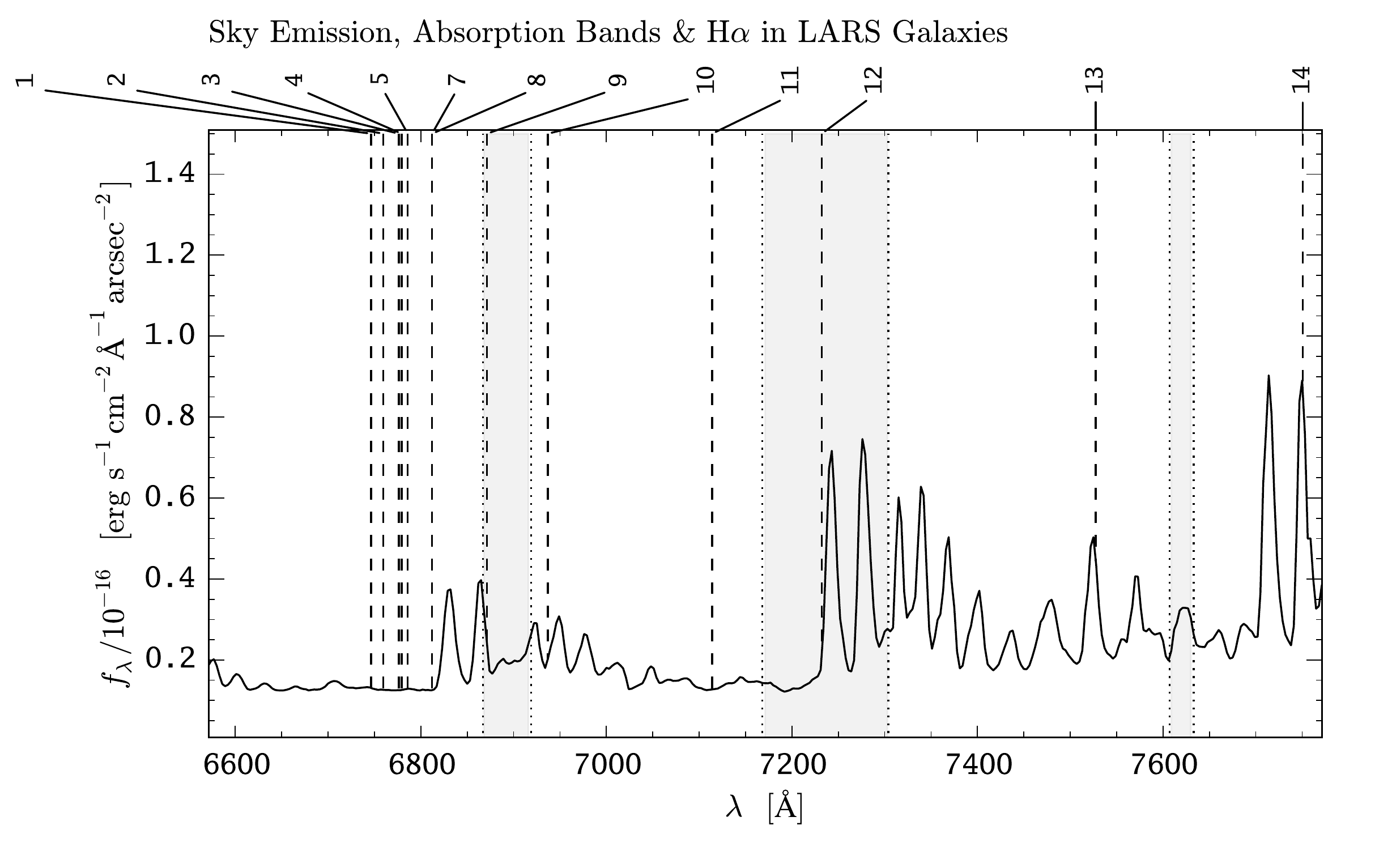}
  \caption{Wavelengths of the redshifted H$\alpha$ lines of the LARS
    galaxies (vertical dashed lines) compared to the night sky
    emission spectrum at Calar Alto \citep{Sanchez2007a}.  Grey
    regions indicate the telluric absorption bands \citep[O$_2$
    B-Band: 6887\,\AA{} - 6919\,\AA{}, O$_2$ A-Band: 7607\,\AA{} -
    7633\,\AA{}, and H$_2$O a-Band: 7168\,\AA{} - 7304\,\AA{}, see
    e.g.][]{Cox2000}.}
  \label{fig:sky}
\end{figure}

In Fig.~\ref{fig:sky} we show the wavelengths of the redshifted
H$\alpha$ line for all LARS galaxies overlaid on the typical night sky
emission spectrum at Calar Alto from \cite{Sanchez2007a}. As can be
seen for two galaxies (LARS 13 and LARS 14) the H$\alpha$ line signal is
contaminated by telluric line emission and two other galaxies (LARS 9 and
LARS 12) have their H$\alpha$ line within an absorption band. This,
however, has no effect on the presented analysis.  For LARS 13
and LARS 14 we could optimally subtract the interfering sky-lines from
the science exposures using the separate sky frames. Moreover, as
we use only spaxels with high signal to noise H$\alpha$ lines in our
analysis (cf. Sect.~\ref{sec:analysis}), the high frequency changes within
the telluric absorption bands do not alter the quantified features 
-- width and peak position -- of the profiles.

\section{PMAS data reduction}
\label{sec:data-reduction}

\subsection{Basic reduction with \texttt{p3d}}
\label{sec:basic-reduction-with}

For reducing the raw data the
\texttt{p3d}-package\footnote{\url{http://p3d.sourceforge.net}}
\citep{Sandin2010,Sandin2012} was utilized. This pipeline covers all
basic steps needed for reducing fiber-fed integral field spectroscopic
data: bias subtraction, flat fielding, cosmic ray removal, tracing and
extraction of the spectra, correction of differential atmospheric
refraction and co-addition of exposures \citep[see
also][]{Turner2010}. We now describe how we applied the tasks of
\texttt{p3d} on our raw data for each of these steps\footnote{As the
  amount of raw data from the observations was substantial, we greatly
  benefited from the scripting capabilites of \texttt{p3d}
  \citep{Sandin2011}. Our shell scripts and example parameter files
  can be found at \url{https://github.com/Knusper/pmas_data_red}}:
\begin{itemize}
\item \emph{Bias subtraction}: Master bias frames were created by
  \texttt{p3d\_cmbias}.  These master bias frames were then subtracted
  from the corresponding science frames.  Visual inspection of the bias
  subtracted frames showed no measurable offsets in unexposed regions
  between the 4 quadrants of the PMAS CCD, which attests optimal bias
  removal.
\item \emph{Flat fielding}: Each of the 256 fibers has its own
  wavelength dependent throughput curve. To determine these the task
  \texttt{p3d\_cflatf} was applied on the twilight flats.  The 
  determined curves are then applied in the extraction step of the
  science frame (see below) to normalize the extracted spectra.
\item \emph{Cosmic ray removal}: Cosmic ray hits on the CCD were
  removed with \texttt{p3d\_cr}. Visual inspections guided by the
  parameter study of \cite{Husemann2012} lead us to apply the
  L.A. Cosmic algorithm \citep{vanDokkum2001} with
  $\sigma_\mathrm{lim} = 5$, $\sigma_\mathrm{frac} = 1$,
  $f_\mathrm{lim} = 15$, a grow radius of 2 and a maximum of 4
  iterations. % TODO: mention MEDIAN FILTER
\item \emph{Tracing and extraction}: Spectra were extracted with the
  modified \cite{Horne1986} optimal extraction algorithm
  \citep{Sandin2010}. The task \texttt{p3d\_ctrace} was used to
  determine the traces and the cross-dispersion profiles of the
  spectra on the detector. Spectra were then extracted with
  \texttt{p3d\_cobjex}, also using the median recentering recommended
  in \cite{Sandin2012}.  
\item \emph{Wavelength calibration}: With \texttt{p3d\_cdmask} we
  obtained dispersion solutions (i.e. mappings from pixel to wavelength
  space) from the HgNe lamp frames using a sixth-order
  polynomial.  The wavelength sampling of the extracted science
  spectra is typically 0.46\,\AA{}\,px$^{-1}$.  We further improved the
  wavelength calibration of the science frames by applying small
  shifts (typically $0.1$\,--\,$0.3$\,px) determined from the
  strong 6300\AA{} and 6364\AA{} [\ion{O}{i}] sky lines (not available within the
  targeted wavelength range for LARS 12, 13, and 14)
\item \emph{Sensitivity function and flux calibration}: Using
  extracted and wavelength-calibrated standard star spectra, we
  created a sensitivity function utilizing \texttt{p3d\_sensfunc}.
  Absorption bands in the standard star spectra as well as telluric
  absorption bands (see Fig.~\ref{fig:sky}) were masked for the fit.
  Extinction curves were created using the empirical formula presented
  in \cite{Sanchez2007} using the extinction in $V$-band as measured
  by the Calar Alto extinction
  monitor\footnote{\url{http://www.caha.es/CAVEX/cavex.php}}.  We flux-calibrated all extracted and wavelength-calibrated science frames
  with the derived sensitivity function and extinction curve using
  \texttt{p3d\_fluxcal}.
\end{itemize}

The final data products from the \texttt{p3d}-pipeline
are flux- and wavelength-calibrated data cubes for all science
exposures, as well as the corresponding error cubes. From here we now
perform the following reduction steps with our custom procedures
written in \texttt{python}\footnote{\url{http://www.python.org}}.

\subsection{Sky subtraction and co-addition}
\label{sec:sky-subtraction-co}

\begin{itemize}
\item \emph{Sky subtraction:} All sky frames were reduced like science
  frames as described in the previous section. These extracted,
  wavelength- and flux-calibrated sky frames were then subtracted from
  the corresponding science frames and errors were propagated
  accordingly.  Unfortunately, no separate sky exposures were taken
  for three targets (LARS 4, LARS 7 and LARS 9,
  cf. Sect.~\ref{sec:observ}). For these targets we created a narrow
  band image of the H$\alpha$-[\ion{N}{ii}] region by summing up the
  relevant layers in the datacube. In this image we selected spaxels
  that do not contain significant amounts of flux. From these spaxels
  then an average sky spectrum was created, which was then
  subtracted from all spaxels. As the spectral resolution varies
  across the FoV (cf.  Sect.~\ref{sec:determ-spectr-resol}), this
  method produces some residuals at the position of the sky lines,
  which however do not touch the H$\alpha$ lines of the affected
  targets.
\item \emph{Stacking:} We co-added all individual flux-calibrated and
  sky-subtracted data cubes using the variance-weighted mean. For this
  calculation the input variances were derived from squaring the
  error cubes. Before co-addition we ensured by visual inspection that
  there are no spatial offsets between individual exposures. 
\end{itemize}

\subsection{Spectral resolving power determination}
\label{sec:determ-spectr-resol}

\begin{figure}[t!]
  \centering
  \includegraphics[width=0.4\textwidth]{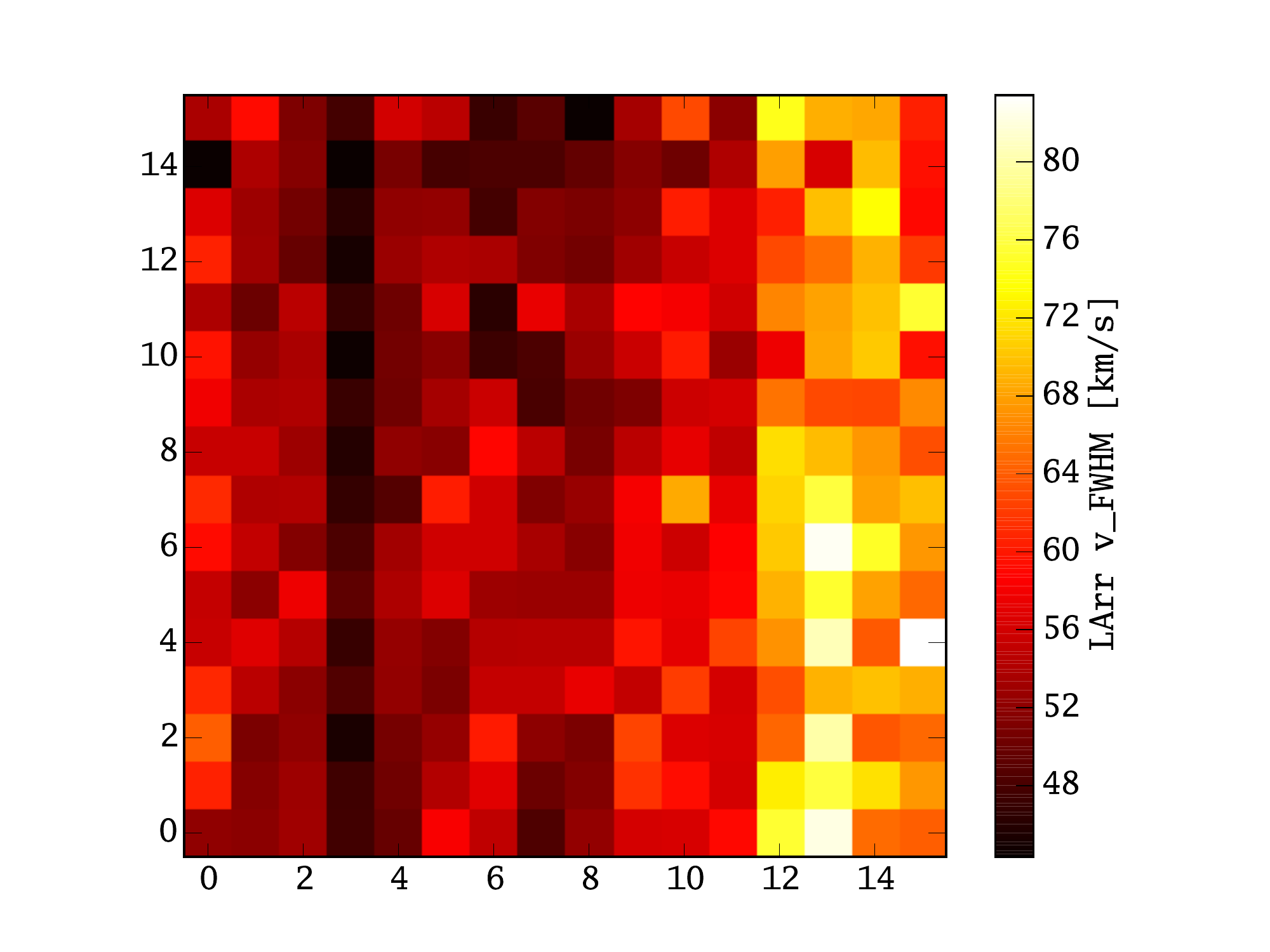}
  \caption{Representative resolving power map for the observation of
    LARS\,1. The resoultion is expressed as $v_\mathrm{FWHM}$ of a 1D
    Gaussian fit. Black spaxels at the positions $(x,y)=(0,14)$ and
    $(8,15)$ are dead fibres. }
  \label{fig:resmapex}
\end{figure}

To determine the intrinsic width of our observed H$\alpha$ lines we
need to correct for instrumental broadening by the spectrograph's line
spread function \citep[the spectrograph's resolving power,
][]{Robertson2013}.  It is known that for PMAS the resolving power
varies from fibre to fibre and with wavelength
\citep[e.g.][]{Sanchez2012}.  To determine the broadening at the
wavelength of the H$\alpha$ line we use the HgNe arc lamp exposures,
which were originally used to wavelength calibrate the on-target
exposures. We first create wavelength-calibrated data cubes from these
arc frames. Next we select two strong arc lamp lines which are nearest
in wavelength to the corresponding galaxy's H$\alpha$ line. Finally,
we fit 1D Gaussians in each spaxel to each of those lines
\citep[e.g. similar as in][]{Alonso-Herrero2009}.  Typically the
separation from a galaxy's H$\alpha$ line to one of those arc lamp
lines is $\sim$50\AA{}. The difference between the full-width half
maximum (FWHM) of the fits to the different lines is typically
$\sim1$km\,s$^{-1}$, hence we take the average of both at the position
of H$\alpha$ as the resolving power for each spaxel.   We
  point out that the arc-lamp lines FWHM is always sampled by more
  than 2 pixels, therfore aliasing effects can be neglected
  \citep{Turner2010}.

This procedure provides us with resolving power maps.  As an example
we show in Fig.~\ref{fig:resmapex} a so derived map for the
observations of LARS 1. We note that the spatial gradient seen in
Fig.~\ref{fig:resmapex} is not universal across our observations (see
\citealt{Sanchez2012} for an explanation). We further note
  that the formal uncertainties on the resolving power determination
  are negligible in comparison to the uncertainties derived on the
  H$\alpha$ profiles in Sect. \ref{sec:analysis}.

In Table~\ref{tab:log} we give the mean resolving power at the
position of H$\alpha$ for each galaxy as $\overline{R}_\mathrm{FWHM}$.
Variations within the FoV typically have an amplitude of $\sim$30
km\,s$^{-1}$. The average resolving power of all our observations is
$R=5764$ or 52 km\,s$^{-1}$, which is consistent with the values given
in the PMAS online grating
manual\footnote{\url{http://www.caha.es/pmas/PMAS_COOKBOOK/TABLES/pmas_gratings.html#4K_1200_1BW}}.

\subsection{Registration on astrometric grid of HST observations}
\label{sec:registr-astr-grid}

To facilitate a comparison of our PMAS observations with the HST
imaging results from \cite{Hayes2014} we have to register our PMAS
data cubes with respect to the LARS HST data products (cf. Sect.~\ref{sec:ancill-lars-datapr}).  As explained in
\cite{Ostlin2014} all HST images are aligned
with respect to each other and have a common pixel scale of
0.04\arcsec{}px$^{-1}$.  We use the continuum-subtracted
H$\alpha$ line image as a reference.  From this image we create
contours that highlight the most prominent morphological features in
H$\alpha$. We then produce a continuum-subtracted H$\alpha$ map from
the PMAS data cubes by subtracting a version of the data cube that is
median-filtered in spectral direction.  Finally, we visually match the
contours from the reference image with the PMAS H$\alpha$ map. This
constrains the position of the PMAS FoV relative to the HST imaging.
We emphasize that we make full use of the FITS header world-coordinate
representation in our method, i.e. the headers of our final data cubes
are equipped with keyword value pairs to determine the position of
each spaxel on the sky \citep{Greisen2002,Calabretta2002}.

For galaxies having a single PMAS pointing we present the final result
of our registration procedure in Fig.~\ref{fig:maps1} (or
Fig.~\ref{fig:lars9_fig} / Fig.~\ref{fig:lars13_fig} for the two
galaxies with two PMAS pointings): For each LARS galaxy the H$\alpha$
line intensity map extracted from our HST observations
(cf. Sect.~\ref{sec:ancill-lars-datapr}) is shown in the top panel and
can be compared to the PMAS H$\alpha$ signal-to-noise ratio (SNR) map
in the panel below. SNR values for each spaxel are calculated by
summation of all flux values within a narrow spectral window centred
on H$\alpha$ and subsequent division by the square root of the sum of
the variance values in that window. The width of the summation window
is taken as twice the H$\alpha$ lines FWHM.  Our display of the HST
H$\alpha$ images uses an asinh-scaling \citep{Lupton2004} cut at 95\%
of the maximum value (see Section 4 and Fig.~3 of \citealt{Hayes2014}
for absolute H$\alpha$ intensities) and we scaled our PMAS H$\alpha$
SNR maps logarithmically.  The inferred final position of the PMAS FoV
is indicated with a white square (or two squares for the galaxies
having two pointings) within the H$\alpha$ panel. Also shown are the
H$\alpha$ contours used for visual matching. As can be seen, most of
the prominent morphological characteristics present in the HST
H$\alpha$ maps are unambiguously identifiable in the lower resolution
PMAS maps, exemplifying the robustness of our registration method.

\section{Ancillary LARS dataproducts}
\label{sec:ancill-lars-datapr}

We compare our PMAS data to the HST imaging results of the LARS
project presented in \cite{Hayes2013} and
\citetalias{Hayes2014}. Specifically we will use the continuum
subtracted H$\alpha$ and Ly$\alpha$ images that were presented in
\citetalias{Hayes2014}. These images were produced from our HST
observation using the LARS extraction software LaXs \citep{Hayes2009}.
Details on the observational strategy, reduction steps and analysis
performed to obtain these HST dataproducts used in the present study are
given in \citetalias{Hayes2014} and \citetalias{Ostlin2014}.

We also compare with the published results of the LARS \ion{H}{i} imaging and
spectroscopy observations that were obtained with the 100m Robert
C. Byrd Green Bank Telescope (GBT) and the Karl G. Jansky Very Large
Array (VLA).  GBT single-dish spectra are present for all systems, but
for the three LARS galaxies with the largest distances (LARS 12, LARS
13 and LARS 14) the \ion{H}{i} signal could not be detected.  VLA
interferometric imaging results are only available for LARS 2, LARS 3,
LARS 4, LARS 8 and LARS 9.  We emphasise that with beam sizes from
59\arcsec{} to 72\arcsec{} (VLA D-configuration) the spatial scale
that is resolved within the VLA \ion{H}{i} images is much larger than our PMAS
observations.  Moreover, the single-dish GBT spectra are sensitive to
\ion{H}{i} at the observed frequency range within 8\arcmin{} of the beam.
Full details on data acquisition, reduction and analysis are presented
in \citetalias{Pardy2014}.

\section{Analysis and results}
\label{sec:analysis}
 
\subsection{H$\alpha$ velocity fields}
\label{sec:halpha-veloc-fields}

\begin{figure}
  \centering
  \includegraphics[width=0.5\textwidth]{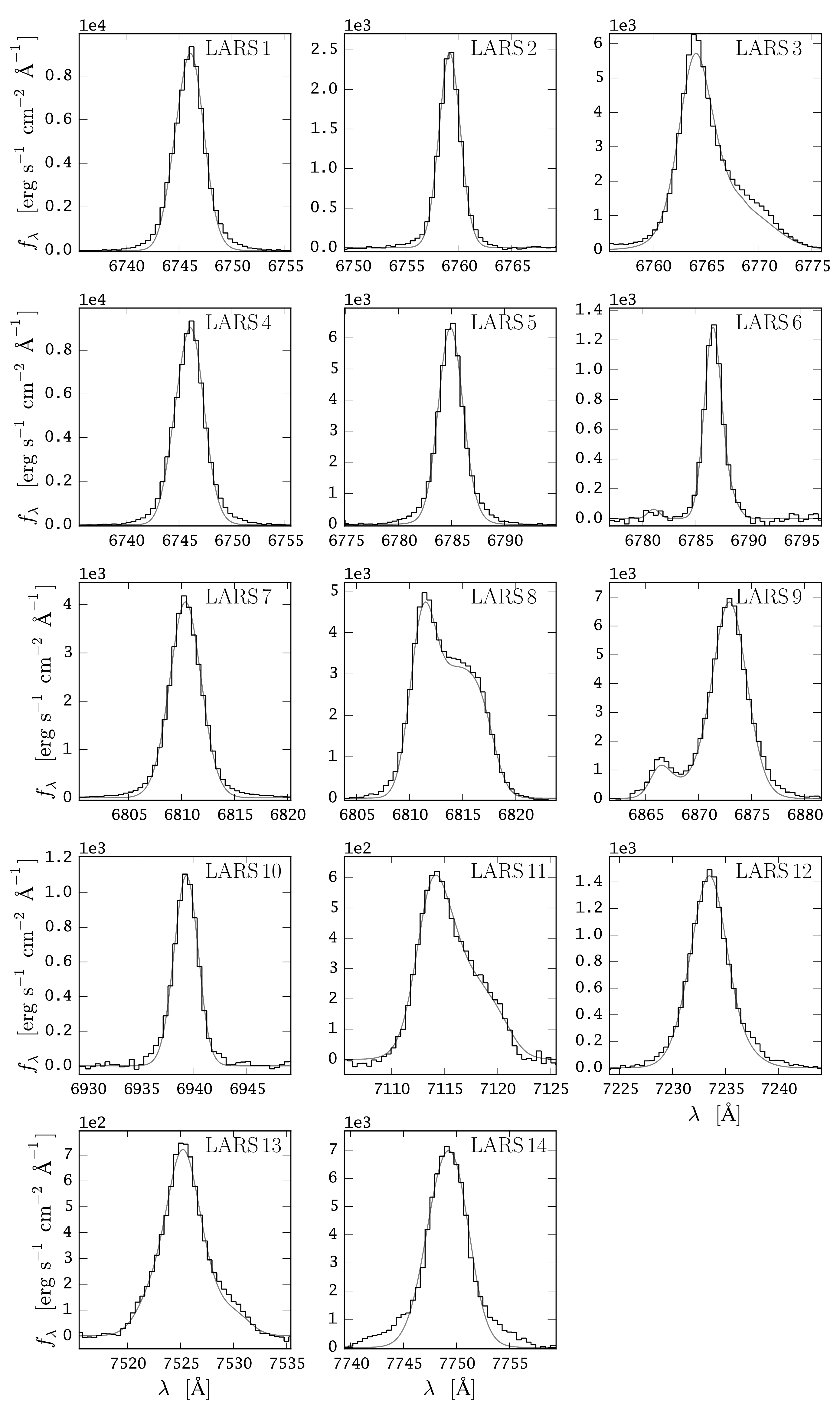}
  \caption{Integrated continuum-subtracted H$\alpha$ profiles of the
    LARS galaxies (black) compared to summed 1D Gaussian model
    profiles (grey) from which radial-velocity and velocity-dispersion
    maps were generated.}
  \label{fig:full_spect}
\end{figure}

\begin{figure*}
  \centering
% publishable version - uncomment when submitting
 % \includegraphics[width=0.45\textwidth]{figures/err_vs_sn.pdf}\includegraphics[width=0.45\textwidth]{figures/err_vs_sn_vrad.pdf}
% temporary compressed version - faster load & display times - comment
% out when submitting
 \includegraphics[width=0.45\textwidth]{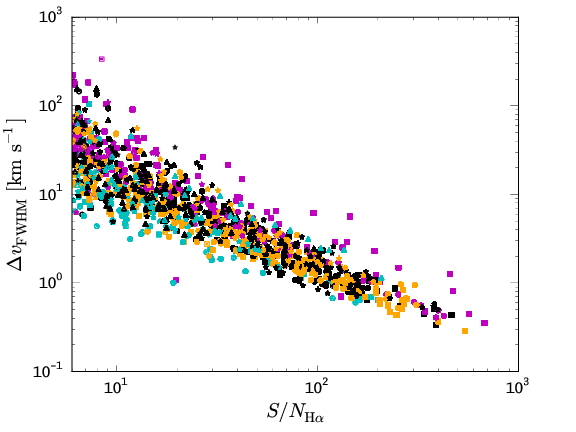}\includegraphics[width=0.45\textwidth]{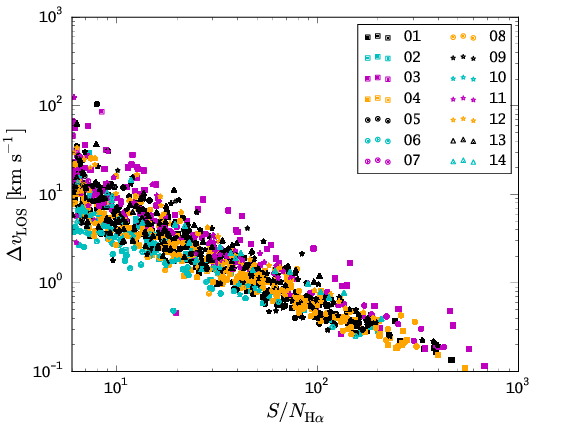}
  \caption{Uncertainties on derived velocity dispersions (\emph{left panel})
    and radial velocities (\emph{right panel}) for Gaussian profile
    fits to H$\alpha$ for all galaxies. All uncertainties for a specific
    galaxy have the same symbol according to the legend in the right panel.}
  \label{fig:error}
\end{figure*}

 \begin{figure*}[h!]\vspace{-1mm} % remove upper border
  \centering
  \includegraphics[width=0.99\textwidth]{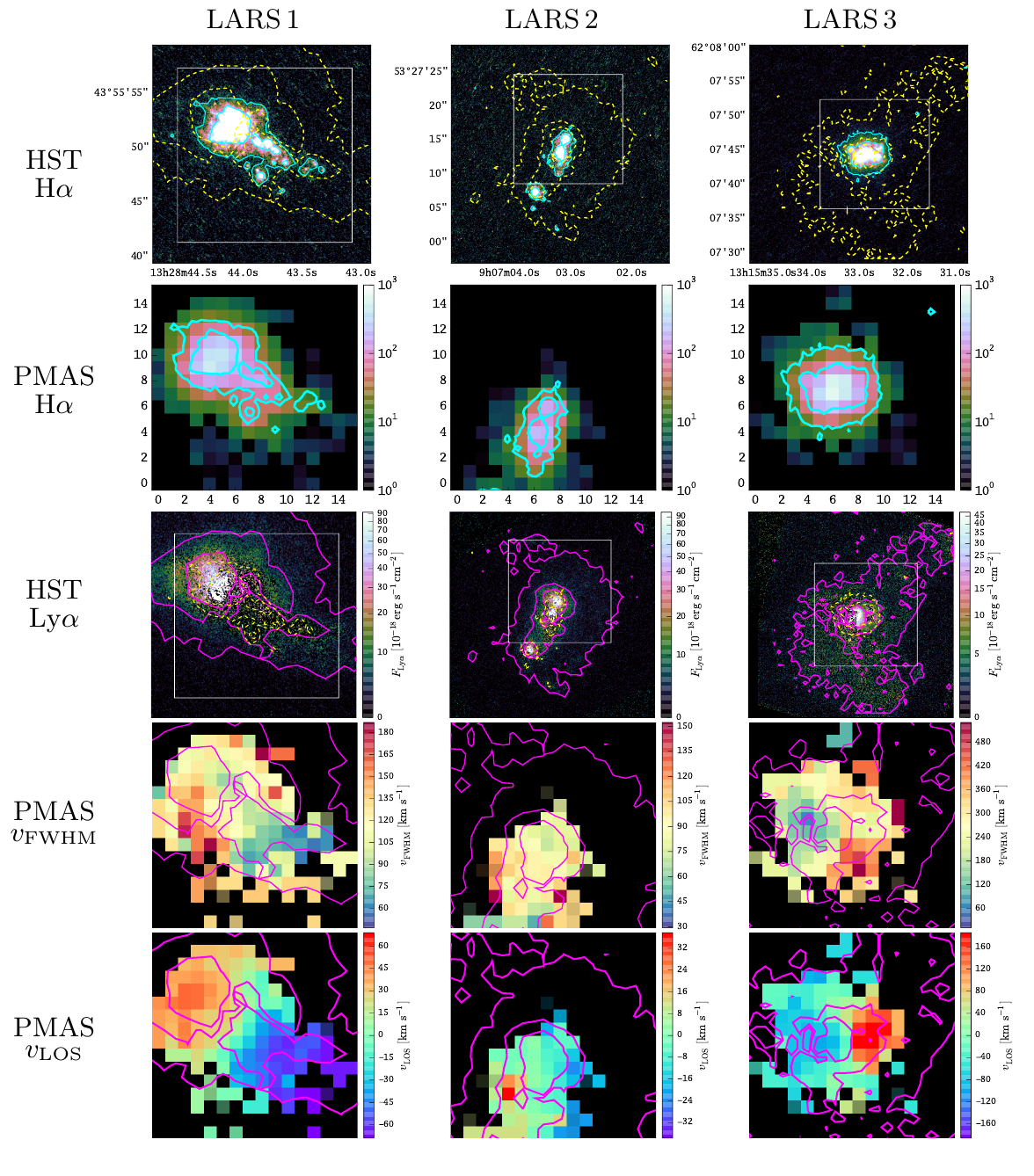}\vspace{-1em}
  \caption{Comparison of LARS HST imaging results of the LARS sample
    to spatially resolved PMAS H$\alpha$ spectroscopy.  North is
    always up and east is always to the left. For each galaxy from top to
    bottom: The \emph{first panel} shows the LaXs H$\alpha$ line
    intensity map; tick labels indicate right-ascension and
    declination and an asinh-scaling is used cut at 95\% of the
    maximum value. The \emph{second panel} shows a SNR map of the
    continuum-subtracted H$\alpha$ signal observed with PMAS. Tick
    labels in the PMAS SNR map are in arc-seconds, the scaling is
    logarithmic from 1 to $10^3$ and only spaxels with $S/N>1$ are
    shown.  The \emph{third panel} shows the LARS Ly$\alpha$ images
    with a colourbar indicating the flux scale in cgs-units; scaling
    is the same as in the H$\alpha$ map.  The \emph{fourth panel}
    shows resolution-corrected H$\alpha$ $v_\mathrm{FWHM}$ maps from
    our PMAS observations and the corresponding H$\alpha$
    $v_\mathrm{LOS}$ maps are displayed in the \emph{fifth panel}. In
    the first and third panel we indicate position and extend of the
    PMAS field of view with a white box.  Cyan contours in the HST
    H$\alpha$ image are contours of constant surface brightness,
    adjusted to highlight the most prominent morphological
    features. Similarly, magenta contours in the HST Ly$\alpha$ images
    indicate the Ly$\alpha$ morphology; these contours are also shown
    in the fourth and fifth panel. To highlight the difference between
    Ly$\alpha$ and H$\alpha$, the Ly$\alpha$ and H$\alpha$ panels
    contain as dashed yellow lines the contours from the H$\alpha$ and
    Ly$\alpha$ panels, respectively.}
  \label{fig:maps1}
\end{figure*}
\begin{figure*}%\vspace{-5mm} % remove upper border
  \centering
\includegraphics[width=0.99\textwidth]{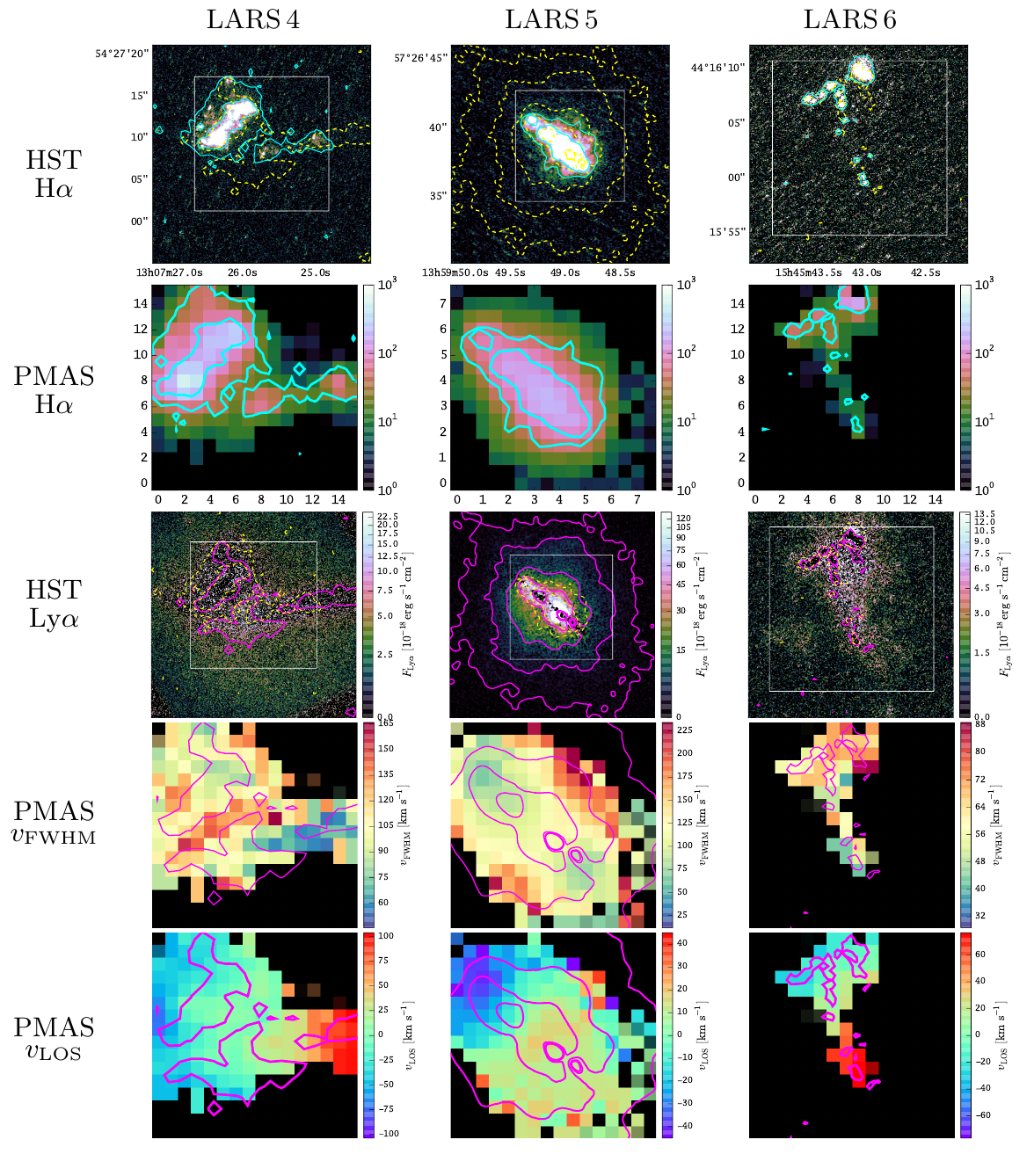}%\vspace{-1em}
  \addtocounter{figure}{-1}\vspace{-1em}
  \caption{Continued.}
\end{figure*}
\begin{figure*}%\vspace{-5mm} % remove upper border
  \centering
  \includegraphics[width=0.99\textwidth]{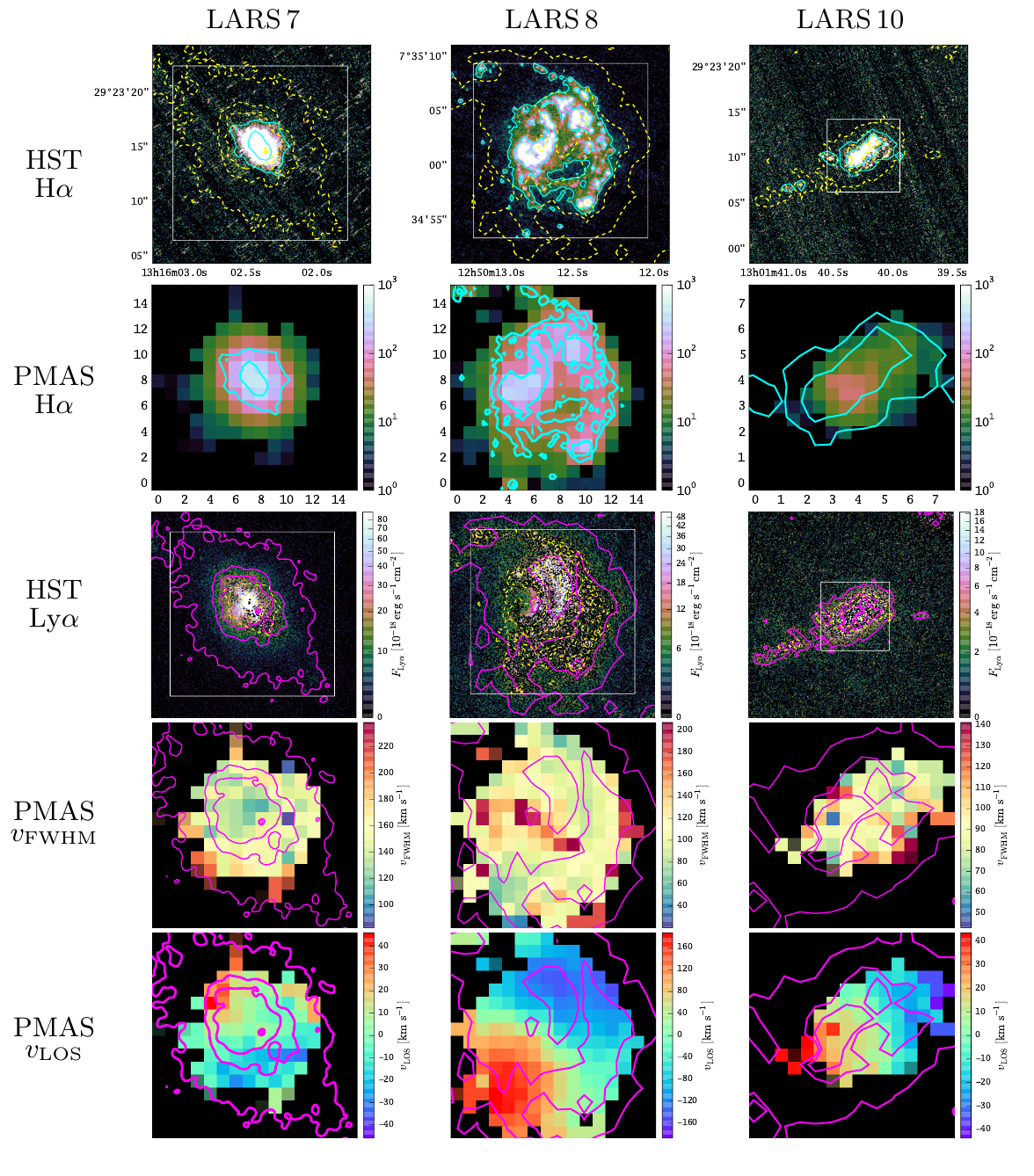}\vspace{-1em}
  \addtocounter{figure}{-1}
  \caption{Continued.}
\end{figure*}
\begin{figure*} % \vspace{-5mm} % remove upper border
  \centering
  \includegraphics[width=0.99\textwidth]{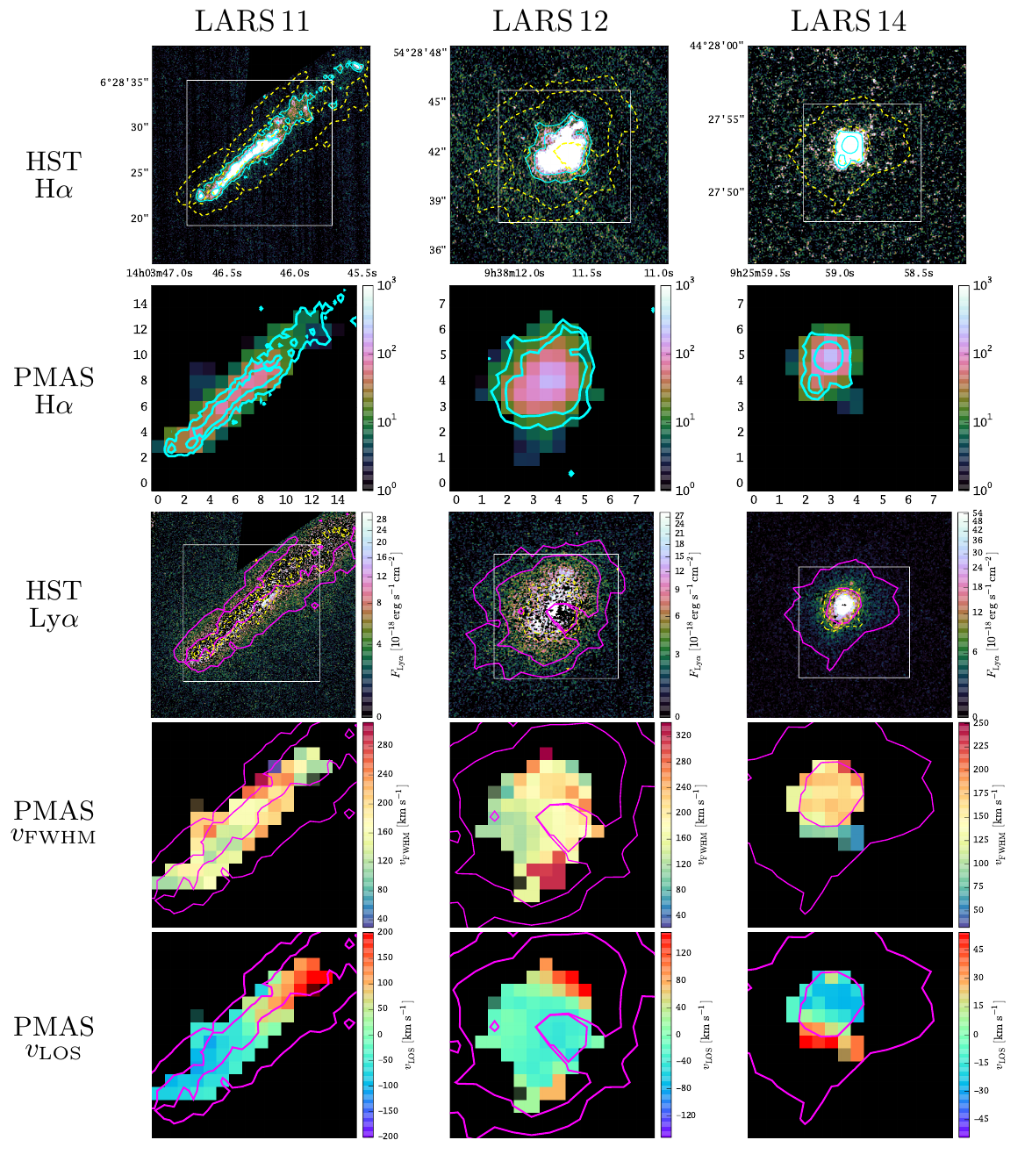}\vspace{-1em}
  \addtocounter{figure}{-1}
  \caption{Continued.}
\end{figure*}
% LARS 9 & LARS 13 in a separate panel
% \begin{figure*}\vspace{-4mm}
%   \centering 
%   \includegraphics[width=\textwidth]{figures/lars9_lars13_figure.pdf}\vspace{-4mm}
%   \caption{Same as Fig.~\ref{fig:maps1}, but for galaxies with 2 PMAS
%     pointings: LARS 9 and LARS 13. Hatched regions in the
%     $v_\mathrm{LOS}$ map of LARS 9 and LARS 13 indicate regions, where
%     the H$\alpha$ emission shows a more complex profile than a simple
%     Gaussian (for details see Appendix~\ref{sec:lars-9} for LARS 9 and
%     Appendix~\ref{sec:lars-13} for LARS 13).}
%   \label{fig:maps_double}
% \end{figure*}
\begin{figure*}
  \centering
  \includegraphics[width=0.9\textwidth]{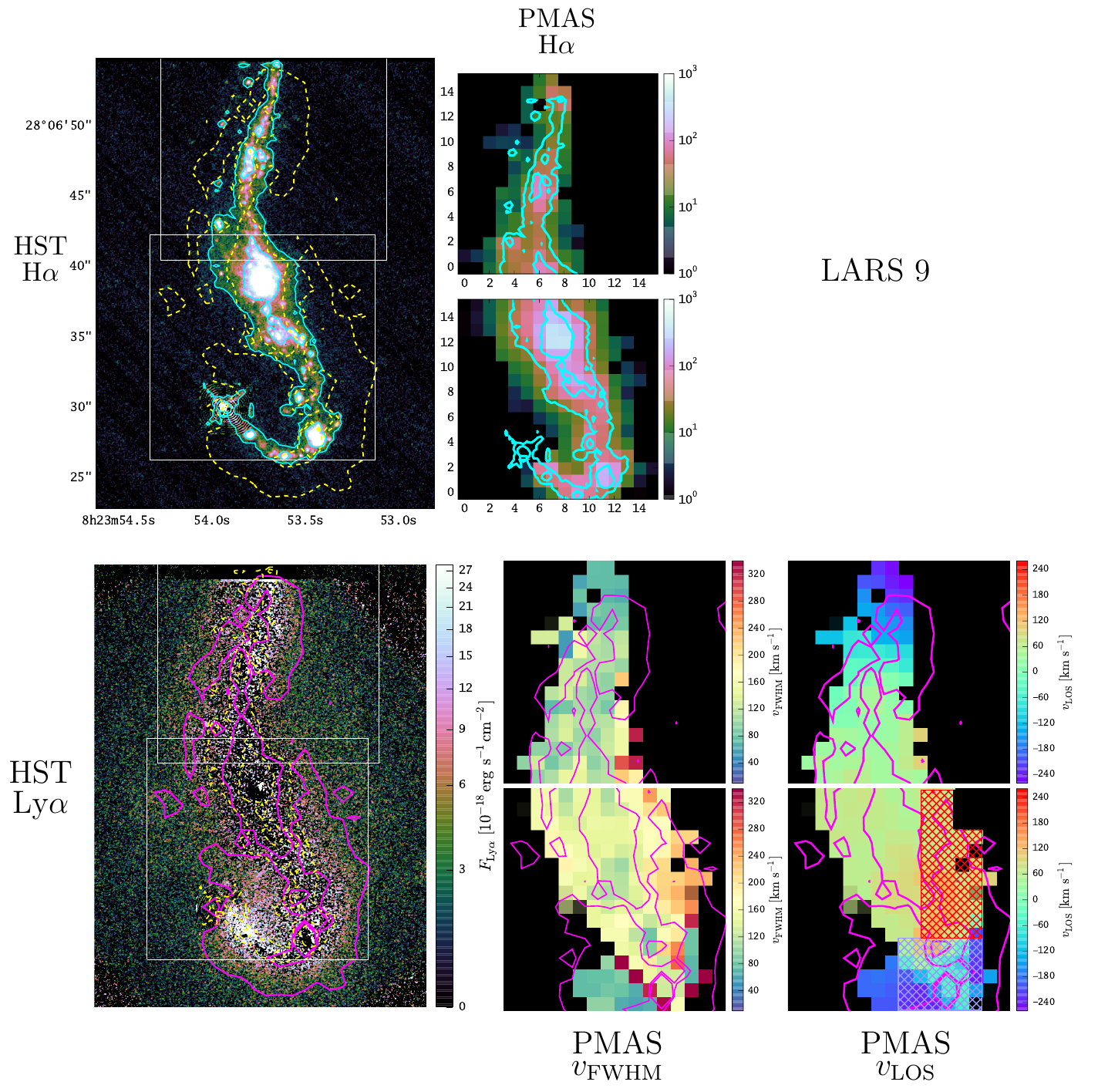}
  \caption{Comparison of LARS HST imaging results to spatially
    resolved PMAS H$\alpha$ spectroscopy for LARS 9. For detailed
    description of individual panels see caption of
    Figure~\ref{fig:maps1}.  This galaxy was covered with 2
    PMAS pointings. Hatched regions in the $v_\mathrm{LOS}$ map 
    indicate regions, where the H$\alpha$ emission shows a more
    complex profile that could not be described by a simple Gaussian
    (cf. Appendix~\ref{sec:lars-9}).}\label{fig:lars9_fig}
\end{figure*}
\begin{figure*}
  \centering
  \includegraphics[width=0.9\textwidth]{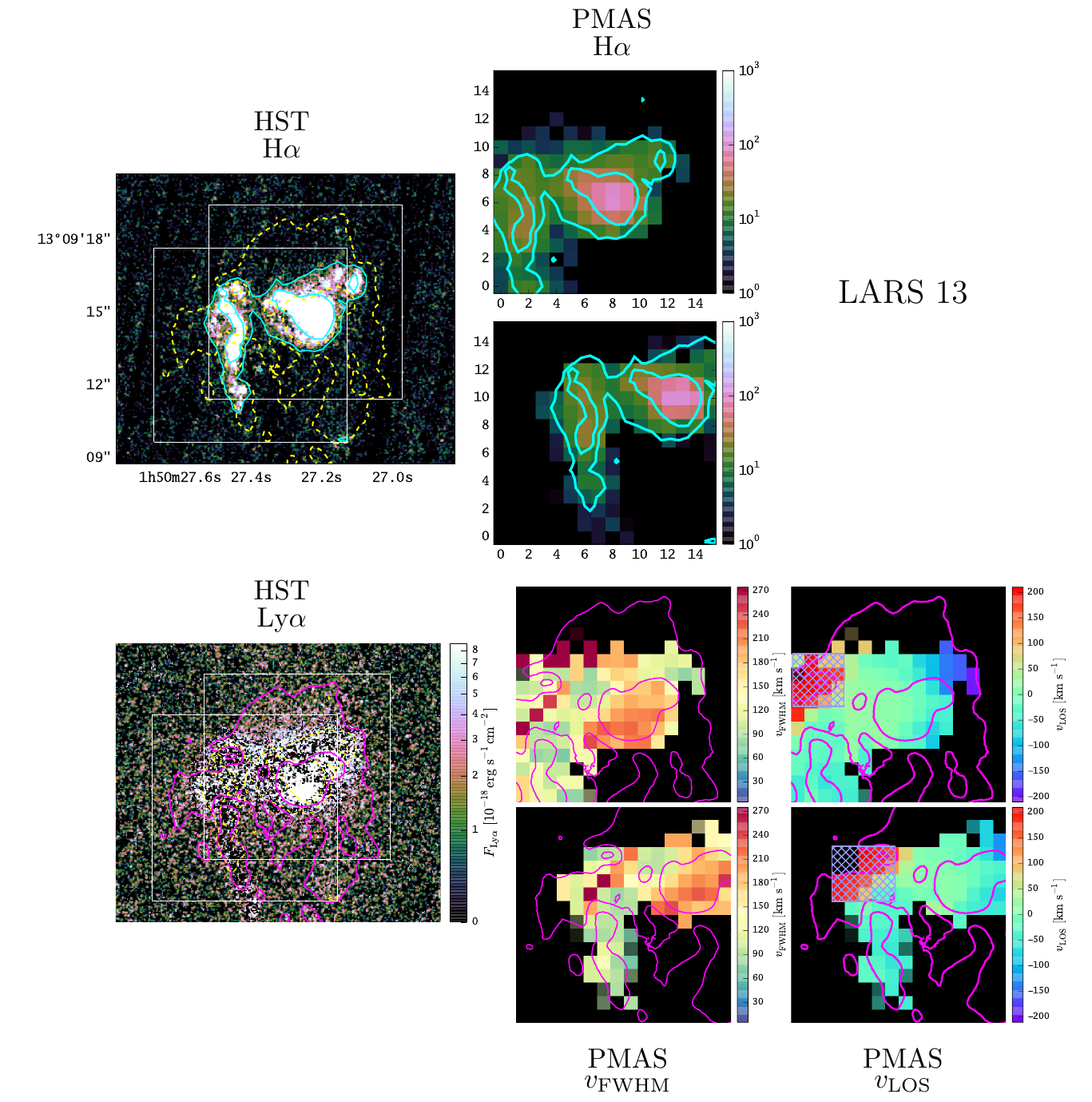}
  \caption{Same as Figure~\ref{fig:lars9_fig}, but for LARS
    13. Hatched region in the $v_\mathrm{LOS}$ map indicates the
    region, where the H$\alpha$ emission shows a more complex profile
    that could not be described by a simple Gaussian
    (cf. Appendix~\ref{sec:lars-13}).}\label{fig:lars13_fig}
\end{figure*}

We condense the kinematical information traced by H$\alpha$ in our
data cubes into two-dimensional maps depicting velocity dispersion and
line of sight radial-velocity at each spaxel. Both quantities are
derived from 1D Gaussian fits to the observed line shapes from the
continuum subtracted cube (see Sect.~\ref{sec:registr-astr-grid}).  We
calculate the radial velocity offset $v_\mathrm{LOS}$ to the central
wavelength value given by the mean of all fits and the FWHM of the
fitted line in velocity space $v_\mathrm{FWHM}$.  To obtain higher
signal to noise ratios in the galaxies' outskirts we use the weighted
Voronoi tessellation binning algorithm by \cite{Diehl2006}, which is a
generalisation of the Voronoi binning algorithm of
\cite{Cappellari2003}. 

We visually scrutinised the fits to the spectra and decided that lines
observed at a SNR of 6 were still reliably fit. Hence, 6 defines the
minimum SNR in the Voronoi binning algorithm.  Furthermore, we adopted
a maximum bin size of 3$\times$3 pixels. In those calculations signal
and noise are defined as in Sect.~\ref{sec:registr-astr-grid}. In what
follows, only results from the fits to bins that meet the minimum
signal to noise requirement are considered.  Most of the bins used are
actually only a single spaxel, and very few bins in the outskirts of
the galaxies meeting the minimum SNR criterion consist of 2 or 3
spaxels.  In practice this means that we are typically sensitive to
regions at H$\alpha$ surface brightness higher than
$5\times10^{-16}$erg\,s$^{-1}$cm$^{-2}$arcsec$^{-2}$. Using the
\cite{Kennicutt1998} conversion and neglecting extinction, this
translates to an detection limit of
$\sim3\times10^{-2}$M$_\odot$yr$^{-1}$kpc$^{-1}$, more than an order
of magnitude deeper than probed by high redshift integral field
spectroscopy studies \citep[$\sim 1$M$_\odot$yr$^{-1}$kpc$^{-1}$, see
e.g.][]{Law2009}.

By visual inspection we also ensured that 1D Gaussian profiles are a
sufficient model of the observed H$\alpha$ profiles seen in the PMAS
LARS data cubes. Only in two galaxies (LARS 9 and LARS 13) we have
some spaxels where a 1D Gaussian certainly fails to reproduce the
complexity of the observed spectral shape. Moreover, in LARS 14 weak
broad wings are visible in the individual H$\alpha$ profiles.  We have
not attempted to model these special cases, but we describe their
qualitative appearance in detail in the
Appendices~\ref{sec:lars-9},~\ref{sec:lars-13} and \ref{sec:lars-14}
and acknowledge their existence in our interpretation in
Sect.~\ref{sec:discussion}. To demonstrate the overall quality of our
H$\alpha$ velocity fields, we compare in Fig.~\ref{fig:full_spect} the
integrated H$\alpha$ profiles of the LARS galaxies to the sum of the
fitted 1D Gaussians used to generate the $v_\mathrm{FWHM}$ and
$v_\mathrm{LOS}$ maps. As can be seen the observed profile and the one
derived from the models are largely in agreement.  The strongest
deviation between data and models is apparent in LARS\,14, where the
broad wings of the observed H$\alpha$ profile are not reproduced at
all by out 1D Gaussian models (cf. Sect.~\ref{sec:local} and
Appendix~\ref{sec:lars-14}).

To estimate formal uncertainties on $v_\mathrm{FWHM}$ and
$v_\mathrm{LOS}$, we use a Monte Carlo technique: We perturb our
spectra 100 times with the noise from our noise cubes and then fit the
1D Gaussian profile in exactly the same way to each of these 100
realisations.  The width of the distribution of all fit results
characterised by their standard deviation gives a robust measure for
the uncertainty on the derived quantities \citep[see also Appendix
B.4. in][]{Davies2011}. The central moment of the distribution is the
final value that will be shown in our maps. In Fig.~\ref{fig:error} we
display for all performed fits the error on the velocity dispersion
$\Delta v_\mathrm{FWHM}$ and on radial velocity
$\Delta v_\mathrm{LOS}$ as a function of the lines SNR.  We
  note, that on average our uncertainties on $v_\mathrm{FWHM}$ and
  $v_\mathrm{LOS}$ follow the scaling laws expected for fitting
  Gaussian profiles to noisy emission-line spectra
  $\Delta v_\mathrm{LOS} \propto \Delta v_\mathrm{FWHM} \propto
  v_\mathrm{FWHM}^{1/2} \mathrm{SNR}^{-1}$
  \citep{Landman1982,Lenz1992}. For reference, at our minimum
$\mathrm{SNR}=6$ we have typical errors of
$\Delta v_\mathrm{FWHM} \approx 30\,$km\,s$^{-1}$ and
$\Delta v_\mathrm{LOS} \approx 10\,$km\,s$^{-1}$, and at the median
(mean) SNR of our data - $\mathrm{SNR}_\mathrm{median} = 22$
($\mathrm{SNR}_\mathrm{mean} = 50$) - we have
$\Delta v_\mathrm{FWHM} \approx 7\,$km\,s$^{-1}$
($\approx 3\,$km\,s$^{-1}$) and
$\Delta v_\mathrm{LOS} \approx 3\,$km\,s$^{-1}$
($\approx 1\,$km\,s$^{-1}$). 

In Fig.~\ref{fig:maps1} we show the resulting $v_\mathrm{FWHM}$ and
$v_\mathrm{LOS}$ maps together with the LaXs Ly$\alpha$ images (lower
three panels).  Maps and images of the galaxies observed with two
pointings are shown in Fig.~\ref{fig:lars9_fig} and
Fig.~\ref{fig:lars13_fig} for LARS 9 and LARS 13, respectively. All
$v_\mathrm{FWHM}$ maps have been corrected for instrumental broadening
using the spectral resolving power maps derived in
Sect.~\ref{sec:determ-spectr-resol}.  In all $v_\mathrm{LOS}$ and
$v_\mathrm{FWHM}$ panels we overlay contours derived from the
Ly$\alpha$ image to emphasise distinctive morphological Ly$\alpha$
properties.

Individual maps are discussed in 
Appendix~\ref{sec:notes-indiv-objects}. Using the classification
scheme introduced by \cite{Flores2006} \citep[see also Sect. 3.3
in ][]{Glazebrook2013} the LARS galaxies can be characterised by the
qualitative appearance of their velocity fields:
\begin{itemize}
\item \emph{Rotating Disks:}  LARS 8 and LARS 11. Both galaxies show a
regular symmetric dipolar velocity field with a steep gradient in the
central regions. This steep gradient is also the reason for
artificially broadened emission near the kinematical
centre. Morphologically both galaxies also bear resemblance to a
disk and the kinematical axis is aligned with the morphological
axis. However, we point out in Appendix~\ref{sec:lars-8} that
LARS 8 can also be classified as a shell galaxy, hinting at a recent
merger event.
\item \emph{Perturbed Rotators:} LARS 1, LARS 3, LARS 5, LARS 7, and
  LARS 10. In those galaxies traces of orbital motion are still
  noticeable, but the maps look either significantly perturbed
  compared to a classical disk case (LARS 3, LARS 5, LARS 7 and LARS
  10) and/or the observed velocity gradient is very weak (LARS 1, LARS
  5 and LARS 10). Based solely on morphology, we previously classified
  two of those galaxies as dwarf edge-on disks \citepalias[LARS 5 and
  LARS 7 - ][]{Hayes2014,Pardy2014}. Moreover, LARS 3 is a member of a
  well known merger of two similar massive disks (cf. Appendix
  ~\ref{sec:lars-3}). From our imaging data alone LARS 7 and LARS 10
  bear resemblance to shell galaxies (cf. Appendices \ref{sec:lars-7}
  and \ref{sec:lars-10}).
\item \emph{Complex Kinematics:} LARS 2, LARS 4, LARS 6, LARS 9, LARS
  12 LARS 13, and LARS 14. These seven galaxies are characterised by
  more chaotic $v_\mathrm{LOS}$ maps and they all have also
  irregular morphologies. Four of the irregulars consist of
  photometrically well separated components that are also
  kinematically distinct (LARS 4, LARS 6, LARS13 and LARS 14), while
  the other three each have their own individual peculiarities (LARS
  2, LARS 9 and most spectacularly LARS 12 -
  cf. Appendices ~\ref{sec:lars-2}, \ref{sec:lars-9} and
  \ref{sec:lars-12}).
\end{itemize}
These kinematic classes are also listed in Table~\ref{tab:kin}.

Notably, in all LARS galaxies, except for LARS 6, we observe
high-velocity dispersions ($v_\mathrm{FWHM}\gtrsim100$\,km\,s$^{-1}$);
the most extreme case is LARS 3 where locally lines as broad as
$v_\mathrm{FWHM} \sim 400$\,km\,s$^{-1}$ are found. The observed width
of the H$\alpha$ line seen in our $v_\mathrm{FWHM}$ maps can be
described as a successive convolution of the natural H$\alpha$ line
(7\,km\,s$^{-1}$ FHWM) with the thermal velocity distribution of the
ionised gas (21.4\,km\,s$^{-1}$ FWHM at 10$^4$K) and with non-thermal
motions in the gas \citep[e.g.][]{Jimenez-Vicente1999}. Since all our
dispersion measurements are significantly higher than the thermally
broadened profile non-thermal motions dominate the observed
line-widths.  Non-thermal motions could be centre-to-centre
dispersions of the individual \ion{H}{ii} regions and turbulent
motions of the ionised gas. The latter appears to be the main driver
for the observed line widths, since the observed velocity dispersions
are highly supersonic \citep[i.e. $\gtrsim10$\,km\,s$^{-1}$ for
\ion{H}{ii} at 10$^4$K, see also Sect. 5.1 in][]{Glazebrook2013}.

\subsection{Global kinematical properties: $v_\mathrm{shear}$,
  $\sigma_0$, $\sigma_\mathrm{tot}$ and $v_\mathrm{shear} / \sigma_0$}
\label{sec:glob-kinem-para}
 
\begin{table*} [t!]
  \centering
  \caption{Global kinematic parameters from H$\alpha$ for the LARS galaxies, calculated as
    described in Sect.~\ref{sec:glob-kinem-para}. For reference we also
    tabulate $\mathrm{EW}_\mathrm{Ly\alpha}$, Ly$\alpha$/H$\alpha$ and $f_\mathrm{esc.}^\mathrm{Ly\alpha}$ 
    from \citetalias{Hayes2014}, as well as $\xi_\mathrm{Ly\alpha}$ from \cite{Hayes2013}.}
  \begin{tabular}{ccccccccccc}  \hline \hline \\[-2ex]
 ID & Class & $\sigma_\mathrm{0}$  & $\sigma_\mathrm{tot}$
    &$v_\mathrm{shear}$  & $v_\mathrm{shear} / \sigma_\mathrm{0}$ & $\mathrm{EW}_{Ly\alpha}$ & Ly$\alpha$/H$\alpha$ & $f_\mathrm{esc.}^\mathrm{Ly\alpha}$ & $\xi_\mathrm{Ly\alpha}$\\ 
    {} & {} & [km\,s$^{-1}$] & [km\,s$^{-1}$] & [km\,s$^{-1}$] & {} & [\AA{}] & {} & {} \\ [.5ex] \hline \\ [-1.5ex]
1  & P &  47.5$\,\pm\,$0.1 & \phantom{1}60.5 &  \phantom{1}56$\,\pm\,$2  & 1.2$\,\pm\,$0.1  & 33.0 & 1.36   & 0.119 & 3.37   \\
2  & C&  38.6$\,\pm\,$0.9  & \phantom{1}46.0 &  \phantom{1}23$\,\pm\,$3  & 0.6$\,\pm\,$0.1  & 81.7 & 4.53   & 0.521 & 2.27   \\
3  & P&  99.5$\,\pm\,$3.7  & 130.9&  138$\,\pm\,$3            & 1.4$\,\pm\,$0.2             & 16.3 & 0.16   & 0.003 & 0.77   \\
4  & C&  44.1$\,\pm\,$0.1  & \phantom{1}55.1 &  \phantom{1}74$\,\pm\,$9  & 1.7$\,\pm\,$0.1  & 0.00 & 0.00   & 0.000 & \dots  \\
5  & P&  46.8$\,\pm\,$0.3  & \phantom{1}54.7 &  \phantom{1}37$\,\pm\,$4  & 0.8$\,\pm\,$0.1  & 35.9 & 2.16   & 0.174 & 2.61   \\
6  & C&  27.2$\,\pm\,$0.3  & \phantom{1}58.9 &  \phantom{1}52$\,\pm\,$7  & 1.9$\,\pm\,$0.2  & 0.00 & 0.00   & 0.000 & \dots  \\
7  & P&  58.7$\,\pm\,$0.3  & \phantom{1}67.4 &  \phantom{1}31$\,\pm\,$3  & 0.5$\,\pm\,$0.1  & 40.9 & 1.94   & 0.100 & 3.37   \\
8  & R&  49.0$\,\pm\,$0.1  & 111.8&  155$\,\pm\,$3            & 3.2$\,\pm\,$0.1             & 22.3 & 0.67   & 0.025 & 1.12   \\
9  & C&  58.6$\,\pm\,$0.1  & 112.7&  182$\,\pm\,$3            & 3.1$\,\pm\,$0.1             & 3.31 & 0.13   & 0.007 & $>$2.85\\
10 & P&  38.2$\,\pm\,$1.0  & \phantom{1}49.2 &  \phantom{1}36$\,\pm\,$3  & 0.9$\,\pm\,$0.2  & 8.90 & 0.47   & 0.026 & 2.08   \\
11 & R&  69.3$\,\pm\,$3.8  & 114.8&  149$\,\pm\,$4            & 2.1$\,\pm\,$0.3             & 7.38 & 0.72   & 0.036 & 2.27   \\
12 & C&  72.7$\,\pm\,$1.0  & \phantom{1}81.3 &  \phantom{1}96$\,\pm\,$3  & 1.3$\,\pm\,$0.1  & 8.49 & 0.48   & 0.009 & 3.48   \\
13 & C&  69.2$\,\pm\,$0.7  & \phantom{1}97.3 &  183$\,\pm\,$4            & 2.5$\,\pm\,$0.2  & 6.06 & 0.29   & 0.010 & 1.74   \\
14 & C&  67.3$\,\pm\,$1.3  & \phantom{1}72.9 &  \phantom{1}40$\,\pm\,$1  & 0.6$\,\pm\,$0.1  & 39.4 & 2.24   & 0.163 & 3.62   \\ [.5ex] \hline
  \end{tabular}
  \tablefoot{Class refers to the three kinematic classes proposed by
    \cite{Flores2006}: C = complex kinematics, P = 
    perturbed rotators and R =  rotating disks.}
  \label{tab:kin}
\end{table*}

We now quantify the global kinematical properties of our H$\alpha$
velocity fields that were qualitatively discussed in the previous
section. Therefore we compute three non-parametric estimators that are
commonly adopted in the literature for this purpose
\citep[e.g.][]{Law2009,Basu-Zych2009,Green2010,Glazebrook2013,Gonccalves2010,Green2014,Wisnioski2015}:
Shearing velocity $v_\mathrm{shear}$, intrinsic velocity dispersion
$\sigma_0$ and their ratio $v_\mathrm{shear}/\sigma_0$. We describe
the physical meaning of those parameters and our method to determine
them in the following three subsections. Furthermore, we also
calculate the integrated velocity dispersion $\sigma_\mathrm{tot}$.
This measure provides a useful comparison for unresolved distant
galaxies (cf. Sect.~\ref{sec:integr-veloc-disp}).  Our results on
$v_\mathrm{shear}$, $\sigma_0$, $\sigma_\mathrm{tot}$ and
$v_\mathrm{shear}/\sigma_0$ are tabulated in Table~\ref{tab:kin}.

\subsubsection{Shearing velocity $v_\mathrm{shear}$}
\label{sec:shear-veloc-v_mathrm}

The shearing velocity $v_\mathrm{shear}$ is a measure for the
large-scale gas bulk motion along the line of sight.  We calculate it
via
\begin{equation}
  \label{eq:1}
v_\mathrm{shear}= \frac{1}{2} \left ( v_\mathrm{max} - v_\mathrm{min}
\right )  \;\text{.}
\end{equation}
Here we take the median of the lower and upper 5 percentile of the
distribution of values in the $v_\mathrm{LOS}$ maps for
$v_\mathrm{min}$ and $v_\mathrm{max}$, respectively. This choice
ensures that the calculation is robust against outliers while sampling
the true extremes of the distribution. We conservatively estimate the
uncertainty by propagating the full width of the upper and lower 5
percentile of the velocity map.  We list our results for
$v_\mathrm{shear}$ in the second column of Table~\ref{tab:kin}.  Our
derived $v_\mathrm{shear}$ values for the LARS sample range from
$30$\,km\,s$^{-1}$ to $180$\,km\,s$^{-1}$, with 65\,km\,s$^{-1}$ being
the median. This is less than half the typical maximum velocity
$v_\mathrm{max}$ of H$\alpha$ rotation curves observed in spiral
galaxies - e.g.  $v_\mathrm{max} = 145$\,km\,s$^{-1}$ is the median of
153 local spirals \citep{Epinat2010}.  In contrast to our
$v_\mathrm{shear}$ measurement, $v_\mathrm{max}$ values are
inclination corrected. In a sample with random inclinations on average
the correction
$v_\mathrm{shear} = \frac{\pi}{4} v_\mathrm{max} \approx 0.79
\,v_\mathrm{max}$
is expected \citep[e.g.][]{Law2009} but even with this correction
classical disks appear still incompatible with most of our
measurements by a factor larger than two. Only for the two
  galaxies that we classified as rotating disks in
  Sect. \ref{sec:halpha-veloc-fields} we find $v_\mathrm{shear}$ values
  compatible with local rotators.  Nevertheless, high
$v_\mathrm{shear}$ values do not necessarily imply the presence of a
disk. Large-scale bulk motions at high amplitude occur also in close
encounters of spatially and kinematically distinct companions
(e.g. LARS 13).

Sensitivity is an important factor in the determination of the
observed $v_\mathrm{shear}$. An observation of less depth will not
detect fainter regions at larger radii, which may have the highest
velocities. Indeed, for most of our galaxies the spatial positions of
$v_\mathrm{min}$ and $v_\mathrm{max}$ values are in those
outer lower surface-brightness regions. Hence observations
of lower sensitivity will be biased to lower $v_\mathrm{shear}$
values.  This sensitivity bias is strongly affecting high-$z$ studies
because such observations often limited only to the brightest regions
\citep{Law2009,Gonccalves2010}.

\subsubsection{Intrinsic velocity dispersion $\sigma_0$}
\label{sec:local-veloc-disp}

The intrinsic velocity dispersion $\sigma_0$ \citep[sometimes also
called resolved velocity dispersion or local velocity dispersion, cf.
][Sect. 1.2.1]{Glazebrook2013} is in our case a measure for the
typical random motions of the ionised gas within a galaxy.  We
calculate it by taking the flux weighted average of the observed
velocity dispersions in all spaxels:
\begin{equation}
  \label{eq:2}
\sigma_0 = \frac{\sum F^\mathrm{H\alpha}_\mathrm{spaxel} \sigma_\mathrm{spaxel}}{
\sum F^\mathrm{H\alpha}_\mathrm{spaxel}  } \; \text{.}
\end{equation}
Here $F^\mathrm{H\alpha}_\mathrm{spaxel}$ is the H$\alpha$ flux in
each spaxel and $\sigma_\mathrm{spaxel}$ is the velocity dispersion
measured in that spaxel; the sum runs over all spaxels with
$\text{SNR}\geq6$ (Sect.~\ref{sec:halpha-veloc-fields}).  Such a
flux-weighted summation $\sigma_0$ has been widely adopted in the
literature to quantify the typical intrinsic velocity dispersion
component
\cite[e.g.][]{Oestlin2001,Law2009,Gonccalves2010,Green2010,Glazebrook2013,Green2014}.

PSF smearing in the presence of strong velocity gradients broadens the
H$\alpha$ line at the position of the gradients\footnote{This effect
  is similar to beam-smearing in radio interferometric imaging
  observations.}. Such a broadening could bias our calculation of the
intrinsic velocity dispersion.  At the distances of the LARS galaxies
our typical seeing-disk PSF FWHM of 1\arcsec{} corresponds to scales
of 0.6\,kpc to 3\,kpc. In most of our observations this average PSF
FWHM is comparable to the size of one PMAS spaxel
(1\arcsec{}$\times$1\arcsec{}) and most of the LARS galaxies exhibit
rather weak velocity gradients. The PSF smearing effect is seen only
in galaxies with strong velocity gradients (e.g. LARS 8, cf. Appendix
~\ref{sec:lars-8}), especially when observed with PMAS'
0.5\arcsec{}$\times$0.5\arcsec{} magnification mode (e.g. LARS 12,
cf. Appendix~\ref{sec:lars-12}).  In principle the effect could be
corrected by subtracting a PSF-smeared kinematical model of the
$v_\mathrm{LOS}$ velocity field.  Given the complexity
of our observed velocity fields, however, this is not practicable. And,
moreover, at $z\sim0.1$ \cite{Green2014} find that for classical
rotating disk kinematics the velocity dispersions corrected by a
kinematical model are, on average, negligible. Comparison between
adaptive-optics derived PSF-smearing unaffected $\sigma_0$ values to
those derived from seeing-limited observations were presented by
\cite{Bassett2014}.  They find that while for one galaxy $\sigma_0$
remains unaffected by an increase in spatial resolution, for the other
galaxy $\sigma_0$ decreases by $\sim$10\,km\,s$^{-1}$.
\cite{Bassett2014} attribute the $\sim$10\,km\,s$^{-1}$ discrepancy in
one of their galaxies to a strong velocity gradient that is co-spatial
with the strongest emission. A similar comparison is possible for
three of our LARS galaxies - LARS 12, LARS 13 and LARS 14. For these
galaxies adaptive optics IFS observations were presented by
\cite{Gonccalves2010} (see Appendices~\ref{sec:lars-12},
\ref{sec:lars-13} and \ref{sec:lars-14}).  \cite{Gonccalves2010}
derive $\sigma_0=67$\,km\,s$^{-1}$, $\sigma_0=74$\,km\,s$^{-1}$ and
$\sigma_0=71$\,km\,s$^{-1}$ for LARS 12, LARS 13 and LARS 14,
respectively. These values are in agreement with our measurements
of $\sigma_0=73$\,km\,s$^{-1}$, $\sigma_0=69$\,km\,s$^{-1}$ and
$\sigma_0=67$\,km\,s$^{-1}$ for those galaxies\footnote{We note that
  for LARS 13 \cite{Gonccalves2010} do not sample the entire galaxy in
  their observations - see Sect.~\ref{sec:lars-13}.}.  From all these
considerations we are certain, that the adopted formalism in
Eq.~(\ref{eq:2}) gives a robust estimate of the intrinsic velocity
dispersion.

We list our derived values for $\sigma_0$ in the third column of
Table~\ref{tab:kin}.  In general, the ionised gas kinematics of the
LARS galaxies are characterised by high intrinsic velocity dispersions
ranging from 40\,km\,s$^{-1}$ to 100\,km\,s$^{-1}$ with
54\,km\,s$^{-1}$ being the median of the sample.  Such high intrinsic
dispersions are in contrast to \ion{H}{ii} velocity dispersions of
$\sim 10 - 50$\,km\,s$^{-1}$ typically found in local spirals
\citep[e.g.][]{Epinat2008,Epinat2008a,Epinat2010,Erroz-Ferrer2015}. In
high-$z$ star-forming galaxies intrinsic dispersions
$\gtrsim50$\,km\,s$^{-1}$ appear to be the norm
\citep[e.g.][]{Puech2006,Genzel2006,Law2009,FoersterSchreiber2009,Erb2014,Wisnioski2015}.
In the local $z\lesssim0.1$ Universe high intrinsic velocity
dispersions are also found in the studies of blue compact galaxies
\citep[BCGs,][]{Oestlin1999,Oestlin2001}, local Lyman break analogues
\citep{Basu-Zych2009,Gonccalves2010} and (ultra-)luminous infrared
galaxies \citep[e.g][]{Monreal-Ibero2010}. Finally, $\sigma_0$ values
of comparable amplitude are also common in a sample of 67 bright
H$\alpha$ emitters
($10^{40.6} \leq L_\mathrm{H\alpha} \leq 10^{42.6}$)
\citep{Green2010,Green2014}. Indeed, \cite{Green2010} and
\cite{Green2014} show that star formation rate is positively
correlated with intrinsic velocity dispersion. We will confirm this
trend among our LARS galaxies and discuss its implications on
Ly$\alpha$ observables in Sect.~\ref{sec:global}.

\subsubsection{$v_\mathrm{shear} / \sigma_0$ ratio}
\label{sec:v_mathrmsh--sigm}

The ratio $v_\mathrm{shear} / \sigma_0$ is a metric to quantify
whether the gas kinematics are dominated by turbulent or ordered (in
some cases orbital) motions. Our results for this quantity are listed
in Table~\ref{tab:kin}.  Given the above
(Sects.~\ref{sec:shear-veloc-v_mathrm} and~\ref{sec:local-veloc-disp})
mentioned differences to disks, it appears cogent that our
$v_\mathrm{shear}/\sigma_0$ ratios (median 1.4, mean 1.6) are much
smaller than typical disk $v_\mathrm{shear}/\sigma_0$ ratios
$\sim 4 - 8 $ (first to last quartile of the \citeauthor{Epinat2010}
sample).

Objects with $v_\mathrm{shear}/\sigma_0 < 1$ are commonly labelled
dispersion dominated galaxies \citep{Newman2013,Glazebrook2013}.  -
According to this criterion five galaxies of the 14 LARS sample are
dispersion dominated: LARS 2, LARS 5, LARS 7, LARS 10 and LARS 14 -
three perturbed rotators and two with complex kinematics.  Dispersion
dominated galaxies are frequently found in kinematic studies of
high-$z$ star forming galaxies.  Local samples with a similar high
fraction ($\sim35$\%) of dispersion dominated galaxies are the blue
compact galaxies studied by \cite{Oestlin2001} or the Lyman break
analogues studied by \cite{Gonccalves2010} . In Sect.~\ref{sec:global}
we will show that dispersion dominated systems are more likely to have
a significant fraction of Ly$\alpha$ photons escaping.

\subsubsection{Integrated velocity dispersion $\sigma_\mathrm{tot}$}
\label{sec:integr-veloc-disp}

For reference we also compute the integrated (spatially averaged)
velocity dispersion $\sigma_\mathrm{tot}$. This measure provides a
useful comparison for studies of unresolved distant galaxies, where no
disentanglement between ordered and disordered, random motions is
possible
\citep[e.g.][]{McLinden2011,Guaita2013,McLinden2014,Rhoads2014,Erb2014}.

We obtain $\sigma_\mathrm{tot}$ by calculating the square root of the
second moment of the integrated H$\alpha$ profiles shown in
Fig.~\ref{fig:full_spect}. The results are given in the fourth column
of Table~\ref{tab:kin}. Our integrated velocity dispersions for the
LARS sample range from 46\,km\,s$^{-1}$ to 115\,km\,s$^{-1}$, with
70\,km\,s$^{-1}$ being the median of the sample.  Given the high SNR
of the integrated profiles the formal uncertainties on this
$\sigma_\mathrm{tot}$ are very small ($\sim 10^{-2}$km\,s$^{-1}$).  As
expected, when large scale motions dominate in the integrated spectrum
(i.e. high $v_\mathrm{shear}/\sigma_0$ ratios), $\sigma_\mathrm{tot}$
is significantly larger than $\sigma_0$, but for dispersion dominated
systems the discrepancy becomes less extreme.

\section{Discussion: Influence of H$\alpha$ kinematics on a galaxy's
  Ly$\alpha$ emission}
\label{sec:discussion}

\subsection{Clues on Ly$\alpha$ escape mechanisms via spatially resolved
  H$\alpha$ kinematics }
\label{sec:local}

Recent theoretical modelling of the Ly$\alpha$ radiative transport
within realistic interstellar medium environments predicts that small
scale interstellar medium physics are an decisive factor in regulating
the Ly$\alpha$ escape from galaxies
\citep{Verhamme2012,Behrens2014}. In particular \cite{Behrens2014}
demonstrate how supernova blown outflow cavities become favoured
escape channels for Ly$\alpha$ radiation. These cavities naturally
occur in zones of enhanced star formation activity, where the input of
kinetic energy from supernovae and stellar winds into the surrounding
interstellar medium is also expected to drive highly turbulent gas
flows. Do we see supporting evidence for this scenario in our IFS
observations of the LARS galaxies?

As described in Sect.~\ref{sec:halpha-veloc-fields}, all our velocity
fields are characterised by broad profiles that are best explained by
turbulent flows of ionised gas. The more violent these flows are, the
more likely they carve holes or bubbles through the interstellar
medium through which the ionised gas eventually flows out into the
galaxy's halo. Close inspection of Fig.~\ref{fig:maps1} reveals, that
in LARS 1, LARS 3 and LARS 13 (Fig.~\ref{fig:lars13_fig})  zones of
high Ly$\alpha$ surface brightness occur near or even cospatial to
zones where for these galaxies the maximum velocity dispersions are
observed. In both LARS 1 and LARS 13 these zones are also co-spatial
with the regions of highest H$\alpha$ surface-brightness, hence here
the star formation rate density is highest. Moreover, in LARS 1 the
H$\alpha$ image shows a diffuse fan-like structure emanating outwards
from this star-forming knot
\citepalias[][]{Hayes2014,Ostlin2014}. Finally, also the analysis of
the COS spectra in \citetalias{Rivera-Thorsen2015} shows blue-shifted
low-ionisation state interstellar absorption lines in LARS 1 and LARS
13 indicative of outflowing neutral gas.  Hence for those two galaxies
together with the high-resolution HST imaging data our IFS data
provides further evidence for the relevance of localised outflows in
shaping the observed Ly$\alpha$ morphology.

However, already in LARS 3 the interpretation is not as
straight-forward.  Here the broadest H$\alpha$ lines do not occur
co-spatial with the highest star formation rate density. Although also
here the COS spectrum shows a bulk outflow of cold neutral gas
\citep[][]{Rivera-Thorsen2015}, the locally enhanced dispersion values
seen in the PMAS maps appear not to be related to this outflow.
Instead, it appears more naturally that the high turbulence within the
ionised gas is not primarily driven by star formation activity, but is
rather the result of the violent interaction with the north-western
companion (see e.g. \citealt{Teyssier2010} for theoretical support).
The interpretation is further complicated by the fact, that the HST
imaging data resolves the local Ly$\alpha$ knot on smaller scales then
which we are probing with our PMAS observations.

Nevertheless, there is another example among the LARS galaxies for
Ly$\alpha$ escape being related to an outflow: LARS 5 (cf. Duval, A\&A
submitted). For this galaxy we presumed already in
\citetalias{Hayes2014} the existence of a starburst driven wind based
solely on the filamentary structure seen in the HST H$\alpha$
image. Our dispersion map now reveals that these filaments fill the
interior of a biconical structure, with its base confined on the most
prominent star-forming clump (cf. Appendix~\ref{sec:lars-5}) -- fully
in accordance with theoretical expectations for an evolved starburst
driven superwind \citep[e.g.][]{Cooper2008}.  Also here Ly$\alpha$
photons could preferentially escape the disc through the cavity blown
by the wind. The strongest Ly$\alpha$ enhancement is found directly
above and below the main star-forming clump and the low surface
brightness Ly$\alpha$ halo is then produced by those photons
scattering on a bipolar shell-like structure of ambient neutral gas
swept up by the superwind. Unfortunately,  the brightest hot-spots in
Ly$\alpha$ above and below the plane are at the base of this
outflowing cone and occur at scales that we cannot resolve in our PMAS
data.  Similar but less spectacular examples of elevated velocity
dispersions above and below a disk can be seen in LARS 7 and LARS 11,
with both galaxies are also being embedded in low-surface brightness
Ly$\alpha$ halos.

Another possible example for the importance of outflow kinematics in
facilitating direct Ly$\alpha$ escape is the most luminous LAE in the
sample: LARS 14 (cf. Appendix~\ref{sec:lars-14}). Here the observed
H$\alpha$ profiles are always characterised by an underlying fainter
broad component (see Fig.~\ref{fig:full_spect}) that is believed to be
directly related to outflowing material
\citep[e.g.][]{Yang2015}. 

We conclude that while in some individual cases a causal connection
between spatially resolved H$\alpha$ kinematics and localised outflow
scenarios might be conjectured, globally this is not a trend seen in
our observations.  We note further, that H$\alpha$ kinematics tell
only one part of the whole story, and the suggested outflow scenarios
should also be traceable by gas with a high degree of ionisation.  At
least for one galaxy with Ly$\alpha$ imaging similar to the LARS
galaxies,  such a connection has been demonstrated:
ESO338-IG04.  Recently, \cite{Bik2015} analysed observations of this
galaxy obtained with the MUSE integral field spectrograph
\citep{Bacon2014}. They show, that the Ly$\alpha$ fuzz seen around the
main star-forming knot of this galaxy \citep{Hayes2005,Ostlin2009} can
be related to an outflow which can be traced both with H$\alpha$
kinematics and by a high degree of ionisation. As a difference to our
observations, \cite{Bik2015} see the fast outflowing material in their
$v_\mathrm{LOS}$ map as significantly redshifted H$\alpha$
emission. No similar prominent effect is apparent in our
$v_\mathrm{LOS}$ maps. Nevertheless, the $v_\mathrm{LOS}$ fields of
LARS 7 and LARS 12 display strongly localised redshifts in H$\alpha$
that are spatially coincident with filamentary H$\alpha$ fingers,
hence here we might see also a fast outflow pointed away from the
observer. But also in those galaxies, no co-spatial relation between
Ly$\alpha$ emissivity and the suspected outflows can be established.

\subsection{Comparison of H$\alpha$ to \ion{H}{i} observations}
\label{sec:comp-hi-observ}

\begin{figure}
  \centering
  \includegraphics[width=0.5\textwidth]{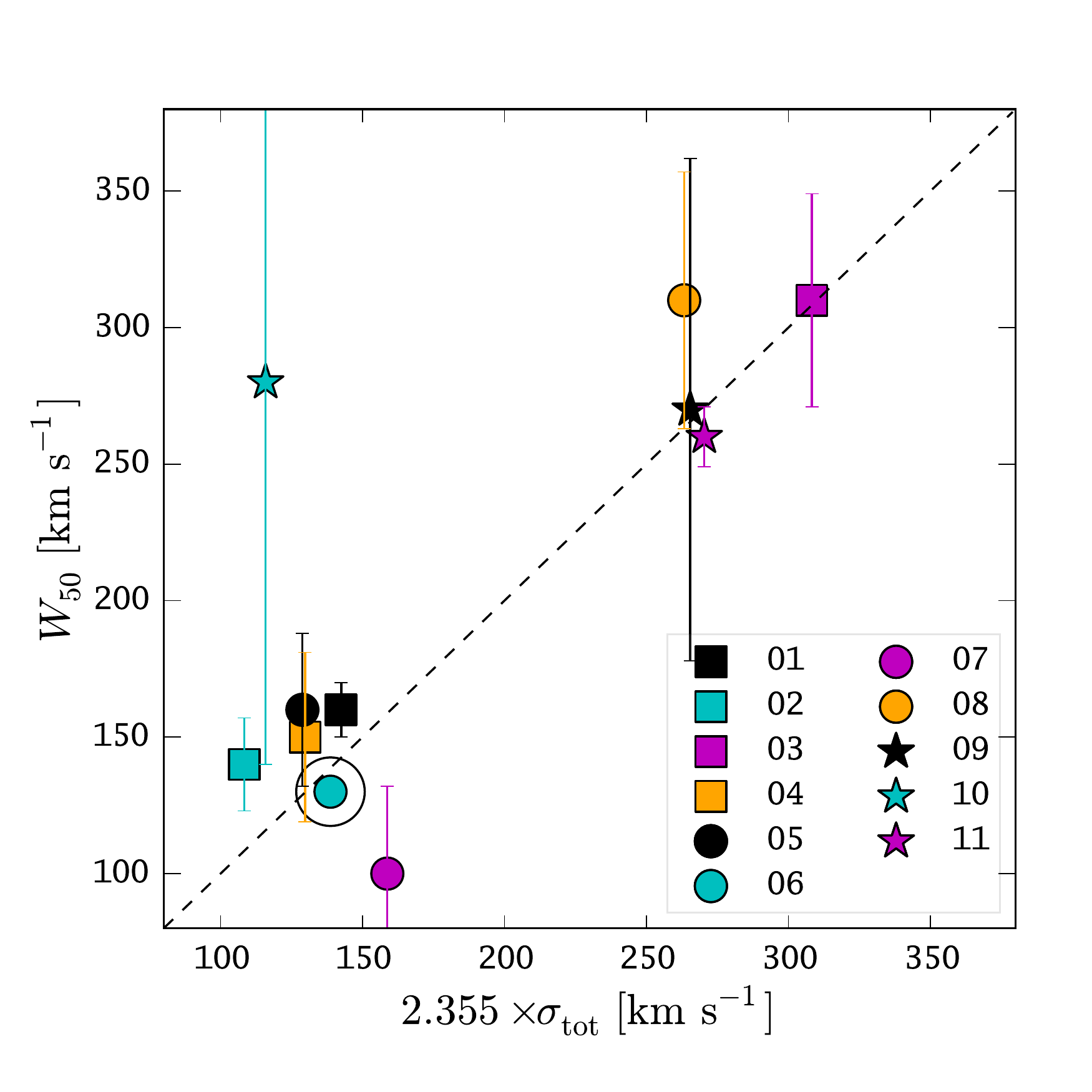}\vspace{-1em}
  \caption{Comparison of integrated \ion{H}{i} linewidth $W_{50}$ to
    FWHM of the H$\alpha$ integrated velocity dispersion
    $2\sqrt{2\ln 2}\times \sigma_\mathrm{tot}$. The dashed line
    indicates the one-to-one relation. We note, that for LARS 6
    (encircled point) we plot as $W_{50}$ a preliminary result based
    upon VLA C-configuration interferometry, as the beam of the GBT
    profile (used to derive $W_\mathrm{50}$ in \citetalias{Pardy2014})
    is too broad to separate LARS 6 from a neighbouring galaxy.
  }
  \label{fig:w50tot}
\end{figure}

Radiative transfer of Ly$\alpha$ photons depends on the relative
velocity differences between scattering \ion{H}{i} atoms and
Ly$\alpha$ sources.  If the Ly$\alpha$ sources are out of resonance
with the bulk of the neutral medium, they are less likely scattered,
and in turn more likely to escape the galaxy.  It is exactly this
interplay between the ionised and neutral ISM phases that is
determining the whole the Ly$\alpha$ radiative transfer.  In
\citetalias{Pardy2014} we studied LARS galaxies in the \ion{H}{i}
21\,cm line, using single-dish GBT and VLA D-configuration
observations (cf. Sect.~\ref{sec:ancill-lars-datapr}).  We now
attempt a comparison between the neutral and ionised gas kinematics in
the LARS galaxies, cognisant of the fact that the spatial scales
probed by both instruments are much larger than our PMAS \ion{H}{ii}
observations.

Generally our GBT \ion{H}{i} spectra have low SNR, and for the
three most distant LARS galaxies -- LARS 12, LARS 13 and LARS 14 -- we
could not detect any significant signal at all. The  profiles
are mostly single- or multiple-peaked, but rarely show a classical
double-horn profile that would be expected for a flat rotation
curve. Hence, qualitatively our observed GBT \ion{H}{i} line profiles
are consistent with our PMAS results that most of the LARS galaxies
are kinematically perturbed or sometimes even strongly interacting
systems.

In \citetalias{Pardy2014} we measured the width of the \ion{H}{i}
lines at 50\% of the line peak from the GBT spectra.  In
Fig.~\ref{fig:w50tot} we compare this quantity, $W_{50}$, to the FWHM
of the integrated H$\alpha$ velocity dispersion $\sigma_\mathrm{tot}$
(Sect.~\ref{sec:integr-veloc-disp}). Notably, two systems deviate
significantly from the one-to-one relation: LARS 7 and LARS 10.  LARS
10 shows the lowest SNR \ion{H}{i} spectrum and moreover our PMAS FoV
does not capture two smaller star-forming clumps in the
south-east. Therefore we believe that observational difficulties are
source for the $\sigma_\mathrm{tot}$--$W_{50}$ difference in this
galaxy.  In LARS 7, however, we suspect the difference to be
genuine. In this galaxy the H$\alpha$ morphology is significantly
puffed up and rounder compared to the disk-like continuum
(cf. Appendix~\ref{sec:lars-7}). Therefore we suspect that the smaller
$W_{50}$ measurement indicates that bulk of \ion{H}{i} is in a
kinematically more quiescent state then the ionised gas. This galaxy
is one of the stronger LAEs in the sample
($f_\mathrm{esc.}^\mathrm{Ly\alpha}=0.14$ and
EW$_\mathrm{Ly\alpha}$=40\AA{}).  That considerable amounts of
Ly$\alpha$ photons escape from LARS 7 was noted as peculiar in
\citetalias{Rivera-Thorsen2015}, since the metal absorption lines
indicated large amounts of neutral gas sitting at the systemic
velocity of the galaxy. Our observations now indicate that the
intrinsic Ly$\alpha$ photons are produced in gas that is less
quiescent than the scattering medium, which therefore is more
transparent for a significant fraction of the intrinsic Ly$\alpha$
photons.

Spatially resolved VLA velocity fields are available  for only a subset
of five LARS galaxies (LARS 2, LARS 3, LARS 4, LARS 8, LARS 9). They
represent rotations, disturbed by interactions with neighbours.  In
four of them, \ion{H}{i} kinematical axes have similar orientations as
those of the PMAS H$\alpha$ velocity field. In the fifth object, the
complex interacting system LARS 9, the H$\alpha$ and \ion{H}{i}
velocity fields show different characteristics, but the complexities
seen in the PMAS maps of this galaxy are on scales far beyond the
resolving power of the VLA D-configuration
(cf. Appendix~\ref{sec:lars-9}).  However, some of the LARS galaxies
have already been observed with the VLA in C- and B-configuration and
the analysis is currently in progress.  These data will allow for the
first time a comparison between H$\alpha$, \ion{H}{i} and Ly$\alpha$
on meaningful physical scales. It is evident, that these comparisons
will provide a critical benchmark for our understanding of Ly$\alpha$
radiative transfer in interstellar and circum-galactic environments.

\subsection{Relations between kinematical properties and galaxy
  parameters}
\label{sec:global}

\begin{figure*}
  \centering
  \includegraphics[width=\textwidth]{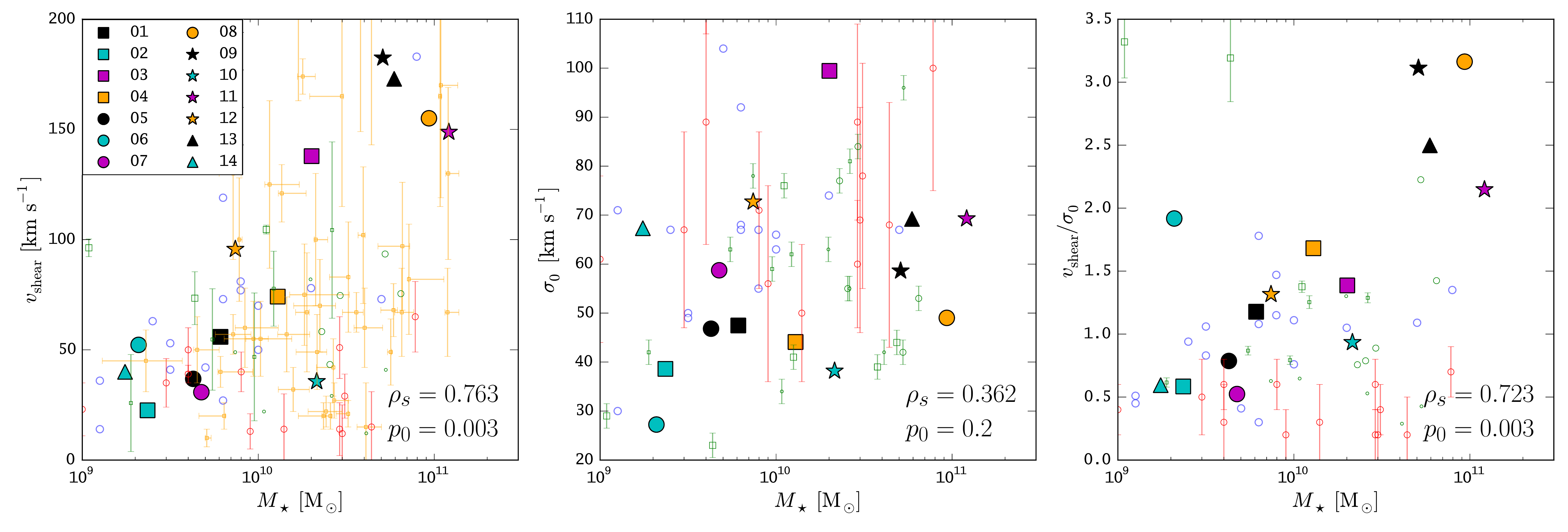}
  \caption{Global H$\alpha$ kinematical parameters $v_\mathrm{shear}$
    (\emph{left panel}), $\sigma_0$ (\emph{middle panel}) and
    $v_\mathrm{shear}/\sigma_0$ (\emph{right panel}) in comparison to
    stellar mass for LARS galaxies \citep[from][]{Hayes2014} and in
    comparison to literature values: DYNAMO $z\sim0.1$ [compact]
    perturbed rotators from \cite{Green2014} as \emph{[small] green
      circles}; DYNAMO $z\sim0.1$ [compact] complex kinematics from
    \cite{Green2014} as \emph{[small] green squares}; Local Lyman
    Break Analogues from \cite{Gonccalves2010} as \emph{blue circles};
    Keck/OSIRIS resolved $z\sim2-3$ star forming galaxies from
    \cite{Law2009} as \emph{red circles}; SINS $z\sim2-3$ galaxies
    from \cite{FoersterSchreiber2009} as \emph{orange squares} (only
    in the left panel).  Uncertainties given where available. The LARS
    galaxies are represented by symbols according to the legend in the
    left panel.  Spearman rank correlation coefficients $\rho_s$ and
    corresponding $p_0$ values for the LARS galaxies are shown in each
    panel.}
  \label{fig:masstrends}
\end{figure*}
\begin{figure*}
  \centering
  \includegraphics[width=\textwidth]{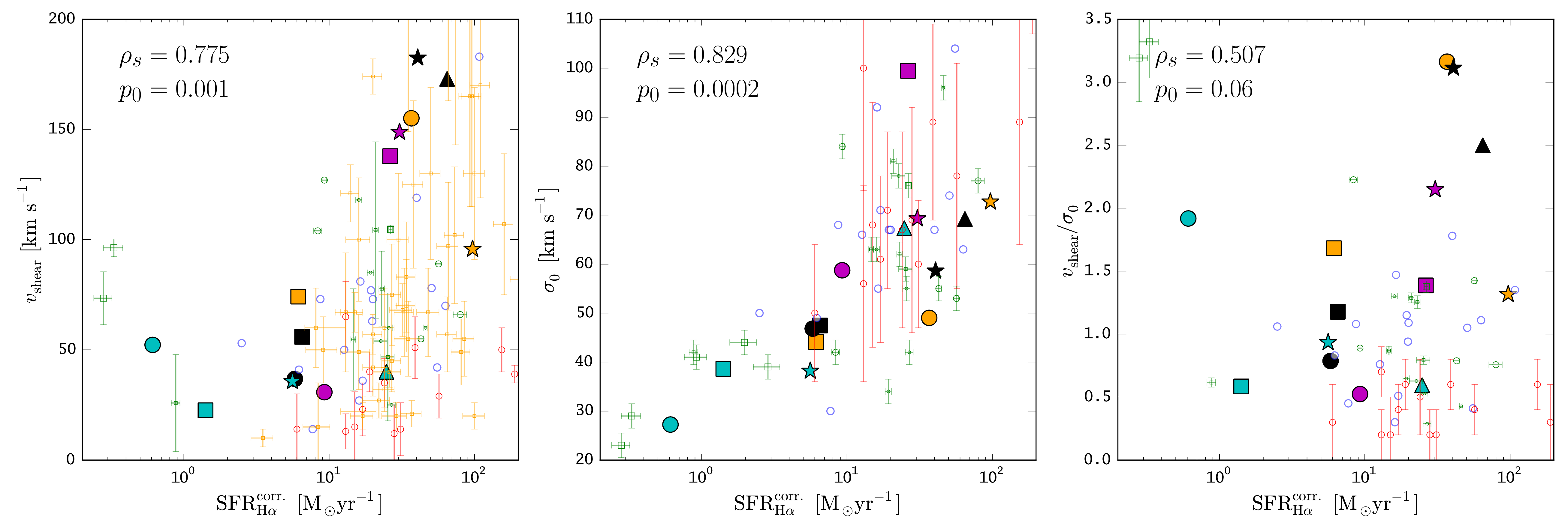}
  \caption{Global H$\alpha$ kinematical parameters $v_\mathrm{shear}$
    (\emph{left panel}), $\sigma_0$ (\emph{middle panel}) and
    $v_\mathrm{shear}/\sigma_0$ (\emph{right panel}) in comparison to
    star formation rate for LARS galaxies \citep[from][]{Hayes2014}
    and in comparison to literature values (same symbols as in
    Fig.~\ref{fig:masstrends}).  Uncertainties given where
    available.  Spearman rank correlation coefficients $\rho_s$ and
    corresponding $p_0$ values for the LARS galaxies are shown in each
    panel.}
  \label{fig:sfrtrends}
\end{figure*}

In Sect.~\ref{sec:glob-kinem-para} we quantified the global
kinematical properties of the LARS galaxies using the non-parametric
estimators $v_\mathrm{shear}$, $\sigma_0$ and
$v_\mathrm{shear} / \sigma_0$. Before linking these observables to
global Ly$\alpha$ properties of the LARS galaxies
(cf. Sect.~\ref{sec:relat-betw-glob}), we need to understand which 
galaxy parameters are encoded in them. 

We find strong correlations between stellar mass $M_\star$ and
$v_\mathrm{shear}$, as well as star formation rate (SFR) and
$\sigma_0$ ($M_\star$ and SFR from
\citetalias{Hayes2014}). Graphically we show these correlations in
Fig.~\ref{fig:masstrends} (left panel) and Fig.~\ref{fig:sfrtrends}
(centre panel). The Spearman rank correlation coefficients
\citep[e.g.][]{Wall1996} are $\rho_s=0.763$ for the
$M_\star$-$v_\mathrm{shear}$ relation and $\rho_s=0.829$ for the
SFR-$\sigma_0$ relation. This corresponds to likelihoods of the
null-hypothesis that no monotonic relation exists between the two
parameters of $p_0=0.02$\% and $p_0=0.3$\% (two-tailed
test\footnote{We consider correlations as significant when the
  likelihood of the null-hypothesis, $p_0$, is smaller than 5\%}) for
the SFR-$\sigma_0$ and $M_\star$-$v_\mathrm{shear}$ relation,
respectively. We discuss these two relations in more detail in
Sect.~\ref{sec:sfr-sigm-corr} and
Sect.~\ref{sec:m_st-v_mathrmsh-corr}, where we also introduce the
comparison samples shown in Fig.~\ref{fig:masstrends} and
Fig.~\ref{fig:sfrtrends}.

Besides this tight correlation between SFR and $\sigma_0$, there is
also a weaker correlation between SFR and $v_\mathrm{shear}$ in our
sample ($\rho_s =0.775$, $p_0=0.1$\%). We show the data in
Fig.~\ref{fig:sfrtrends} (left panel).  While not shown graphically,
we note that the DYNAMO disks would scatter over the whole
SFR-$v_\mathrm{shear}$ plane, also filling the upper left corner in
this diagram. Finally, in Fig.~\ref{fig:sfrtrends} (right panel) we
find that $v_\mathrm{shear}/\sigma_0$ is not significantly correlated
with the SFR in LARS ($\rho_s=0.507$, $p_0=6$\%).

We also check in Fig.~\ref{fig:masstrends} (centre panel) for a
$M_\star$-$\sigma_0$ correlation, but with $\rho_s = 0.362$ the
null-hypothesis that the variables are uncorrelated can not be
rejected ($p_0=20$\%). Similar low correlation-coefficients are found
for the comparison samples in Fig.~\ref{fig:masstrends} (centre
panel). Since there is likely a monotonic relation between
$M_\star$-$v_\mathrm{shear}$ and 
$M_\star$-$\sigma_0$ appear uncorrelated, the monotonic relation 
$M_\star$-$v_\mathrm{shear}/\sigma_0$ ($\rho_s=0.723$, $p_0=0.3$\%) -
Fig.~\ref{fig:masstrends} (right panel) - is expected. 

From these results we conclude, that dispersion-dominated galaxies in
our sample are preferentially low mass systems with
$M_\star \lesssim 10^{10}$M$_\odot$. This result is also commonly
found in samples of high-$z$ star-forming galaxies
\citep{Law2009,FoersterSchreiber2009,Newman2013}.

\subsubsection{SFR-$\sigma_0$ correlation}
\label{sec:sfr-sigm-corr}

The highly significant correlation between $\sigma_0$ and SFR has long
been established for giant \ion{H}{ii} regions \citep{Terlevich1981}. Recently
it has been shown that it extents over a large dynamical range in star
formation and mass, not only locally but also at high redshifts
\citep{Green2010, Green2014}. The physical nature of the $\sigma_0$-SFR
relation might be that star formation feedback powers turbulence in
the interstellar medium. On the other hand, the processes that lead to
a high SFR might also be responsible to produce a high $\sigma_0$. In
particular inflows of cold gas that feed the star formation processes
are expected to stir up the interstellar medium and thus lead to
turbulent flows \citep[e.g.][and references therein]{Wisnioski2015}. 

Graphically we compare in Fig.~\ref{fig:sfrtrends} (centre panel) the
LARS SFR-$\sigma_0$ points to the values from the DYNAMO galaxies
\citep{Green2014} with complex kinematics and the DYNAMO perturbed
rotators. To put this result in context with high-$z$ studies, we also
show in Fig.~\ref{fig:sfrtrends} the SFR-$\sigma_0$ points from the
Keck/OSIRIS $z\sim2-3$ star-forming galaxies by \cite{Law2009}, and
the local Lyman break analogues by \cite{Gonccalves2010}. An
exhaustive compilation of is presented in \cite{Green2014}, and the
LARS SFR-$\sigma_0$ points do not deviate from this relation.

\subsubsection{$M_\star$-$v_\mathrm{shear}$ correlation}
\label{sec:m_st-v_mathrmsh-corr}

Given the complexity of the ionised gas velocity fields seen in the
LARS galaxies, a tight relation between our measured,
inclination-uncorrected $v_\mathrm{shear}$ values with stellar mass
appears surprising.  It indicates that, at least in a statistical
sense, $v_\mathrm{shear}$ is tracing systemic rotation in our systems
and that the scatter in our relation is dominated by the unknown
inclination correction \citep[see also][]{Law2009}.

To put our data in context to other studies, we compare our
$M_\star$-$v_\mathrm{shear}$ data points in Fig.~\ref{fig:masstrends}
(left panel) to the \cite{Green2014} DYNAMO sample and to the
\cite{Gonccalves2010} local Lyman break analogues. Our high-$z$
comparison samples are the Keck/OSIRIS $z\sim2-3$ star-forming
galaxies \citeauthor{Law2009} and the $z\sim1-3$ SINS sample
by \cite{FoersterSchreiber2009}.  We note that for the
rotation-dominated and perturbed rotators in the DYNAMO sample
\citeauthor{Green2014} tabulate rotation velocity at 2.2 disc scale
lengths obtained from fitting model discs to their velocity fields.
We convert these to an inclination-uncorrected value by multiplication
with $\sin i$.  Again, for the DYNAMO sample we compare only to their
objects with complex kinematics or their perturbed rotators, as they
are dominant in our sample.

It is apparent from Fig.~\ref{fig:masstrends} (left panel) that while
our data points line up well with the $z\sim0.1$ galaxies, a
significant number of high-$z$ galaxies shows lower $v_\mathrm{shear}$
values at given stellar mass in the range
$10^{10}$M$_\odot$-$10^{11}$M$_\odot$.  Reason is, that high-$z$
studies are not sensitive enough to reach the outer faint isophotes,
hence they are biased towards lower $v_\mathrm{shear}$ values (see
Sect.~\ref{sec:shear-veloc-v_mathrm}). Indeed \cite{Law2009} report a
detection limit of $\sim 1$\,M$_\odot$yr$^{-1}$kpc$^{-1}$, while
\cite{FoersterSchreiber2009} report
$\approx 0.03$\,M$_\odot$yr$^{-1}$kpc$^{-1}$ as an average for their
sample, which is comparable to our depth
(Sect.~\ref{sec:halpha-veloc-fields}).

\subsection{Relations between global kinematical properties and
  Ly$\alpha$ observables}
\label{sec:relat-betw-glob}

\begin{figure*}
  \centering
  \includegraphics[width=\textwidth]{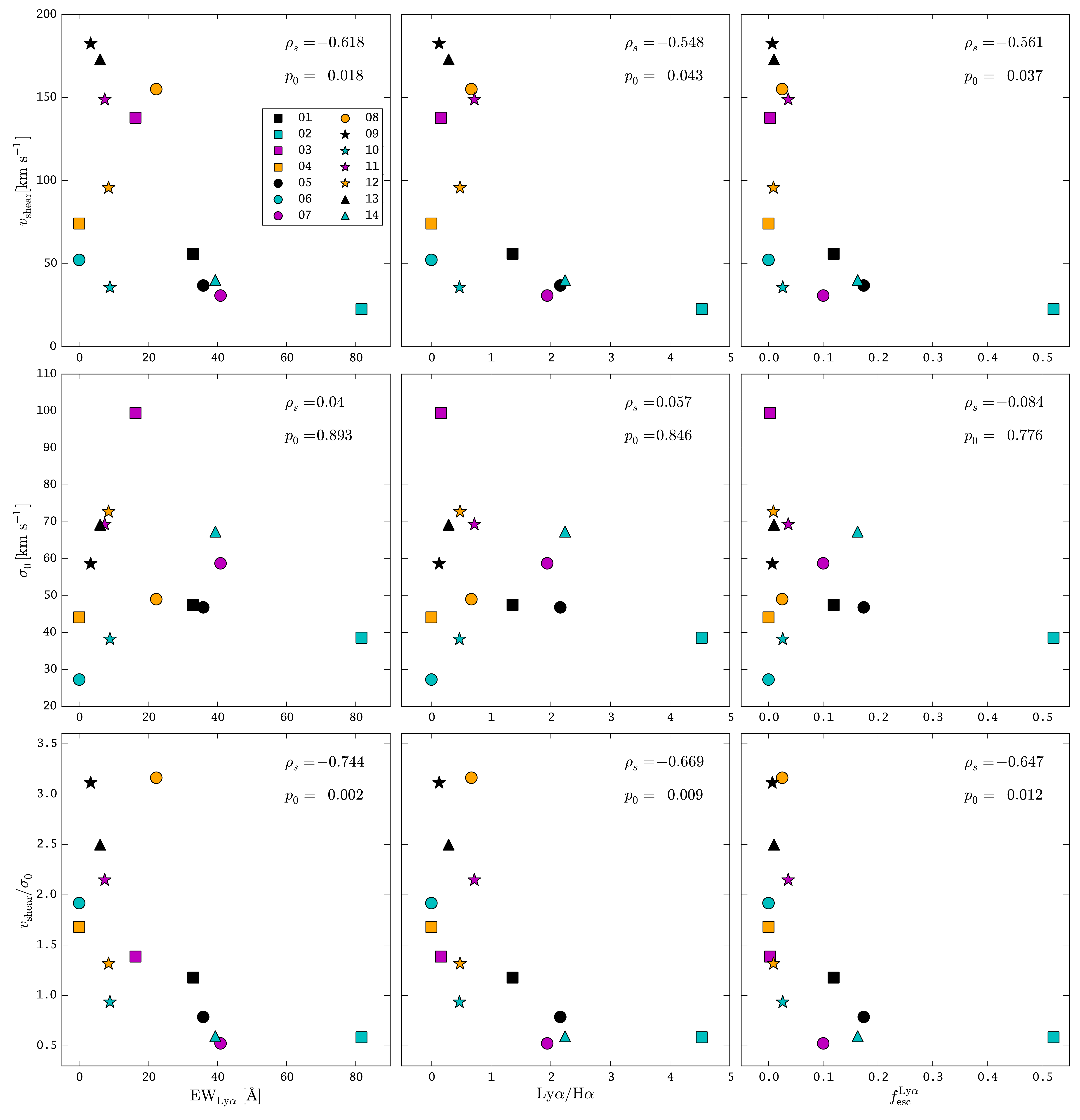}\vspace{-1em}
  \caption{Relations between global Ly$\alpha$ properties of the LARS
    galaxies to global kinematical parameters
    $v_\mathrm{shear}$, $\sigma_0$ and $v_\mathrm{shear}/\sigma_0$}
  \label{fig:ratios}
\end{figure*}

We now explore trends between global kinematical properties derived
from the ionised gas in the LARS galaxies and their Ly$\alpha$
observables determined in \citetalias{Hayes2014}, namely Ly$\alpha$
escape fraction $f_\mathrm{esc}^\mathrm{Ly\alpha}$, ratio of
Ly$\alpha$ to H$\alpha$ flux $\mathrm{Ly}\alpha/\mathrm{H}\alpha$, and
Ly$\alpha$ equivalent width $\mathrm{EW}_\mathrm{Ly\alpha}$. Fore
reference, we list these quantities here again in Table\ref{tab:kin}.
We recall that Ly$\alpha$/H$\alpha$ is the observed flux ratio, while
$f_\mathrm{esc}$ is determined from the intrinsic luminosities,
i.e. after correcting the fluxes for dust-reddening.

Regarding the qualitative classification of the $v_\mathrm{LOS}$ maps
in Sect.~\ref{sec:halpha-veloc-fields} into rotating disks, perturbed
rotators and objects showing complex kinematics there is no preference
in any of the globally integrated Ly$\alpha$ observables.  Both the
maximum and the minimum of these observables occur in complex
kinematics class. Therefore, just the qualitative appearance of the
velocity field seems not to predict whether the galaxy is a LAE or
not.

In the previous section we showed that within the LARS sample higher
SFRs are found in galaxies that have higher $\sigma_0$ and higher
$v_\mathrm{shear}$ measurements and that galaxies of higher mass show
also higher $v_\mathrm{shear}$ values and higher
$v_\mathrm{shear}/\sigma_0$ ratios (Fig.~\ref{fig:masstrends} and
Fig.~\ref{fig:sfrtrends}).  We now compare our $v_\mathrm{shear}$,
$\sigma_0$ and $v_\mathrm{shear}/\sigma_0$-ratios to the aperture
integrated Ly$\alpha$ observables $\mathrm{EW}_\mathrm{Ly\alpha}$,
Ly$\alpha$/H$\alpha$ and $f_\mathrm{esc}^\mathrm{Ly\alpha}$ from
\citetalias{Hayes2014}.  We do this in form of a graphical 3$\times$3
matrix in Fig.~\ref{fig:ratios}. In each panel we include the
Spearman rank correlation coefficient $\rho_s$ and the likelihood
$p_0$ to reject the null-hypothesis.

From the centre row in Fig. \ref{fig:ratios} it is obvious that none
of the Ly$\alpha$ observables correlates with the averaged intrinsic
velocity dispersion $\sigma_0$. As $\sigma_0$ correlates positively
with SFR, this signifies that the observed Ly$\alpha$ emission is a
bad star formation rate calibrator. In Fig.~\ref{fig:ratios} it is
also evident, that galaxies with higher shearing velocities
($v_\mathrm{shear}\gtrsim50$\,km\,s$^{-1}$) have preferentially lower
$\mathrm{EW}_\mathrm{Ly\alpha}$, lower Ly$\alpha$/H$\alpha$ and lower
$f_\mathrm{esc}^\mathrm{Ly\alpha}$.  Therefore, according to the above
presented $M_\star$-$v_\mathrm{shear}$ relation
(Sect.~\ref{sec:m_st-v_mathrmsh-corr}) Ly$\alpha$ emitters are
preferentially found among the systems with
$M_\star \lesssim 10^{10}$M$_\odot$ in LARS. This deficiency of strong
Ly$\alpha$ emitters among high mass systems was already noted in
\citetalias{Hayes2014}. Here we see this trend now from a kinematical
perspective. Also, a low $v_\mathrm{shear}$ value, although a
necessary condition, seems not to be sufficient to have significant
amounts Ly$\alpha$ photons escaping (e.g LARS 6). Again, this shows
that Ly$\alpha$ escape from galaxies is a complex multi-parametric
problem.

Although the LARS sample is small, the result that 7 of 8 non-LAEs
have $v_\mathrm{shear}/\sigma_0 > 1$ and 4 of 6 LAEs have
$v_\mathrm{shear}/\sigma_0 < 1$ signals that dispersion-dominated
kinematics are an important ingredient in Ly$\alpha$ escape. Again,
lower $v_\mathrm{shear}/\sigma_0$ ratios are found in lower $M_\star$
objects in our sample and $\sigma_0$ is uncorrelated with
$M_\star$. Therefore the correlation between Ly$\alpha$ observables
and $v_\mathrm{shear}/\sigma_0$ is a consequence of the correlation
between Ly$\alpha$ observables and $M_\star$.  Nevertheless, low
$v_\mathrm{shear}/\sigma_0$ ratios also state that the ionised gas in
those galaxies must be in a turbulent state. High SFR appears to be
connected to an increase turbulence, but it is currently not clear
whether the processes that cause star formation or feedback from star
formation are responsible for the increased turbulence
(Sect.~\ref{sec:sfr-sigm-corr}). Regardless of the causal relationship
between SFR and $\sigma_0$, our results support that Ly$\alpha$ escape
is being favoured in low-mass systems undergoing an intense star
formation episode. Similarly \cite{Cowie2010} found that in a sample
of UV bright $z\sim0.3$ galaxies LAEs are primarily the young
galaxies that have recently become strongly star forming.  In the
local Universe dispersion-dominated systems with
$v_\mathrm{shear} / \sigma_0 < 1$ are rare, but they become much more
prevalent at higher redshifts \citep{Wisnioski2015}. Coincidentally,
the number density of LAEs rises towards higher redshifts
\citep{Wold2014}. Therefore we speculate, that dispersion dominated
kinematics are indeed a necessary requirement for a galaxy to have a
significant amount of Ly$\alpha$ photons escaping.

\subsection{Ly$\alpha$ extension and H$\alpha$ kinematics}
\label{sec:lyman-alpha-extens}

\begin{figure}\vspace{-1em}
  \centering
  \includegraphics[width=0.4\textwidth]{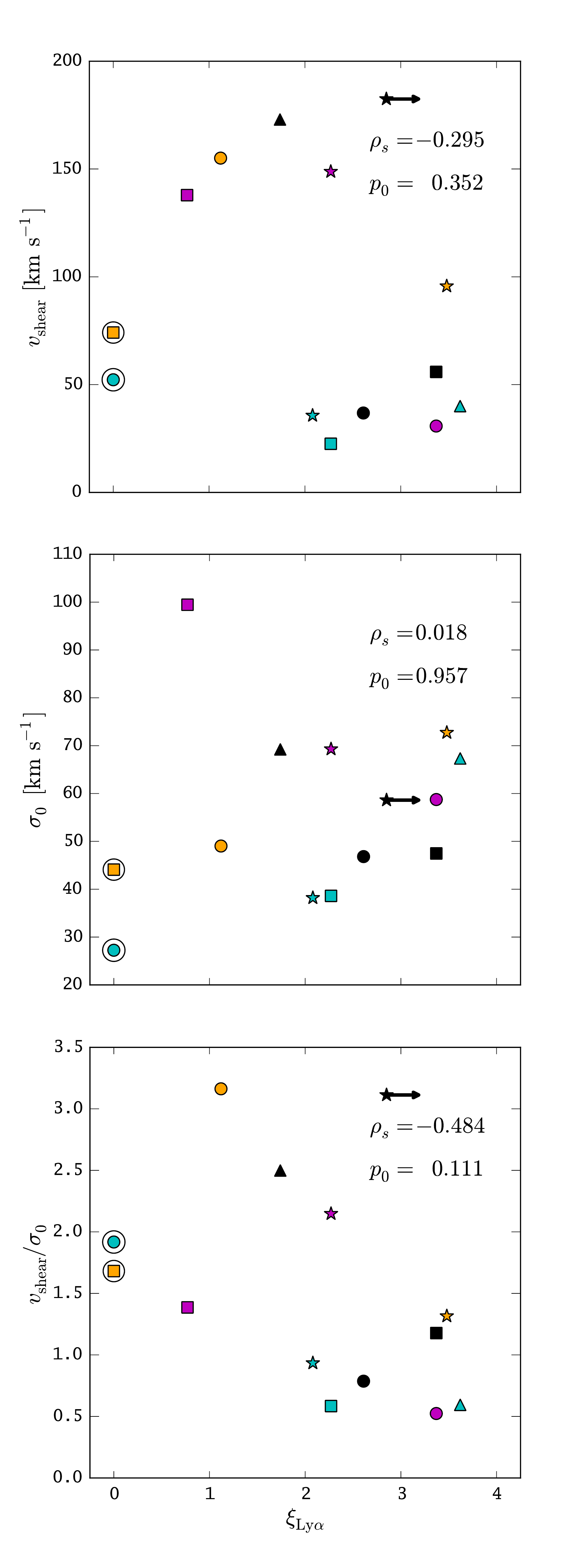}\vspace{-1em}
  \caption{Relations between $\xi_\mathrm{Ly\alpha}$ -- the
    \emph{relative Petrosian extension of Ly$\alpha$} as defined in
    \cite{Hayes2013} -- and global kinematical parameters
    $v_\mathrm{shear}$, $\sigma_0$ and $v_\mathrm{shear}/\sigma_0$
    with symbols according to the legend in
    Fig.~\ref{fig:masstrends}. For galaxies with no Ly$\alpha$
    emission (LARS 4 and LARS 6, circled symbols)
    $\xi_\mathrm{Ly\alpha}$ is defined as zero and for LARS 9 the
    measured $\xi_\mathrm{Ly\alpha}$ presents a lower limit
    \citep[cf.][]{Hayes2013}.  Spearman rank correlation coefficients
    $\rho_s$ and corresponding $p_0$ values are calculated excluding
    the LARS 4 and LARS 6.}
  \label{fig:xicorr}
\end{figure}

All of the LARS LAEs show a significantly more extended
  Ly$\alpha$ morphology compared to their appearance in H$\alpha$ or
  UV continuum.  These large-scale Ly$\alpha$ haloes appear to
  completely encompass the star-forming regions. LARS 1, LARS 2, LARS
  5, LARS 7, LARS 12 and LARS 14 are the most obvious examples of
  extended Ly$\alpha$ emission, but the phenomenon is visible in all
  our objects -- even for those galaxies that show Ly$\alpha$ in
  absorption (Fig.~\ref{fig:maps1}, \ref{fig:lars9_fig} and
  \ref{fig:lars13_fig}; cf. \cite{Hayes2013} and
  \citetalias{Hayes2014}). At high redshift the ubiquity of extended
  Ly$\alpha$ haloes around LAEs was recently revealed by
  \cite{Wisotzki2015} on an individual object-by-object
  basis. However, in contrast to the LARS Ly$\alpha$ haloes, the
  high-$z$ haloes appear to have $\sim$10$\times$ larger extents at a
  given continuum radius \citep[][their Fig.~12]{Wisotzki2015}. In
  order to quantify the spatial extent of observed relative to
  intrinsic Ly$\alpha$ emission we defined in \cite{Hayes2013} the
  \emph{Relative Petrosian Extension} $\xi_\mathrm{Ly\alpha}$ as the
  ratio of the Petrosian radii at $\eta=20\%$ \citep{Petrosian1976}
  measured in Ly$\alpha$ and H$\alpha$:
  $\xi_\mathrm{Ly\alpha} = R_\mathrm{P20}^\mathrm{Ly\alpha} /
  R_\mathrm{P20}^\mathrm{H\alpha}$.
  For reference we list the $\xi_\mathrm{Ly\alpha}$ values of the LARS
  galaxies here again in Table~\ref{tab:kin}. Based on an
  anti-correlation between $\xi_\mathrm{Ly\alpha}$ and UV slope we
  conjectured that smaller Ly$\alpha$ haloes occur in galaxies that
  have converted more of their circum-galactic neutral gas into
  forming stars and subsequently dust \citep{Hayes2013}.  In such a
  scenario the larger extent of high-$z$ Ly$\alpha$ haloes could be
  related to the circum-galactic gas-reservoirs being larger in the
  early universe.

In Fig.~\ref{fig:xicorr} we show that there are no significant
correlations between $\xi_\mathrm{Ly\alpha}$ and the global
kinematical H$\alpha$ parameters $v_\mathrm{shear}$, $\sigma_0$ and
$v_\mathrm{shear} / \sigma_0$.  Therefore we reason that the
kinematics of the interstellar medium do not strongly influence the
appearance of the haloes and that the Ly$\alpha$ halo phenomenon is
only related to the presence of a full gas-reservoir. Further 21cm HI
imaging of the LARS galaxies at high spatial resolution is needed to
test this scenario.

\section{Summary and conclusions}
\label{sec:summary-conclusions}

We obtained the following results From our integral field
spectroscopic observations of the H$\alpha$ line in the LARS galaxies:
\begin{enumerate}
\item Half of the LARS galaxies show complex H$\alpha$ kinematics.
  Only two galaxies have kinematical properties consistent with a
  rotating disk and in five a disturbed rotational signature is
  apparent. With respect to Ly$\alpha$ escape we find no preference of
  high EW$_\mathrm{Ly\alpha}$ or high
  $f_\mathrm{esc.}^\mathrm{Ly\alpha}$ values towards any of those
  classes but the minimum and maximum values occur both in objects
  showing complex kinematics in our sample.
\item A common feature in all LARS galaxies are high H$\alpha$
  velocity dispersions. With
  $v_\mathrm{FWHM} \gtrsim 100$\,km\,s$^{-1}$ our measurements are in
  contrast to values typically seen in local spirals, but in high-$z$
  star-forming galaxies such high values appear to be the norm.
\item While we could not infer a direct relation between spatially
  resolved kinematics of the ionised gas and photometric Ly$\alpha$
  properties for all LARS galaxies, in individual cases our maps
  appear qualitatively consistent with outflow scenarios that promote
  Ly$\alpha$ escape from high-density regions.
\item Currently a spatially resolved comparison of our \ion{H}{ii}
  velocity fields to our \ion{H}{i} data is severely limited, since
  the scales resolved by the radio observations are significantly
  larger. However, for one galaxy (LARS 7) the difference between
  globally integrated velocity dispersion of ionised and neutral gas
  offers a viable explanation for the escape of significant amounts of
  Ly$\alpha$ photons.
\item From our H$\alpha$ velocity maps we derive the non-parametric
  statistics $v_\mathrm{shear}$, $\sigma_0$ and
  $v_\mathrm{shear}/\sigma_0$ to quantify the kinematics of the LARS
  galaxies globally. Our $v_\mathrm{shear}$ values range from
  30\,km\,s$^{-1}$ to 180\,km\,s$^{-1}$ and our $\sigma_0$ values
  range from 40\,km\,s$^{-1}$ to 100\,km\,s$^{-1}$. For our ratios
  $v_\mathrm{shear}/\sigma_0$ we find a median of 1.4 and five of the
  LARS galaxies are dispersion-dominated systems with
  $v_\mathrm{shear}/\sigma_0<1$.
\item Fully consistent with other IFS studies $\sigma_0$ is positively
  correlated with star formation rate in the LARS galaxies. We also
  find strong correlations between $M_\star$ and $v_\mathrm{shear}$,
  $M_\star$ and $v_\mathrm{shear}/\sigma_0$, and SFR and
  $v_\mathrm{shear}$. In view of the $M_\star$-$v_\mathrm{shear}$ the
  SFR- $v_\mathrm{shear}$ correlation implies that more massive
  galaxies have higher overall star formation rates in our sample.
\item The Ly$\alpha$ properties EW$_\mathrm{Ly\alpha}$,
  Ly$\alpha$/H$\alpha$ and $f_\mathrm{esc.}^\mathrm{Ly\alpha}$ do not
  correlate with $\sigma_0$, but they correlate with $v_\mathrm{shear}$ and
  $v_\mathrm{shear}/\sigma_0$ (Fig.~\ref{fig:ratios}). We find no
  correlation between the global kinematical statistics  and the extent
  of the Ly$\alpha$ halo.
\item 4 of 6 LARS LAEs are dispersion-dominated systems with
  $v_\mathrm{shear}/\sigma_0 < 1$ and 7 of 8 non-LAEs have
  $v_\mathrm{shear}/\sigma_0 > 1$.
\end{enumerate}

Observational studies of Ly$\alpha$ emission in local-universe
galaxies have so far focused on imaging and UV spectroscopy \citep[for
a recent review see][]{Hayes2015}. Here we provided for the first time
empirical results from IFS observations for a sample of galaxies with
known Ly$\alpha$ observables. In our pioneering study we focused on
the spectral and spatial properties of the intrinsic Ly$\alpha$
radiation field as traced by H$\alpha$.  We found a direct relation
between the global non-parametric kinematical statistics of the
ionised gas and the Ly$\alpha$ observables
$f_\mathrm{esc.}^\mathrm{Ly\alpha}$ and
$\mathrm{EW}_\mathrm{Ly\alpha}$, and our main result is that
dispersion dominated systems favour Ly$\alpha$ escape.  The prevalence
of LAEs among dispersion dominated galaxies could be simply a
consequence of these systems being the lower mass systems in our
sample -- a result already found in \citetalias{Hayes2014}.  However,
the observed turbulence in actively star-forming systems should be
related to ISM conditions that ease Ly$\alpha$ radiative transfer out
of high density environments.  In particular, if turbulence is a
direct consequence of star formation, then the energetic input from
the star formation episode might also be powerful enough to drive
cavities through the neutral medium into the lower density
circum-galactic environments.  Of course, the kinematics of the
ionised gas offer only a limited view of the processes at play and in
future studies we attempt to connect our kinematic measurements to
spatial mappings of the ISMs ionisation state. The synergy between IFS
data, space based UV imaging and high resolution \ion{H}{i}
observations of nearby star forming galaxies will be vital to build a
coherent observational picture of Ly$\alpha$ radiative transport in
galaxies.

\begin{acknowledgements} 
  E.C.H. especially thanks Sebastian Kamann and Bernd Husemann for
  teaching him how to operate the PMAS instrument.  We thank the
  support staff at Calar Alto observatory for help with the
  visitor-mode observations. All plots in this paper were created
  using \texttt{matplotlib} \citep{Hunter2007}.  Intensity related
  images use the the \texttt{cubhelix} colour scheme by
  \cite{Green2011}. This research made extensive use of the
  \texttt{astropy} pacakge \citep{AstropyCollaboration2013}.
  M.H. acknowledges the support of the Swedish Research Council,
  Vetenskapsr\aa{}det and the Swedish National Space Board (SNSB) and
  is Academy Fellow of the Knut and Alice Wallenberg Foundation. HOF
  is currently granted by a C\'atedra CONACyT para J\'ovenes
  Investigadores. I.O. has been supported by the Czech Science
  Foundation grant GACR 14-20666P. DK is funded by the Centre National
  d'\'{E}tudes Spatiales (CNES). FD is grateful for financial support
  from the Japan Society for the Promotion of Science (JSPS) fund. PL
  acknowledges support from the ERC-StG grant EGGS-278202.
\end{acknowledgements}

\appendix

\section{Notes on individual objects}
\label{sec:notes-indiv-objects}

In the following we detail the observed H$\alpha$ velocity fields for
each galaxy. We compare our maps to the photometric Ly$\alpha$
properties derived from the HST images from \citetalias{Hayes2014} and \ion{H}{i}
observations from \citetalias{Pardy2014}
(see Sect.~\ref{sec:ancill-lars-datapr}).

\subsection{LARS 1 (\object{Mrk 259})}
\label{sec:lars-1}

With $L_\mathrm{Ly\alpha} = 8 \times 10^{41}$\,erg\,s$^{-1}$ and
$\mathrm{EW}_\mathrm{Ly\alpha}=33$\,\AA{} LARS 1 is a strong
Ly$\alpha$ emitting galaxy. A detailed description of the photometric
properties of this galaxy was presented in \citetalias{Ostlin2014}. If
LARS 1 would be at high-$z$, it would easily be selected in conventional
narrow-band imaging surveys \citepalias{Guaita2015}. The galaxy shows
a highly irregular morphology. Its main feature is UV-bright complex
in the north-east that harbours the youngest stellar population.  As
noted already in \citetalias{Hayes2014} filamentary structure
emanating from the north-eastern knot is seen in H$\alpha$.  Towards
the south-east numerous smaller star-forming complexes are found that
blend in with an older stellar population.

Contrasting the irregular appearance, the line of sight velocity field
of LARS 1 is rather symmetric. The velocity field is consistent with a
rotating galaxy. The kinematical centre is close to centre of the PMAS
FoV and the kinematical axis appears to run from the north-west to the
south-east. We measure $v_\mathrm{shear}=56\pm1$ km\,s$^{-1}$. The GBT
single dish \ion{H}{i} spectrum of this source is reminiscent of a classic
double-horn profile, also an obvious sign of rotation, with a peak
separation consistent with our $v_\mathrm{shear}$ measurement.

The highest velocity dispersions ($v_\mathrm{FWHM} \approx 150$
km\,s$^{-1}$) are found in the north-eastern region. Here LARS 1 also
shines strong in Ly$\alpha$. But while a fraction of Ly$\alpha$
appears to escape directly towards us, a more extended Ly$\alpha$ halo around
this part is indicative of resonant scatterings in the
circum-galactic gas \citep[cf. also][]{Ostlin2014}.  Contrary, in the
south-western part of the galaxy, where we observe the lowest velocity
dispersions ($v_\mathrm{FWHM} \approx 70$ km\,s$^{-1}$), Ly$\alpha$
photons do not escape along the line of sight.

\subsection{LARS 2 (\object{Shoc 240})}
\label{sec:lars-2}

Everywhere in LARS 2 where H$\alpha$ photons are produced, Ly$\alpha$
photons emerge along the line of sight. LARS 2 is the galaxy with the
highest global escape fraction of Ly$\alpha$ photons
($f^{\mathrm{Ly}\alpha}_\mathrm{esc.}=52.1$\%) and the highest
Ly$\alpha$/H$\alpha$ ratio (Ly$\alpha$/H$\alpha=4.53$).

With $v_\mathrm{shear} = 23\pm2$ km\,s$^{-1}$, LARS 2 shows the
smallest $v_\mathrm{shear}$ in our sample. However, we consider this
value as a lower limit since we missed in our pointing the
southernmost star forming knot. Our observed velocity field appears to
be disturbed, but we could envision a kinematical axis orthogonal to
the photometric major axis - i.e. roughly from west to east. This idea
is supported by our VLA imaging of this source \citepalias[see
Fig.~7 in][]{Pardy2014}. The \ion{H}{i} velocity field also indicates
$v_\mathrm{shear} \approx 30$ km\,s$^{-1}$, meaning that our
incomplete measurement gives a sensible lower limit.
Within the observed region our velocity dispersion map is rather
uniform, with $v_\mathrm{VFWHM} \approx 100$ km\,s$^{-1}$.

\subsection{LARS 3 (\object{Arp 238})}
\label{sec:lars-3}

LARS 3 is the south-eastern nucleus of the violently interacting pair
of similar sized spiral galaxies \object{Arp 238}. With a global
$f_\mathrm{esc}^\mathrm{Ly\alpha} = 0.1$ and
$L_\mathrm{Ly\alpha}=10^{41}$erg\,s$^{-1}$ this dust rich luminous
infrared galaxy is a relatively weak Ly$\alpha$ emitter. VLA \ion{H}{i}
imaging reveals an extended tail towards the west. This tidal tail is
much larger than the optical dimensions of the pair \object{Arp 238}
\citep[see also][for similar extended tidal \ion{H}{i} structures around the
Ly$\alpha$ emitting star bursts \object{Tol 1924-416} and \object{IRAS
  08339+6517}]{Cannon2004}.

In H$\alpha$ the main kinematical axis runs from west to east and the
radial velocity field appears symmetric, although slightly disturbed
towards the north and the south.  We measure
$v_\mathrm{shear} \approx 140$\,km\,s$^{-1}$, a value which is in the
domain of typical maximum velocities of inclination corrected
H$\alpha$ rotation curves of spiral galaxies
\citep[e.g.][]{Epinat2008,Epinat2010,Erroz-Ferrer2015}.  With
$v_\mathrm{FWHM} \gtrsim 300$\,km\,s$^{-1}$ highest velocity
dispersions are observed in the western part. The largest Ly$\alpha$
surface brightness is also observed in the western part of the
nucleus. Ly$\alpha$ appears in absorption in the eastern
region, where we also see a minimum in velocity dispersions with
$v_\mathrm{FWHM} \approx 70$\,km\,s$^{-1}$.

\subsection{LARS 4 (\object{SDSS J130728.45+542652.3})}
\label{sec:lars-4}

Although LARS 4 is similar to LARS 1 in terms of dust content and star
formation rate, globally this galaxy shows Ly$\alpha$ in
absorption. The galaxy can be photometrically decomposed into two main
components: an elongated lower surface brightness structure in the
west and a more puffed up and luminous companion in the east. The
eastern structure is tilted at $\approx40^\circ$ with respect to the
western one.

Compared to the highly irregular morphology in UV and H$\alpha$, the
radial velocity field appears rather regular, with a moderate
amplitude of $v_\mathrm{shear} \approx 75$\,km\,s$^{-1}$ along a well
defined axis from west to east.  The kinematic centre appears to be
right between the two photometric components, indicating that the
observed shearing is the velocity difference between the merging
components and not a consequence of rotation.  The lack of a well
defined simple and organised disk is further supported by
the absence of a double-horn profile in the GBT single-dish \ion{H}{i}
profile. Nevertheless, higher sensitivity VLA imaging results show a
coherent velocity field on larger scales with an amplitude comparable
to $v_\mathrm{shear}$ \citepalias[][their Fig.~9]{Pardy2014}.

In the dispersion map, the two components are clearly distinguishable,
with the eastern component showing higher velocity dispersions
($v_\mathrm{FWHM} \approx 100$\,km\,s$^{-1}$) than the western
($v_\mathrm{FWHM} \approx 65$\,km\,s$^{-1}$). In both regions,
however, Ly$\alpha$ is seen in absorption.  The highest velocity
dispersions are observed at the boundary region where the two
components are separated photometrically - but also here only little
Ly$\alpha$ is escaping. 

\subsection{LARS 5 (\object{Mrk 1486})}
\label{sec:lars-5}

In the HST images LARS 5 appears as a small (i.e. projected diameter
$d\approx 5$\,kpc) highly inclined edge-on disk. Within the apparent
disk Ly$\alpha$  appears in absorption but the galaxy shows an
extended halo, which is azimuthally symmetric at low surface
brightness isophotes, while brighter isophotes resemble more the
elongated H$\alpha$ shape.  Most prominent in the H$\alpha$ image are
filamentary ``fingers'' extending below and above the disk,
reminiscent of an outflowing wind.

Although the profile of our single-dish \ion{H}{i} observations shows multiple
peaks close to the noise level that might correspond to the peaks of a
faint double horn profile \citepalias{Pardy2014}, our H$\alpha$ kinematics
appear incompatible with a typical disk scenario: Firstly, our
measured $v_\mathrm{shear} = 37$\,km\,s$^{-1}$ indicates a very low
rotation curve amplitude - untypical even for a small disk
\citep[e.g. in the largest H$\alpha$ disk sample of ][for
$d\leq5$\,kpc the maximum rotation curve velocity is on average
95\,km\,s$^{-1}$]{Epinat2010}.  Secondly, there is apparent asymmetry
in the spatial distribution of the $v_\mathrm{LOS}$ values, i.e. a
large sector in the south-west characterised by similar line of sight
velocities ($v_\mathrm{LOS} \sim + 20 $\,km\,s$^{-1}$), while only a
small sector in the north-east shows blue-shifted H$\alpha$ emission
($v_\mathrm{LOS} \sim -40 $\,km\,s$^{-1}$).

The most prominent feature in the velocity dispersion map is a
biconical zone of increasing velocity dispersions (from
$\sim$120\,km\,s$^{-1}$ to $\sim$220\,km\,s$^{-1}$) with increasing
distance from the centre. The base of this zone coincides the
brightest region in UV and H$\alpha$ in the south-western sector. We
note that the seeing PSF FWHM for this observation is 1.3\arcsec{},
thus $\sim$2$\times$ the extend of the
0.5\arcsec{}$\times$0.5\arcsec{} spaxels used on LARS 5. However, as
the kinematical centre is not co-spatial with the base of the cones,
and since the velocity gradient around the base of the cone is very
weak we are confident that this feature is real and not caused by PSF
smearing effects.

\subsection{LARS 6 (\object{KISSR 2019})}
\label{sec:lars-6}

With $\lesssim 1$\,M$_\odot$yr$^{-1}$ conversion of gas into stars
LARS 6 shows the lowest star formation rate in the sample
\citep{Hayes2014}. The main star forming knot is in the north with a
tail of much smaller and less luminous knots extending to the
south. In our seeing limited data cubes we can not disentangle these
individual knots photometrically.  Ly$\alpha$ is seen in absorption,
even on the smallest scales.

Shearing is observed between the main component and the tail, although
the amplitude is moderate ($v_\mathrm{shear} = 52$\,km\,s$^{-1}$).
 The
main component also shows higher velocity dispersions
($v_\mathrm{VFWHM} \approx 70$\,km\,s$^{-1}$) than the tail
($v_\mathrm{VFWHM} \approx 50$\,km\,s$^{-1}$). Overall this galaxy has
the lowest observed velocity dispersion in the sample. 

Our GBT single-dish \ion{H}{i} measurements are severely contaminated by the
nearby field spiral UGC\,10028. However, newly obtained but as of yet
unpublished VLA D-configuration images allow a first-order separation
from LARS 6 and this companion. From these data, coherent rotation is
present in the \ion{H}{i} gas associated with LARS 6 at a level roughly
consistent with the $v_\mathrm{shear}$ estimate we obtain from PMAS.

\subsection{LARS 7 (\object{IRAS F13136+2938 })}
\label{sec:lars-7}

In the continuum LARS 7 resembles a highly inclined disk comparable to
LARS 5.  But, in H$\alpha$ an almost azimuthally symmetric structure
emerges that is highly distended with respect to the elongated
continuum morphology. Moreover, two extended low surface brightness
red lobes are found at the opposite ends of the inclined disk,
reminiscent of a shell like structure. This is suggestive of a recent
merger event \citep[e.g.][]{Bettoni2011}.  In Ly$\alpha$ images LARS 7
appears even more extended, with the bright isophotes following the
H$\alpha$ shape and low-surface brightness isophotes resembling a more
scaled up version of the apparent disk.

For similar reasons as outlined for LARS 5 in Sect.~\ref{sec:lars-5},
our H$\alpha$ kinematics argue against a typical disk scenario: We
observe a low shearing amplitude; with
$v_\mathrm{shear}=31$\,km\,s$^{-1}$ even lower than in LARS
5. Moreover, the line of sight velocity maps appears disturbed
compared to that expected for a classical disk. Our GBT single dish
\ion{H}{i} observations reveal a single broad ($92$\,km\,s$^{-1}$)
line. LARS 7 is the only object where the GBT line-width is
incompatible to its integrated H$\alpha$ linewidth
(cf. Sect.~\ref{sec:comp-hi-observ}).  A main kinematical axis can be
envisioned to run from north east to south west along the continuum
major axis.  But in the north west and south east the $v_\mathrm{LOS}$
values do not follow this weak gradient at all. Overall the galaxy
shows high velocity dispersion in H$\alpha$
($v_\mathrm{FWHM} \gtrsim 160$\,km\,s$^{-1}$), with the lowest values
occurring in north east ($v_\mathrm{FWHM} \approx 120$\,km\,s$^{-1}$),
where the brightest fluxes in Ly$\alpha$ and the highest
Ly$\alpha$/H$\alpha$ ratios are observed.

\subsection{LARS 8 (\object{SDSS-J125013.50+073441.5})}
\label{sec:lars-8}

LARS 8 appears a face-on disk with the highest metalicity of the
sample.  The apparent disk also possesses a highly dust obscured
nucleus. Nevertheless, there is a high degree of irregularity compared
to classical disks and spiral arms are not well defined. As traces of
shell-like structures are visible in the inner parts and in the
outskirts of the galaxy (cf. Fig.~6 in \citetalias{Ostlin2014}), LARS
8 can be classified as a shell system hinting at a recent merger event
\citep[e.g.]{Bettoni2011}.  The Ly$\alpha$ light distribution of LARS
8 does not resemble that seen in H$\alpha$. The brightest Ly$\alpha$ zone
is in the north of the disk, while in the south Ly$\alpha$ appears
only in absorption.  At lower surface brightness isophotes a
Ly$\alpha$ halo becomes visible. With an equivalent width
$\mathrm{EW}_\mathrm{Ly\alpha}=20.3$\AA{} this galaxy would be
selected as LAE in conventional narrow-band surveys.

Our $v_\mathrm{LOS}$ map seems disk like. Assuming an infinitely thin
disk the ellipticity of 0.2 \citep{Guaita2015} implies an inclination
of 40$^\circ$, hence our observed
$v_\mathrm{shear} = 155$\,km\,s$^{-1}$ translates to
$v_\mathrm{max} = 241$\,km\,s$^{-1}$. The GBT single-dish \ion{H}{i} profile
shows a broad, possibly double peaked profile. Although these peaks
are not significantly separated from the overall \ion{H}{i} signal, the
implied velocity difference of $\sim$300\,km\,s$^{-1}$ appears
consistent with our $v_\mathrm{shear}$ measurement. The orientation of the
velocity field from VLA interferometric \ion{H}{i} observations is
qualitatively consistent with our PMAS H$\alpha$ observations,
although at the VLA D-configuration beam size of 72\arcsec{} LARS 8
appears only marginally resolved.

Elevated velocity dispersions with
$v_\mathrm{FWHM}\gtrsim160$\,km\,s$^{-1}$ are apparent to extend
orthogonal to the kinematical axis along the 0 \,km\,s$^{-1}$ iso
velocity contour. In the other parts of the $v_\mathrm{FWHM}$ map is
rather flat, typically with $v_\mathrm{FWHM} \sim 100$\,km\,s$^{-1}$ -
a value that is also still commonly observed within the sample of 153
local spirals by \cite{Epinat2010}. Given the strong gradient in the
velocity field, the observed elevated velocity dispersions are likely a
result of PSF smearing effects.

\subsection{LARS 9 (\object{IRAS 08208+2816 })}
\label{sec:lars-9}

\begin{figure*}
  \centering
  \includegraphics[width=\textwidth]{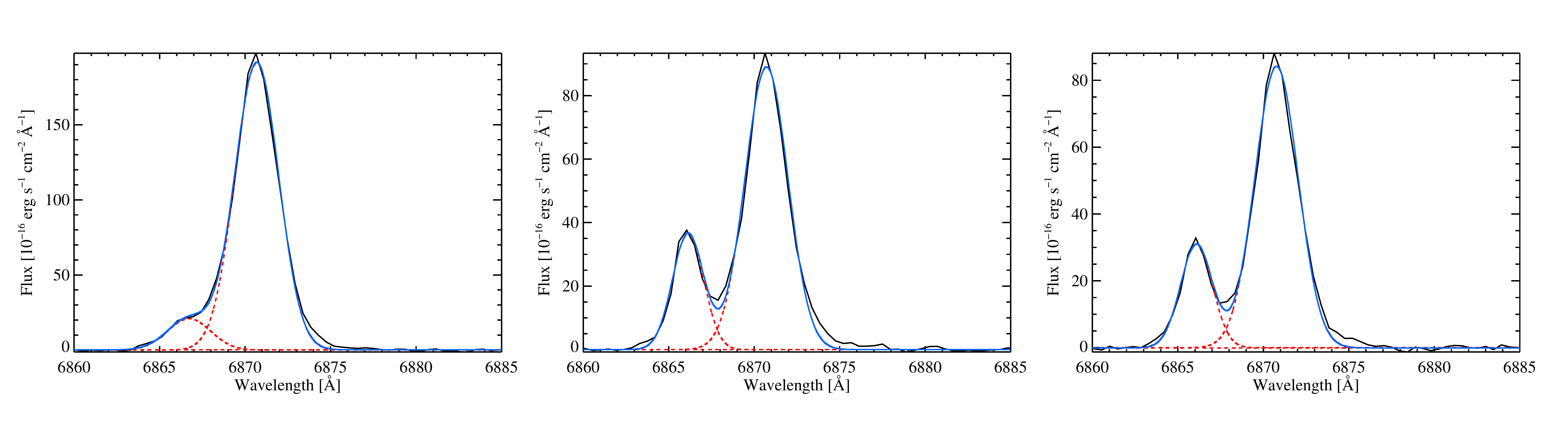}\vspace{-1em}
  \caption{Representative LARS 9 H$\alpha$ line profiles in blue
    hatched region of Fig.~\ref{fig:lars9_fig} described by a fit
    using two Gaussian components. The \emph{red dashed lines} show
    the individual components and the \emph{blue line} shows the sum,
    while the \emph{black line} shows the profile as observed. For
    all profiles shown the single component fit used to create the map
    shown in Fig.~\ref{fig:lars9_fig} converged on the stronger red
    component. }
  \label{fig:lars9_double}
\end{figure*}
\begin{figure*}
  \centering
  \includegraphics[width=\textwidth]{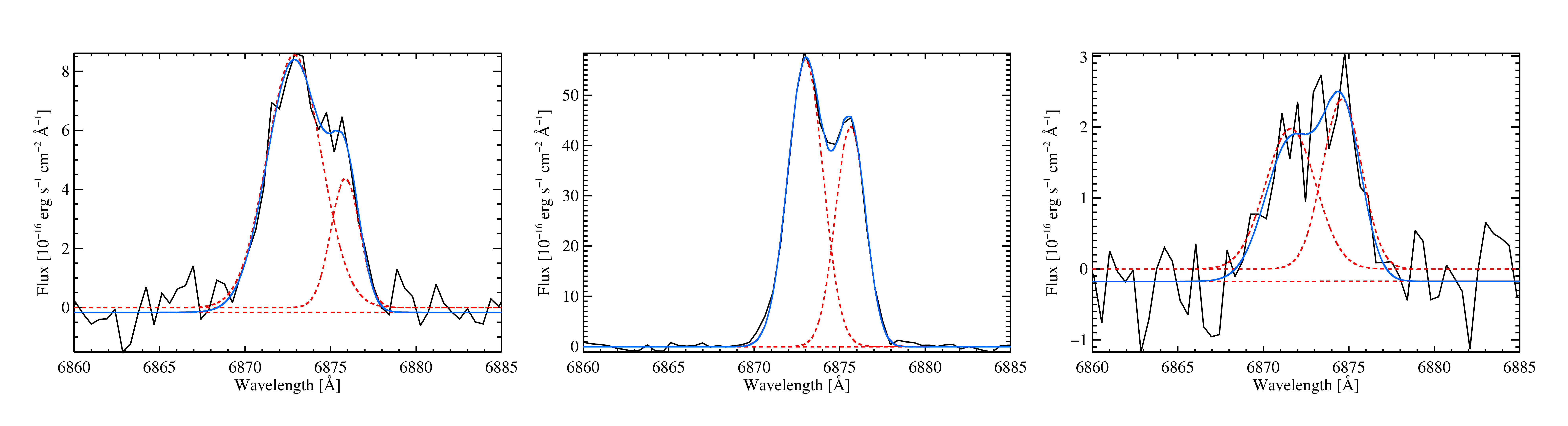}\vspace{-1em}
  \caption{Representative LARS 9 H$\alpha$ line profiles in red
    hatched region of Fig.~\ref{fig:lars9_fig}, similar to
    Fig.~\ref{fig:lars9_double}. The single component fit used to
    create the map shown in Fig.~\ref{fig:lars9_fig} converged on the
    stronger blue component in the centre panel, but for the profiles
    shown in the left and right panel the single component fit is
    artificially broadened. }
  \label{fig:lars9_double2}
\end{figure*}

This luminous infrared galaxy has highly irregular morphology.
Numerous star forming knots are visible in H$\alpha$ along two arms of
large extend that are connected to central, very bright H$\alpha$
nucleus. We note that at the end of the southern-tail a foreground
star appears in projection; contributions from this star have been
masked out in the PMAS data.  In LARS 9 Ly$\alpha$ is almost
everywhere absorbed along the line of sight towards the star forming
regions. Nevertheless, this galaxy is embedded in a faint extended
Ly$\alpha$. Morphologically, this fuzz broadly traces a scaled up
version of its optical and H$\alpha$ shape. Having a globally
integrated Ly$\alpha$ luminosity of
$L_\mathrm{Ly\alpha}\approx3\times10^{41}$\,erg\,s$^{-1}$ and an
equivalent width of $EW_\mathrm{Ly\alpha} \approx 8$\AA{}, this galaxy
would not be selected by its Ly$\alpha$ emission at high-$z$ in
conventional narrow-band surveys.

Since LARS 9 extends more than $\sim0.5$\arcmin{} on the sky, we
needed to cover it with two PMAS 16\arcsec{}$\times$16\arcsec{}
pointings (Fig.~\ref{fig:lars9_fig}). The line of sight velocity field
for this galaxy is very peculiar. From the north to the centre a weak
($\sim100$\,km\,s$^{-1}$) gradient from blue-shifted to systemic
velocity is apparent. From the centre this weak gradient continues
towards the south. The southern tail then shows an opposite gradient
from red- to blue-shifts. We note that in this south- and
south-western zone a single Gaussian often is not an optimal
representation of the observed H$\alpha$ lines. In this zone the
spectral profile is sometimes asymmetric with an extended red or blue
wing, but often it shows also clearly double peaked morphology (for
examples see Fig.~\ref{fig:lars9_double}).  In some these regions our
fit tries to capture both lines, hence the obtained line of sight
velocity is centred between the peaks and the linewidth appears very
broad (e.g. left and right panel of Fig.~\ref{fig:lars9_double2}). In
some other spaxels, when the secondary peak is very weak or when only
an extended wing is seen, the fit converges to the stronger line.  The
affected zones have hatched rectangles overlaid in
Fig.~\ref{fig:lars9_fig}. These zones harbour a kinematical distinct
secondary component. This secondary component does not follow the
large scale motions of the northern component, to which also the
nucleus belongs. We indicate by red- and blue-hatchings in
Fig.~\ref{fig:lars9_fig} the regions of this component that are red-
and blue-shifted with respect to the systemic velocity (as given by
the nucleus).  As a natural outcome from this qualitative analysis the
emission in south-eastern tail of LARS 9 is solely coming from this
secondary component. We note that also only at the end of this tail a
significant fraction of Ly$\alpha$ photons appear to escape towards
the observer.

From our PMAS observation we conclude that LARS 9 is a closely
interacting pair of galaxies in an advanced stage of merging.  The
interaction scenario is also supported by our 21cm observations. The
integrated GBT spectrum shows a broad single line, indicating that the
bulk of the \ion{H}{i} is not partaking in an ordered flat rotation.
LARS 9 was only marginally resolved in our VLA \ion{H}{i} maps
\citepalias{Pardy2014}. However, as of yet unpublished VLA C
configuration observations (K. Fitzgibbon, in prep.) show extended
\ion{H}{i} towards the west, enclosing the galaxy
SDSS\,J082353.65+280622.2. For this galaxy no spectroscopic redshift
is available, but the photometric redshift agrees with being
associated to LARS 9.  These observations therefore strongly suggests
that a third system is significantly involved in the interaction.
Moreover, the VLA C-configuration observations of LARS 9 show a peak
in the second-moment maps that trace the random motions of \ion{H}{i}
that coincides spatially with the zones of multiple \ion{H}{ii}
components in our PMAS maps . Hence, at these locations (hatched
regions in Fig.~\ref{fig:lars9_fig}) Ly$\alpha$ photons of both
distinct kinematical components are likely scattered by \ion{H}{i} at
resonance.

\subsection{LARS 10 (\object{Mrk 0061})}
\label{sec:lars-10}

Similar to LARS 1 this galaxy appears morphologically to be in an
advanced merger state. It possesses a large UV bright core in the
north-west and a spur of smaller, less luminous star forming regions
towards the south-east. Several redder low-surface brightness
structures are seen outside the main body of the galaxy, reminiscent
of the appearance of shell galaxies
\citep[e.g.][]{Marino2009,Bettoni2011}.  From the central star-forming
parts Ly$\alpha$ is only seen in absorption, while at larger radii a
faint halo emerges. When considering solely the central parts the
galaxy would remain undetected in high-$z$ LAE surveys since
$\mathrm{EW_\mathrm{Ly\alpha}}=8$\AA{}. But, when integrating over the
lower-surface brightness emission $\mathrm{EW_\mathrm{Ly\alpha}}$
rises to 31\AA{} and with its luminosity of
$L_\mathrm{Ly\alpha}=2\times10^{41}$erg\,s$^{-1}$ it would be within
the sensitivity of the deepest contemporary LAE surveys
\citep{Rauch2008,Bacon2015}.

The line of sight velocity field is symmetric and consistent with
rotation and contrasts the irregular continuum morphology. The
velocity gradient running from the south-east to the north-west is
weak and we measure $v_\mathrm{shear} = 36$\,km\,s$^{-1}$. The seeing
for this observation was quite substantial (1.5\arcsec{} FWHM, i.e
3$\times$ the size of the $0.5$\arcsec{}$\times$0.5\arcsec{}
spaxels). Despite the weak gradient, this leads to non-negligible PSF
smearing effects in the dispersion map. Moreover, due to the short
exposure time and the rather low H$\alpha$ flux this galaxy has the
lowest SNR per spaxel of the whole sample; this manifests in a 
noisy $v_\mathrm{FWHM}$ map. Both of these effects make essentially
local disturbances in the velocity dispersion untraceable in our LARS
10 dataset, but our flux weighted global measurement of
$\sigma_0\approx 40$\,km\,s$^{-1}$ is robust against these nuisances.

The GBT integrated \ion{H}{i} spectrum shows a broad single line profile
($\sim$280\,km\,s$^{-1}$ FWHM), however at low signal to noise.  This
appears hard to reconcile with the H$\alpha$ results. If real, it
might indicate that a large fraction of neutral gas in and around LARS
10 is kinematically in a different state than the gas around the
galaxy's star forming regions. 

\subsection{LARS 11 (\object{SDSS J140347.22+062812.1})}
\label{sec:lars-11}

This galaxy appears as a highly inclined edge-on disk in the continuum
and UV that is slightly thicker when seen in H$\alpha$.
According to the H$\alpha$ and UV emission, stronger star formation
occurs in the south-eastern zone of the projected disk. In Ly$\alpha$
a mildly extended halo above and below the plane is visible, with the
isophotal contours approximately preserving the axis ratio of
H$\alpha$. Within the disk Ly$\alpha$ occurs exclusively in
absorption.

Running along the disk from the north-west to the south-east a strong
gradient in the $v_\mathrm{LOS}$ map is apparent. The observed
shearing amplitude $v_\mathrm{shear}=150$\,km\,s$^{-1}$ is consistent
with this gradient being caused by rotation
\citep[e.g.][]{Epinat2008,Epinat2010,Erroz-Ferrer2015}. Unfortunately
the beam of our GBT single-dish \ion{H}{i} observations is likely
contaminated by other sources. Nevertheless, within the multiple peaks
in the GBT spectrum a double-horn profile at a velocity separation
consistent with our H$\alpha$ $v_\mathrm{shear}$ measurement is
visible.  Although the projected height of the disk is similar to the
PSF FWHM, the dispersion map shows traces of higher velocity
dispersions above and below the disk
($v_\mathrm{VFWHM}\approx200$\,km\,s$^{-1}$) than within
($v_\mathrm{VFWHM}\approx150$\,km\,s$^{-1}$), especially around the
star-forming regions in the south east.  We note, however, that the
elevated dispersions near the kinematic centre are caused by PSF
smearing of the strong gradient in the $v_\mathrm{LOS}$ field.

\subsection{LARS 12 (\object{LEDA 27453})}
\label{sec:lars-12}

Because of its modest $\mathrm{EW}_\mathrm{Ly\alpha}=13$\AA{} this UV
bright merger would not be selected as a LAE at high-$z$. At its
brightest knot in UV and H$\alpha$ the galaxy shows Ly$\alpha$ only in
absorption. Only at larger radii, where a number of fainter star
forming regions can be appreciated, Ly$\alpha$ appears in
emission. Moreover, LARS 12 is embedded in a faint, low surface
brightness Ly$\alpha$ halo. 

The non-regular H$\alpha$ velocity field indicates complex
kinematics. In the north-western corner a strong
$\sim140$\,km\,s$^{-1}$ gradient over $\sim1.5$\arcsec{}
($\sim 2.8$\,kpc at $d_\mathrm{LARS 12} = 470.5$\,Mpc) is observed,
but from there towards the south-west the velocity field is
essentially flat. The high velocity and velocity dispersion in the
values seen in the south are not significant, as the low SNR of this
broad line in these binned spaxels lead to a 50\% error on the
determined $v_\mathrm{LOS}$ and $\sigma_0$ value.  We note that
because of PSF smearing effects the strong velocity gradient is
responsible for the region of elevated velocity dispersions in the
north-western corner.

For LARS 12 adaptive optics Paschen $\alpha$ IFS observations have
been presented in \cite{Gonccalves2010} - Object ScUVLG 093813 in
their nomenclature. Their line-of-sight velocity field is
qualitatively fully consistent with ours, but their higher resolution
data allows to pinpoint the highest redshifted values directly to the
two filaments that extent towards the north-west. Since they are not
affected by PSF smearing their dispersion map is not contaminated by
high-velocity dispersions in the north-west. Of course, our artifact
in the south is also absent from their map. Nevertheless, their
$\sigma_0$ value of 62\,km\,s$^{-1}$ is in good agreement with
our measurement of 72\,km\,s$^{-1}$.

\subsection{LARS 13 (\object{IRAS 01477+1254})}
\label{sec:lars-13}

\begin{figure*}
  \centering
  \includegraphics[width=\textwidth]{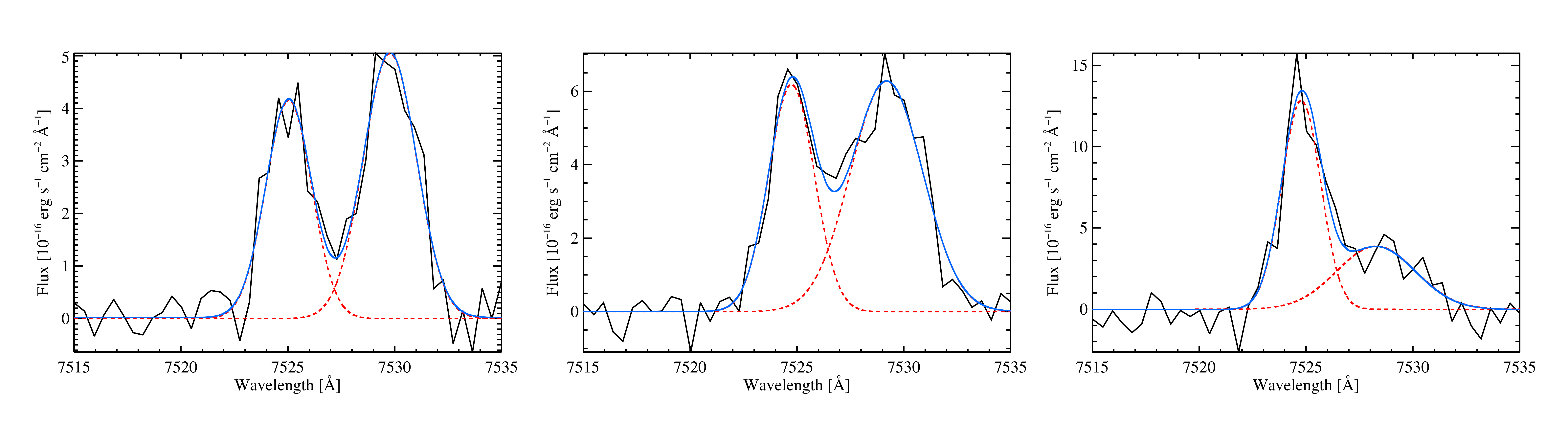}\vspace{-1em}
  \caption{Representative LARS 13 H$\alpha$ line profiles in blue
    hatched region of Fig.~\ref{fig:lars13_fig}, similar to
    Fig.~\ref{fig:lars9_double}. The single component fit used to
    create the map shown in Fig.~\ref{fig:lars13_fig} converged on the
    stronger blue component in the right panel, but for the profiles
    shown in the left and canter panel the single component fit is
    artificially broadened. }
  \label{fig:lars13_double}
\end{figure*}

Based on its highly irregular morphology this starburst system is
clearly interacting.  As LARS 13 shows a narrow Ly$\alpha$ equivalent
width ($\mathrm{EW}_\mathrm{Ly\alpha} = 6$\,\AA{}), it would not be
selected as a LAE at high redshift in conventional narrow-band
surveys. Overall only 1\% of Ly$\alpha$ photons are escaping and the
resulting Ly$\alpha$ luminosity is
$L_\mathrm{Ly\alpha}=7\times10^{41}$\,erg\,s$^{-1}$.

The complex $v_\mathrm{LOS}$ map of this galaxy is characterised by a
gradient from blue- to redshifts along the east-western axis for the
northern component, as well as an gradient from blue-to-redshifts from
south to north for the eastern component. In the north-western zone
where the longitudinally aligned structure overlaps with the
latitudinally aligned one, our single component Gaussian is not an
optimal representation of the observed H$\alpha$ line profiles. In
this region the profile is often asymmetric with an extended blue wing
(e.g. right panel of Fig.~\ref{fig:lars13_double}). Some spaxels show
a double peaked H$\alpha$ profile (e.g. left and centre panel of
Fig.~\ref{fig:lars13_double}). The affected zone has a hatched
rectangle overlaid in Fig.~\ref{fig:lars13_fig}. In the region where
two equally strong peaks are prominent in the spectrum, our fit tries
to fit a broad line that encloses both lines; these spaxels can be
seen in the east running as a diagonal line of broad velocity
dispersions from north-west to south-east. Spaxels east from this
demarcation show an extended blue wing. We interpret this as that we
are seeing separate components of ionised gas kinematics, where one
component belongs to the latitudinal elongated structure while the
other belongs to the longitudinal elongated one. The spatial and
spectral proximity of both regions indicates that the galaxy is still
in an ongoing interaction with both progenitors not having fully
coalesced.

For LARS 13 adaptive optics Paschen $\alpha$ IFS observations have been
presented in \cite{Gonccalves2010} - object ScUVLG 015028 in their
nomenclature. However, their small FoV allows them only to sample
emission from the strongest star forming region in the western part,
so they do not have any information on the longitudinally aligned
eastern part. Consequently, their observed
$v_\mathrm{shear}=78$\,km\,s$^{-1}$ is lower than our value
($v_\mathrm{shear}=173$\,km\,s$^{-1}$), since the north-eastern zone
with highest $v_\mathrm{LOS}$ values in our map is not present in
their data. In the zone where they have signal, there is good
qualitative and quantitative agreement between our maps and their
maps. In particular, they found high velocity dispersions
$v_\mathrm{FWHM}\gtrsim200$\,km\,s$^{-1}$ and a weak velocity gradient
in the north-western star-forming knot. We note that in this region of
high velocity dispersions Ly$\alpha$ appears to escape directly
towards the observer.

\subsection{LARS 14 (\object{SDSS J092600.41+442736.1})}
\label{sec:lars-14}

LARS 14 is the most luminous LAE in the sample
($L_\mathrm{Ly\alpha} = 4.2 \times 10^{42}$erg\,s$^{-1}$).  It is also
classified as a green pea galaxy
\citep[][see also their Fig.~7]{Cardamone2009}. Green pea galaxies show
commonly strong Ly$\alpha$ emission and are also thought to be Lyman
continuum leaking galaxies
\citep{Jaskot2014,Henry2015,Yang2015}. Visible in the HST images is a
small companion in the south and tidal tails extending to the east and
south \citep{Cardamone2009}. Our seeing limited data does not provide
enough resolution to disentangle this structure photometrically from
the main component within the datacube. Kinematically, however, we
observe that the southern region shears at
$\approx 60 - 70$\,km\,s$^{-1}$ with respect to the northern, with
not much velocity substructure seen.  In the main part we observe
high velocity dispersions with
$v_\mathrm{FWHM} \approx 175$\,km\,s$^{-1}$, dropping to
$80 - 90$\,km\,s$^{-1}$ towards the companion.  Our modelling of the
H$\alpha$ profile with a simple 1D Gaussian did not capture the
extended broad wings seen in the H$\alpha$ profiles. These broad wings
likely trace outflowing ionised gas, and are a feature regularly
observed in pea galaxies \citep[][]{Yang2015}.

For this galaxy adaptive optics Paschen $\alpha$ IFS observation have
been presented in \cite{Gonccalves2010} and \cite{Basu-Zych2009}
(object ScUVLG 92600 in their nomenclature).  They report slightly
higher values, both for the shear and the dispersion. Nevertheless,
our derived $v_\mathrm{shear} / \sigma_0 = 0.6 \pm 0.1$ is consistent
with their value of 0.51. This ratio would be lowered by about a
factor of two, if only the northern main component is considered,
since its velocity offset relative to the companion dominates the
shear. Noteworthy, the main component as such therefore presents the
structure with the lowest $v_\mathrm{shear} / \sigma_0$ of our whole
sample.

% References
\bibliographystyle{aa}
\bibliography{phdbibliography.bib}

\begin{thebibliography}{140}
\expandafter\ifx\csname natexlab\endcsname\relax\def\natexlab#1{#1}\fi

\bibitem[{{Adams} {et~al.}(2011){Adams}, {Blanc}, {Hill}, {Gebhardt}, {Drory},
  {Hao}, {Bender}, {Byun}, {Ciardullo}, {Cornell}, {Finkelstein}, {Fry},
  {Gawiser}, {Gronwall}, {Hopp}, {Jeong}, {Kelz}, {Kelzenberg}, {Komatsu},
  {MacQueen}, {Murphy}, {Odoms}, {Roth}, {Schneider}, {Tufts}, \&
  {Wilkinson}}]{Adams2011}
{Adams}, J.~J., {Blanc}, G.~A., {Hill}, G.~J., {et~al.} 2011, \apjs, 192, 5

\bibitem[{{Ahn} {et~al.}(2003){Ahn}, {Lee}, \& {Lee}}]{Ahn2003}
{Ahn}, S.-H., {Lee}, H.-W., \& {Lee}, H.~M. 2003, \mnras, 340, 863

\bibitem[{{Alonso-Herrero} {et~al.}(2009){Alonso-Herrero},
  {Garc{\'{\i}}a-Mar{\'{\i}}n}, {Monreal-Ibero}, {Colina}, {Arribas},
  {Alfonso-Garz{\'o}n}, \& {Labiano}}]{Alonso-Herrero2009}
{Alonso-Herrero}, A., {Garc{\'{\i}}a-Mar{\'{\i}}n}, M., {Monreal-Ibero}, A.,
  {et~al.} 2009, \aap, 506, 1541

\bibitem[{{Astropy Collaboration} {et~al.}(2013){Astropy Collaboration},
  {Robitaille}, {Tollerud}, {Greenfield}, {Droettboom}, {Bray}, {Aldcroft},
  {Davis}, {Ginsburg}, {Price-Whelan}, {Kerzendorf}, {Conley}, {Crighton},
  {Barbary}, {Muna}, {Ferguson}, {Grollier}, {Parikh}, {Nair}, {Unther},
  {Deil}, {Woillez}, {Conseil}, {Kramer}, {Turner}, {Singer}, {Fox}, {Weaver},
  {Zabalza}, {Edwards}, {Azalee Bostroem}, {Burke}, {Casey}, {Crawford},
  {Dencheva}, {Ely}, {Jenness}, {Labrie}, {Lim}, {Pierfederici}, {Pontzen},
  {Ptak}, {Refsdal}, {Servillat}, \& {Streicher}}]{AstropyCollaboration2013}
{Astropy Collaboration}, {Robitaille}, T.~P., {Tollerud}, E.~J., {et~al.} 2013,
  \aap, 558, A33

\bibitem[{{Bacon} {et~al.}(2015){Bacon}, {Brinchmann}, {Richard}, {Contini},
  {Drake}, {Franx}, {Tacchella}, {Vernet}, {Wisotzki}, {Blaizot}, {Bouch{\'e}},
  {Bouwens}, {Cantalupo}, {Carollo}, {Carton}, {Caruana}, {Cl{\'e}ment},
  {Dreizler}, {Epinat}, {Guiderdoni}, {Herenz}, {Husser}, {Kamann}, {Kerutt},
  {Kollatschny}, {Krajnovic}, {Lilly}, {Martinsson}, {Michel-Dansac},
  {Patricio}, {Schaye}, {Shirazi}, {Soto}, {Soucail}, {Steinmetz}, {Urrutia},
  {Weilbacher}, \& {de Zeeuw}}]{Bacon2015}
{Bacon}, R., {Brinchmann}, J., {Richard}, J., {et~al.} 2015, \aap, 575, A75

\bibitem[{{Bacon} {et~al.}(2014){Bacon}, {Vernet}, {Borisiva}, {Bouch{\'e}},
  {Brinchmann}, {Carollo}, {Carton}, {Caruana}, {Cerda}, {Contini}, {Franx},
  {Girard}, {Guerou}, {Haddad}, {Hau}, {Herenz}, {Herrera}, {Husemann},
  {Husser}, {Jarno}, {Kamann}, {Krajnovic}, {Lilly}, {Mainieri}, {Martinsson},
  {Palsa}, {Patricio}, {P{\'e}contal}, {Pello}, {Piqueras}, {Richard},
  {Sandin}, {Schroetter}, {Selman}, {Shirazi}, {Smette}, {Soto}, {Streicher},
  {Urrutia}, {Weilbacher}, {Wisotzki}, \& {Zins}}]{Bacon2014}
{Bacon}, R., {Vernet}, J., {Borisiva}, E., {et~al.} 2014, The Messenger, 157,
  13

\bibitem[{{Barnes} {et~al.}(2011){Barnes}, {Haehnelt}, {Tescari}, \&
  {Viel}}]{Barnes2011}
{Barnes}, L.~A., {Haehnelt}, M.~G., {Tescari}, E., \& {Viel}, M. 2011, \mnras,
  416, 1723

\bibitem[{{Bassett} {et~al.}(2014){Bassett}, {Glazebrook}, {Fisher}, {Green},
  {Wisnioski}, {Obreschkow}, {Cooper}, {Abraham}, {Damjanov}, \&
  {McGregor}}]{Bassett2014}
{Bassett}, R., {Glazebrook}, K., {Fisher}, D.~B., {et~al.} 2014, \mnras, 442,
  3206

\bibitem[{{Basu-Zych} {et~al.}(2009){Basu-Zych}, {Gon{\c c}alves}, {Overzier},
  {Law}, {Schiminovich}, {Heckman}, {Martin}, {Wyder}, \&
  {O'Dowd}}]{Basu-Zych2009}
{Basu-Zych}, A.~R., {Gon{\c c}alves}, T.~S., {Overzier}, R., {et~al.} 2009,
  \apjl, 699, L118

\bibitem[{{Behrens} \& {Braun}(2014)}]{Behrens2014}
{Behrens}, C. \& {Braun}, H. 2014, \aap, 572, A74

\bibitem[{{Behrens} {et~al.}(2014){Behrens}, {Dijkstra}, \&
  {Niemeyer}}]{Behrens2014a}
{Behrens}, C., {Dijkstra}, M., \& {Niemeyer}, J.~C. 2014, \aap, 563, A77

\bibitem[{{Bettoni} {et~al.}(2011){Bettoni}, {Galletta}, {Rampazzo}, {Marino},
  {Mazzei}, \& {Buson}}]{Bettoni2011}
{Bettoni}, D., {Galletta}, G., {Rampazzo}, R., {et~al.} 2011, \aap, 534, A24

\bibitem[{{Bik} {et~al.}(2015){Bik}, {{\"O}stlin}, {Hayes}, {Adamo},
  {Melinder}, \& {Amram}}]{Bik2015}
{Bik}, A., {{\"O}stlin}, G., {Hayes}, M., {et~al.} 2015, \aap, 576, L13

\bibitem[{{Calabretta} \& {Greisen}(2002)}]{Calabretta2002}
{Calabretta}, M.~R. \& {Greisen}, E.~W. 2002, \aap, 395, 1077

\bibitem[{{Cannon} {et~al.}(2004){Cannon}, {Skillman}, {Kunth}, {Leitherer},
  {Mas-Hesse}, {{\"O}stlin}, \& {Petrosian}}]{Cannon2004}
{Cannon}, J.~M., {Skillman}, E.~D., {Kunth}, D., {et~al.} 2004, \apj, 608, 768

\bibitem[{{Cappellari} \& {Copin}(2003)}]{Cappellari2003}
{Cappellari}, M. \& {Copin}, Y. 2003, \mnras, 342, 345

\bibitem[{{Cardamone} {et~al.}(2009){Cardamone}, {Schawinski}, {Sarzi},
  {Bamford}, {Bennert}, {Urry}, {Lintott}, {Keel}, {Parejko}, {Nichol},
  {Thomas}, {Andreescu}, {Murray}, {Raddick}, {Slosar}, {Szalay}, \&
  {Vandenberg}}]{Cardamone2009}
{Cardamone}, C., {Schawinski}, K., {Sarzi}, M., {et~al.} 2009, \mnras, 399,
  1191

\bibitem[{{Cassata} {et~al.}(2011){Cassata}, {Le F{\`e}vre}, {Garilli},
  {Maccagni}, {Le Brun}, {Scodeggio}, {Tresse}, {Ilbert}, {Zamorani},
  {Cucciati}, {Contini}, {Bielby}, {Mellier}, {McCracken}, {Pollo},
  {Zanichelli}, {Bardelli}, {Cappi}, {Pozzetti}, {Vergani}, \&
  {Zucca}}]{Cassata2011}
{Cassata}, P., {Le F{\`e}vre}, O., {Garilli}, B., {et~al.} 2011, \aap, 525,
  A143

\bibitem[{{Cassata} {et~al.}(2015){Cassata}, {Tasca}, {Le F{\`e}vre}, {Lemaux},
  {Garilli}, {Le Brun}, {Maccagni}, {Pentericci}, {Thomas}, {Vanzella},
  {Zamorani}, {Zucca}, {Amorin}, {Bardelli}, {Capak}, {Cassar{\`a}},
  {Castellano}, {Cimatti}, {Cuby}, {Cucciati}, {de la Torre}, {Durkalec},
  {Fontana}, {Giavalisco}, {Grazian}, {Hathi}, {Ilbert}, {Moreau}, {Paltani},
  {Ribeiro}, {Salvato}, {Schaerer}, {Scodeggio}, {Sommariva}, {Talia},
  {Taniguchi}, {Tresse}, {Vergani}, {Wang}, {Charlot}, {Contini}, {Fotopoulou},
  {Koekemoer}, {L{\'o}pez-Sanjuan}, {Mellier}, \& {Scoville}}]{Cassata2015}
{Cassata}, P., {Tasca}, L.~A.~M., {Le F{\`e}vre}, O., {et~al.} 2015, \aap, 573,
  A24

\bibitem[{{Chonis} {et~al.}(2013){Chonis}, {Blanc}, {Hill}, {Adams},
  {Finkelstein}, {Gebhardt}, {Kollmeier}, {Ciardullo}, {Drory}, {Gronwall},
  {Hagen}, {Overzier}, {Song}, \& {Zeimann}}]{Chonis2013}
{Chonis}, T.~S., {Blanc}, G.~A., {Hill}, G.~J., {et~al.} 2013, \apj, 775, 99

\bibitem[{{Christensen} {et~al.}(2012){Christensen}, {Laursen}, {Richard},
  {Hjorth}, {Milvang-Jensen}, {Dessauges-Zavadsky}, {Limousin}, {Grillo}, \&
  {Ebeling}}]{Christensen2012}
{Christensen}, L., {Laursen}, P., {Richard}, J., {et~al.} 2012, \mnras, 427,
  1973

\bibitem[{{Ciardullo} {et~al.}(2012){Ciardullo}, {Gronwall}, {Wolf},
  {McCathran}, {Bond}, {Gawiser}, {Guaita}, {Feldmeier}, {Treister}, {Padilla},
  {Francke}, {Matkovi{\'c}}, {Altmann}, \& {Herrera}}]{Ciardullo2012}
{Ciardullo}, R., {Gronwall}, C., {Wolf}, C., {et~al.} 2012, \apj, 744, 110

\bibitem[{{Cooper} {et~al.}(2008){Cooper}, {Bicknell}, {Sutherland}, \&
  {Bland-Hawthorn}}]{Cooper2008}
{Cooper}, J.~L., {Bicknell}, G.~V., {Sutherland}, R.~S., \& {Bland-Hawthorn},
  J. 2008, \apj, 674, 157

\bibitem[{{Cowie} {et~al.}(2010){Cowie}, {Barger}, \& {Hu}}]{Cowie2010}
{Cowie}, L.~L., {Barger}, A.~J., \& {Hu}, E.~M. 2010, \apj, 711, 928

\bibitem[{{Cox}(2000)}]{Cox2000}
{Cox}, A.~N. 2000, {Allen's astrophysical quantities}

\bibitem[{{Davies} {et~al.}(2011){Davies}, {F{\"o}rster Schreiber}, {Cresci},
  {Genzel}, {Bouch{\'e}}, {Burkert}, {Buschkamp}, {Genel}, {Hicks}, {Kurk},
  {Lutz}, {Newman}, {Shapiro}, {Sternberg}, {Tacconi}, \& {Wuyts}}]{Davies2011}
{Davies}, R., {F{\"o}rster Schreiber}, N.~M., {Cresci}, G., {et~al.} 2011,
  \apj, 741, 69

\bibitem[{{Diehl} \& {Statler}(2006)}]{Diehl2006}
{Diehl}, S. \& {Statler}, T.~S. 2006, \mnras, 368, 497

\bibitem[{{Dijkstra}(2014)}]{Dijkstra2014}
{Dijkstra}, M. 2014, \pasa, 31, 40

\bibitem[{{Dijkstra} {et~al.}(2006){Dijkstra}, {Haiman}, \&
  {Spaans}}]{Dijkstra2006}
{Dijkstra}, M., {Haiman}, Z., \& {Spaans}, M. 2006, \apj, 649, 14

\bibitem[{{Duval} {et~al.}(2014){Duval}, {Schaerer}, {{\"O}stlin}, \&
  {Laursen}}]{Duval2014}
{Duval}, F., {Schaerer}, D., {{\"O}stlin}, G., \& {Laursen}, P. 2014, \aap,
  562, A52

\bibitem[{{Epinat} {et~al.}(2010){Epinat}, {Amram}, {Balkowski}, \&
  {Marcelin}}]{Epinat2010}
{Epinat}, B., {Amram}, P., {Balkowski}, C., \& {Marcelin}, M. 2010, \mnras,
  401, 2113

\bibitem[{{Epinat} {et~al.}(2008{\natexlab{a}}){Epinat}, {Amram}, \&
  {Marcelin}}]{Epinat2008}
{Epinat}, B., {Amram}, P., \& {Marcelin}, M. 2008{\natexlab{a}}, \mnras, 390,
  466

\bibitem[{{Epinat} {et~al.}(2008{\natexlab{b}}){Epinat}, {Amram}, {Marcelin},
  {Balkowski}, {Daigle}, {Hernandez}, {Chemin}, {Carignan}, {Gach}, \&
  {Balard}}]{Epinat2008a}
{Epinat}, B., {Amram}, P., {Marcelin}, M., {et~al.} 2008{\natexlab{b}}, \mnras,
  388, 500

\bibitem[{{Erb} {et~al.}(2014){Erb}, {Steidel}, {Trainor}, {Bogosavljevi{\'c}},
  {Shapley}, {Nestor}, {Kulas}, {Law}, {Strom}, {Rudie}, {Reddy}, {Pettini},
  {Konidaris}, {Mace}, {Matthews}, \& {McLean}}]{Erb2014}
{Erb}, D.~K., {Steidel}, C.~C., {Trainor}, R.~F., {et~al.} 2014, \apj, 795, 33

\bibitem[{Erroz-Ferrer {et~al.}(2015)Erroz-Ferrer, Knapen, Leaman, Cisternas,
  Font, Beckman, Sheth, Muñoz-Mateos, Díaz-García, Bosma, Athanassoula,
  Elmegreen, Ho, Kim, Laurikainen, Martinez-Valpuesta, Meidt, \&
  Salo}]{Erroz-Ferrer2015}
Erroz-Ferrer, S., Knapen, J.~H., Leaman, R., {et~al.} 2015
  [\eprint{1504.06282}]

\bibitem[{{Finkelstein} {et~al.}(2013){Finkelstein}, {Papovich}, {Dickinson},
  {Song}, {Tilvi}, {Koekemoer}, {Finkelstein}, {Mobasher}, {Ferguson},
  {Giavalisco}, {Reddy}, {Ashby}, {Dekel}, {Fazio}, {Fontana}, {Grogin},
  {Huang}, {Kocevski}, {Rafelski}, {Weiner}, \& {Willner}}]{Finkelstein2013}
{Finkelstein}, S.~L., {Papovich}, C., {Dickinson}, M., {et~al.} 2013, \nat,
  502, 524

\bibitem[{{Flores} {et~al.}(2006){Flores}, {Hammer}, {Puech}, {Amram}, \&
  {Balkowski}}]{Flores2006}
{Flores}, H., {Hammer}, F., {Puech}, M., {Amram}, P., \& {Balkowski}, C. 2006,
  \aap, 455, 107

\bibitem[{{F{\"o}rster Schreiber} {et~al.}(2009){F{\"o}rster Schreiber},
  {Genzel}, {Bouch{\'e}}, {Cresci}, {Davies}, {Buschkamp}, {Shapiro},
  {Tacconi}, {Hicks}, {Genel}, {Shapley}, {Erb}, {Steidel}, {Lutz},
  {Eisenhauer}, {Gillessen}, {Sternberg}, {Renzini}, {Cimatti}, {Daddi},
  {Kurk}, {Lilly}, {Kong}, {Lehnert}, {Nesvadba}, {Verma}, {McCracken},
  {Arimoto}, {Mignoli}, \& {Onodera}}]{FoersterSchreiber2009}
{F{\"o}rster Schreiber}, N.~M., {Genzel}, R., {Bouch{\'e}}, N., {et~al.} 2009,
  \apj, 706, 1364

\bibitem[{{Genzel} {et~al.}(2006){Genzel}, {Tacconi}, {Eisenhauer},
  {F{\"o}rster Schreiber}, {Cimatti}, {Daddi}, {Bouch{\'e}}, {Davies},
  {Lehnert}, {Lutz}, {Nesvadba}, {Verma}, {Abuter}, {Shapiro}, {Sternberg},
  {Renzini}, {Kong}, {Arimoto}, \& {Mignoli}}]{Genzel2006}
{Genzel}, R., {Tacconi}, L.~J., {Eisenhauer}, F., {et~al.} 2006, \nat, 442, 786

\bibitem[{{Glazebrook}(2013)}]{Glazebrook2013}
{Glazebrook}, K. 2013, \pasa, 30, 56

\bibitem[{{Gon{\c c}alves} {et~al.}(2010){Gon{\c c}alves}, {Basu-Zych},
  {Overzier}, {Martin}, {Law}, {Schiminovich}, {Wyder}, {Mallery}, {Rich}, \&
  {Heckman}}]{Gonccalves2010}
{Gon{\c c}alves}, T.~S., {Basu-Zych}, A., {Overzier}, R., {et~al.} 2010, \apj,
  724, 1373

\bibitem[{{Green} {et~al.}(2010){Green}, {Glazebrook}, {McGregor}, {Abraham},
  {Poole}, {Damjanov}, {McCarthy}, {Colless}, \& {Sharp}}]{Green2010}
{Green}, A.~W., {Glazebrook}, K., {McGregor}, P.~J., {et~al.} 2010, \nat, 467,
  684

\bibitem[{{Green} {et~al.}(2014){Green}, {Glazebrook}, {McGregor}, {Damjanov},
  {Wisnioski}, {Abraham}, {Colless}, {Sharp}, {Crain}, {Poole}, \&
  {McCarthy}}]{Green2014}
{Green}, A.~W., {Glazebrook}, K., {McGregor}, P.~J., {et~al.} 2014, \mnras,
  437, 1070

\bibitem[{{Green}(2011)}]{Green2011}
{Green}, D.~A. 2011, Bulletin of the Astronomical Society of India, 39, 289

\bibitem[{{Greisen} \& {Calabretta}(2002)}]{Greisen2002}
{Greisen}, E.~W. \& {Calabretta}, M.~R. 2002, \aap, 395, 1061

\bibitem[{{Gronke} {et~al.}(2015){Gronke}, {Bull}, \& {Dijkstra}}]{Gronke2015}
{Gronke}, M., {Bull}, P., \& {Dijkstra}, M. 2015, \apj, 812, 123

\bibitem[{{Gronke} \& {Dijkstra}(2014)}]{Gronke2014}
{Gronke}, M. \& {Dijkstra}, M. 2014, \mnras, 444, 1095

\bibitem[{{Gronwall} {et~al.}(2007){Gronwall}, {Ciardullo}, {Hickey},
  {Gawiser}, {Feldmeier}, {van Dokkum}, {Urry}, {Herrera}, {Lehmer}, {Infante},
  {Orsi}, {Marchesini}, {Blanc}, {Francke}, {Lira}, \&
  {Treister}}]{Gronwall2007}
{Gronwall}, C., {Ciardullo}, R., {Hickey}, T., {et~al.} 2007, \apj, 667, 79

\bibitem[{{Grove} {et~al.}(2009){Grove}, {Fynbo}, {Ledoux}, {Limousin},
  {M{\o}ller}, {Nilsson}, \& {Thomsen}}]{Grove2009}
{Grove}, L.~F., {Fynbo}, J.~P.~U., {Ledoux}, C., {et~al.} 2009, \aap, 497, 689

\bibitem[{{Guaita} {et~al.}(2013){Guaita}, {Francke}, {Gawiser}, {Bauer},
  {Hayes}, {{\"O}stlin}, \& {Padilla}}]{Guaita2013}
{Guaita}, L., {Francke}, H., {Gawiser}, E., {et~al.} 2013, \aap, 551, A93

\bibitem[{{Guaita} {et~al.}(2015){Guaita}, {Melinder}, {Hayes}, {{\"O}stlin},
  {Gonzalez}, {Micheva}, {Adamo}, {Mas-Hesse}, {Sandberg},
  {Ot{\'{\i}}-Floranes}, {Schaerer}, {Verhamme}, {Freeland}, {Orlitov{\'a}},
  {Laursen}, {Cannon}, {Duval}, {Rivera-Thorsen}, {Herenz}, {Kunth}, {Atek},
  {Puschnig}, {Gruyters}, \& {Pardy}}]{Guaita2015}
{Guaita}, L., {Melinder}, J., {Hayes}, M., {et~al.} 2015, \aap, 576, A51

\bibitem[{{Hashimoto} {et~al.}(2013){Hashimoto}, {Ouchi}, {Shimasaku}, {Ono},
  {Nakajima}, {Rauch}, {Lee}, \& {Okamura}}]{Hashimoto2013}
{Hashimoto}, T., {Ouchi}, M., {Shimasaku}, K., {et~al.} 2013, \apj, 765, 70

\bibitem[{{Hashimoto} {et~al.}(2015){Hashimoto}, {Verhamme}, {Ouchi},
  {Shimasaku}, {Schaerer}, {Nakajima}, {Shibuya}, {Rauch}, {Ono}, \&
  {Goto}}]{Hashimoto2015}
{Hashimoto}, T., {Verhamme}, A., {Ouchi}, M., {et~al.} 2015, \apj, 812, 157

\bibitem[{{Hayes}(2015)}]{Hayes2015}
{Hayes}, M. 2015, \pasa, 32, 27

\bibitem[{{Hayes} {et~al.}(2014){Hayes}, {{\"O}stlin}, {Duval}, {Sandberg},
  {Guaita}, {Melinder}, {Adamo}, {Schaerer}, {Verhamme}, {Orlitov{\'a}},
  {Mas-Hesse}, {Cannon}, {Atek}, {Kunth}, {Laursen}, {Ot{\'{\i}}-Floranes},
  {Pardy}, {Rivera-Thorsen}, \& {Herenz}}]{Hayes2014}
{Hayes}, M., {{\"O}stlin}, G., {Duval}, F., {et~al.} 2014, \apj, 782, 6

\bibitem[{{Hayes} {et~al.}(2009){Hayes}, {{\"O}stlin}, {Mas-Hesse}, \&
  {Kunth}}]{Hayes2009}
{Hayes}, M., {{\"O}stlin}, G., {Mas-Hesse}, J.~M., \& {Kunth}, D. 2009, \aj,
  138, 911

\bibitem[{{Hayes} {et~al.}(2005){Hayes}, {{\"O}stlin}, {Mas-Hesse}, {Kunth},
  {Leitherer}, \& {Petrosian}}]{Hayes2005}
{Hayes}, M., {{\"O}stlin}, G., {Mas-Hesse}, J.~M., {et~al.} 2005, \aap, 438, 71

\bibitem[{{Hayes} {et~al.}(2010){Hayes}, {{\"O}stlin}, {Schaerer}, {Mas-Hesse},
  {Leitherer}, {Atek}, {Kunth}, {Verhamme}, {de Barros}, \&
  {Melinder}}]{Hayes2010}
{Hayes}, M., {{\"O}stlin}, G., {Schaerer}, D., {et~al.} 2010, \nat, 464, 562

\bibitem[{{Hayes} {et~al.}(2013){Hayes}, {{\"O}stlin}, {Schaerer}, {Verhamme},
  {Mas-Hesse}, {Adamo}, {Atek}, {Cannon}, {Duval}, {Guaita}, {Herenz}, {Kunth},
  {Laursen}, {Melinder}, {Orlitov{\'a}}, {Ot{\'{\i}}-Floranes}, \&
  {Sandberg}}]{Hayes2013}
{Hayes}, M., {{\"O}stlin}, G., {Schaerer}, D., {et~al.} 2013, \apjl, 765, L27

\bibitem[{{Henry} {et~al.}(2015){Henry}, {Scarlata}, {Martin}, \&
  {Erb}}]{Henry2015}
{Henry}, A., {Scarlata}, C., {Martin}, C.~L., \& {Erb}, D. 2015, \apj, 809, 19

\bibitem[{{Hong} {et~al.}(2014){Hong}, {Dey}, \& {Prescott}}]{Hong2014}
{Hong}, S., {Dey}, A., \& {Prescott}, M.~K.~M. 2014, \pasp, 126, 1048

\bibitem[{{Horne}(1986)}]{Horne1986}
{Horne}, K. 1986, \pasp, 98, 609

\bibitem[{{Hu} {et~al.}(1998){Hu}, {Cowie}, \& {McMahon}}]{Hu1998}
{Hu}, E.~M., {Cowie}, L.~L., \& {McMahon}, R.~G. 1998, \apjl, 502, L99

\bibitem[{Hunter(2007)}]{Hunter2007}
Hunter, J.~D. 2007, Computing In Science \& Engineering, 9, 90

\bibitem[{{Husemann} {et~al.}(2013){Husemann}, {Jahnke}, {S{\'a}nchez},
  {Barrado}, {Bekerait{\.e}}, {Bomans}, {Castillo-Morales},
  {Catal{\'a}n-Torrecilla}, {Cid Fernandes}, {Falc{\'o}n-Barroso},
  {Garc{\'{\i}}a-Benito}, {Gonz{\'a}lez Delgado}, {Iglesias-P{\'a}ramo},
  {Johnson}, {Kupko}, {L{\'o}pez-Fernandez}, {Lyubenova}, {Marino}, {Mast},
  {Miskolczi}, {Monreal-Ibero}, {Gil de Paz}, {P{\'e}rez}, {P{\'e}rez},
  {Rosales-Ortega}, {Ruiz-Lara}, {Schilling}, {van de Ven}, {Walcher}, {Alves},
  {de Amorim}, {Backsmann}, {Barrera-Ballesteros}, {Bland-Hawthorn}, {Cortijo},
  {Dettmar}, {Demleitner}, {D{\'{\i}}az}, {Enke}, {Florido}, {Flores},
  {Galbany}, {Gallazzi}, {Garc{\'{\i}}a-Lorenzo}, {Gomes}, {Gruel}, {Haines},
  {Holmes}, {Jungwiert}, {Kalinova}, {Kehrig}, {Kennicutt}, {Klar}, {Lehnert},
  {L{\'o}pez-S{\'a}nchez}, {de Lorenzo-C{\'a}ceres}, {M{\'a}rmol-Queralt{\'o}},
  {M{\'a}rquez}, {Mendez-Abreu}, {Moll{\'a}}, {del Olmo}, {Meidt}, {Papaderos},
  {Puschnig}, {Quirrenbach}, {Roth}, {S{\'a}nchez-Bl{\'a}zquez}, {Spekkens},
  {Singh}, {Stanishev}, {Trager}, {Vilchez}, {Wild}, {Wisotzki}, {Zibetti}, \&
  {Ziegler}}]{Husemann2013}
{Husemann}, B., {Jahnke}, K., {S{\'a}nchez}, S.~F., {et~al.} 2013, \aap, 549,
  A87

\bibitem[{{Husemann} {et~al.}(2012){Husemann}, {Kamann}, {Sandin},
  {S{\'a}nchez}, {Garc{\'{\i}}a-Benito}, \& {Mast}}]{Husemann2012}
{Husemann}, B., {Kamann}, S., {Sandin}, C., {et~al.} 2012, \aap, 545, A137

\bibitem[{{Iye}(2011)}]{Iye2011a}
{Iye}, M. 2011, Proceeding of the Japan Academy, Series B, 87, 575

\bibitem[{{Jaskot} \& {Oey}(2014)}]{Jaskot2014}
{Jaskot}, A.~E. \& {Oey}, M.~S. 2014, \apjl, 791, L19

\bibitem[{{Jim{\'e}nez-Vicente} {et~al.}(1999){Jim{\'e}nez-Vicente},
  {Battaner}, {Rozas}, {Casta{\~n}eda}, \& {Porcel}}]{Jimenez-Vicente1999}
{Jim{\'e}nez-Vicente}, J., {Battaner}, E., {Rozas}, M., {Casta{\~n}eda}, H., \&
  {Porcel}, C. 1999, \aap, 342, 417

\bibitem[{{Kennicutt}(1998)}]{Kennicutt1998}
{Kennicutt}, Jr., R.~C. 1998, \araa, 36, 189

\bibitem[{{Kunth} {et~al.}(1998){Kunth}, {Mas-Hesse}, {Terlevich}, {Terlevich},
  {Lequeux}, \& {Fall}}]{Kunth1998}
{Kunth}, D., {Mas-Hesse}, J.~M., {Terlevich}, E., {et~al.} 1998, \aap, 334, 11

\bibitem[{{Landman} {et~al.}(1982){Landman}, {Roussel-Dupre}, \&
  {Tanigawa}}]{Landman1982}
{Landman}, D.~A., {Roussel-Dupre}, R., \& {Tanigawa}, G. 1982, \apj, 261, 732

\bibitem[{{Laursen} {et~al.}(2013){Laursen}, {Duval}, \&
  {{\"O}stlin}}]{Laursen2013}
{Laursen}, P., {Duval}, F., \& {{\"O}stlin}, G. 2013, \apj, 766, 124

\bibitem[{{Laursen} \& {Sommer-Larsen}(2007)}]{Laursen2007}
{Laursen}, P. \& {Sommer-Larsen}, J. 2007, \apjl, 657, L69

\bibitem[{{Laursen} {et~al.}(2009){Laursen}, {Sommer-Larsen}, \&
  {Andersen}}]{Laursen2009a}
{Laursen}, P., {Sommer-Larsen}, J., \& {Andersen}, A.~C. 2009, \apj, 704, 1640

\bibitem[{{Law} {et~al.}(2009){Law}, {Steidel}, {Erb}, {Larkin}, {Pettini},
  {Shapley}, \& {Wright}}]{Law2009}
{Law}, D.~R., {Steidel}, C.~C., {Erb}, D.~K., {et~al.} 2009, \apj, 697, 2057

\bibitem[{{Lenz} \& {Ayres}(1992)}]{Lenz1992}
{Lenz}, D.~D. \& {Ayres}, T.~R. 1992, \pasp, 104, 1104

\bibitem[{{Lupton} {et~al.}(2004){Lupton}, {Blanton}, {Fekete}, {Hogg},
  {O'Mullane}, {Szalay}, \& {Wherry}}]{Lupton2004}
{Lupton}, R., {Blanton}, M.~R., {Fekete}, G., {et~al.} 2004, \pasp, 116, 133

\bibitem[{{Malhotra} \& {Rhoads}(2004)}]{Malhotra2004}
{Malhotra}, S. \& {Rhoads}, J.~E. 2004, \apjl, 617, L5

\bibitem[{{Mallery} {et~al.}(2012){Mallery}, {Mobasher}, {Capak}, {Kakazu},
  {Masters}, {Ilbert}, {Hemmati}, {Scarlata}, {Salvato}, {McCracken},
  {LeFevre}, \& {Scoville}}]{Mallery2012}
{Mallery}, R.~P., {Mobasher}, B., {Capak}, P., {et~al.} 2012, \apj, 760, 128

\bibitem[{{Marino} {et~al.}(2009){Marino}, {Iodice}, {Tantalo}, {Piovan},
  {Bettoni}, {Buson}, {Chiosi}, {Galletta}, {Rampazzo}, \& {Rich}}]{Marino2009}
{Marino}, A., {Iodice}, E., {Tantalo}, R., {et~al.} 2009, \aap, 508, 1235

\bibitem[{{Martin} {et~al.}(2014){Martin}, {Chang}, {Matuszewski}, {Morrissey},
  {Rahman}, {Moore}, \& {Steidel}}]{Martin2014}
{Martin}, D.~C., {Chang}, D., {Matuszewski}, M., {et~al.} 2014, \apj, 786, 106

\bibitem[{{Mas-Hesse} {et~al.}(2003){Mas-Hesse}, {Kunth}, {Tenorio-Tagle},
  {Leitherer}, {Terlevich}, \& {Terlevich}}]{Mas-Hesse2003}
{Mas-Hesse}, J.~M., {Kunth}, D., {Tenorio-Tagle}, G., {et~al.} 2003, \apj, 598,
  858

\bibitem[{{McLinden} {et~al.}(2011){McLinden}, {Finkelstein}, {Rhoads},
  {Malhotra}, {Hibon}, {Richardson}, {Cresci}, {Quirrenbach}, {Pasquali},
  {Bian}, {Fan}, \& {Woodward}}]{McLinden2011}
{McLinden}, E.~M., {Finkelstein}, S.~L., {Rhoads}, J.~E., {et~al.} 2011, \apj,
  730, 136

\bibitem[{{McLinden} {et~al.}(2014){McLinden}, {Rhoads}, {Malhotra},
  {Finkelstein}, {Richardson}, {Smith}, \& {Tilvi}}]{McLinden2014}
{McLinden}, E.~M., {Rhoads}, J.~E., {Malhotra}, S., {et~al.} 2014, \mnras, 439,
  446

\bibitem[{{Monreal-Ibero} {et~al.}(2010){Monreal-Ibero}, {Arribas}, {Colina},
  {Rodr{\'{\i}}guez-Zaur{\'{\i}}n}, {Alonso-Herrero}, \&
  {Garc{\'{\i}}a-Mar{\'{\i}}n}}]{Monreal-Ibero2010}
{Monreal-Ibero}, A., {Arribas}, S., {Colina}, L., {et~al.} 2010, \aap, 517, A28

\bibitem[{{Neufeld}(1990)}]{Neufeld1990}
{Neufeld}, D.~A. 1990, \apj, 350, 216

\bibitem[{{Newman} {et~al.}(2013){Newman}, {Genzel}, {F{\"o}rster Schreiber},
  {Shapiro Griffin}, {Mancini}, {Lilly}, {Renzini}, {Bouch{\'e}}, {Burkert},
  {Buschkamp}, {Carollo}, {Cresci}, {Davies}, {Eisenhauer}, {Genel}, {Hicks},
  {Kurk}, {Lutz}, {Naab}, {Peng}, {Sternberg}, {Tacconi}, {Wuyts}, {Zamorani},
  \& {Vergani}}]{Newman2013}
{Newman}, S.~F., {Genzel}, R., {F{\"o}rster Schreiber}, N.~M., {et~al.} 2013,
  \apj, 767, 104

\bibitem[{{Oesch} {et~al.}(2015){Oesch}, {van Dokkum}, {Illingworth},
  {Bouwens}, {Momcheva}, {Holden}, {Roberts-Borsani}, {Smit}, {Franx},
  {Labb{\'e}}, {Gonz{\'a}lez}, \& {Magee}}]{Oesch2015}
{Oesch}, P.~A., {van Dokkum}, P.~G., {Illingworth}, G.~D., {et~al.} 2015,
  \apjl, 804, L30

\bibitem[{{Oke}(1990)}]{Oke1990}
{Oke}, J.~B. 1990, \aj, 99, 1621

\bibitem[{{Ono} {et~al.}(2012){Ono}, {Ouchi}, {Mobasher}, {Dickinson},
  {Penner}, {Shimasaku}, {Weiner}, {Kartaltepe}, {Nakajima}, {Nayyeri},
  {Stern}, {Kashikawa}, \& {Spinrad}}]{Ono2012}
{Ono}, Y., {Ouchi}, M., {Mobasher}, B., {et~al.} 2012, \apj, 744, 83

\bibitem[{{{\"O}stlin} {et~al.}(2001){{\"O}stlin}, {Amram}, {Bergvall},
  {Masegosa}, {Boulesteix}, \& {M{\'a}rquez}}]{Oestlin2001}
{{\"O}stlin}, G., {Amram}, P., {Bergvall}, N., {et~al.} 2001, \aap, 374, 800

\bibitem[{{{\"O}stlin} {et~al.}(1999){{\"O}stlin}, {Amram}, {Masegosa},
  {Bergvall}, \& {Boulesteix}}]{Oestlin1999}
{{\"O}stlin}, G., {Amram}, P., {Masegosa}, J., {Bergvall}, N., \& {Boulesteix},
  J. 1999, \aaps, 137, 419

\bibitem[{{{\"O}stlin} {et~al.}(2014){{\"O}stlin}, {Hayes}, {Duval},
  {Sandberg}, {Rivera-Thorsen}, {Marquart}, {Orlitov{\'a}}, {Adamo},
  {Melinder}, {Guaita}, {Atek}, {Cannon}, {Gruyters}, {Herenz}, {Kunth},
  {Laursen}, {Mas-Hesse}, {Micheva}, {Ot{\'{\i}}-Floranes}, {Pardy}, {Roth},
  {Schaerer}, \& {Verhamme}}]{Ostlin2014}
{{\"O}stlin}, G., {Hayes}, M., {Duval}, F., {et~al.} 2014, \apj, 797, 11

\bibitem[{{{\"O}stlin} {et~al.}(2009){{\"O}stlin}, {Hayes}, {Kunth},
  {Mas-Hesse}, {Leitherer}, {Petrosian}, \& {Atek}}]{Ostlin2009}
{{\"O}stlin}, G., {Hayes}, M., {Kunth}, D., {et~al.} 2009, \aj, 138, 923

\bibitem[{{Ouchi} {et~al.}(2008){Ouchi}, {Shimasaku}, {Akiyama}, {Simpson},
  {Saito}, {Ueda}, {Furusawa}, {Sekiguchi}, {Yamada}, {Kodama}, {Kashikawa},
  {Okamura}, {Iye}, {Takata}, {Yoshida}, \& {Yoshida}}]{Ouchi2008}
{Ouchi}, M., {Shimasaku}, K., {Akiyama}, M., {et~al.} 2008, \apjs, 176, 301

\bibitem[{{Pardy} {et~al.}(2014){Pardy}, {Cannon}, {{\"O}stlin}, {Hayes},
  {Rivera-Thorsen}, {Sandberg}, {Adamo}, {Freeland}, {Herenz}, {Guaita},
  {Kunth}, {Laursen}, {Mas-Hesse}, {Melinder}, {Orlitov{\'a}},
  {Ot{\'{\i}}-Floranes}, {Puschnig}, {Schaerer}, \& {Verhamme}}]{Pardy2014}
{Pardy}, S.~A., {Cannon}, J.~M., {{\"O}stlin}, G., {et~al.} 2014, \apj, 794,
  101

\bibitem[{{Partridge} \& {Peebles}(1967)}]{Partridge1967}
{Partridge}, R.~B. \& {Peebles}, P.~J.~E. 1967, \apj, 147, 868

\bibitem[{{Petrosian}(1976)}]{Petrosian1976}
{Petrosian}, V. 1976, \apjl, 209, L1

\bibitem[{{Puech} {et~al.}(2006){Puech}, {Hammer}, {Flores}, {{\"O}stlin}, \&
  {Marquart}}]{Puech2006}
{Puech}, M., {Hammer}, F., {Flores}, H., {{\"O}stlin}, G., \& {Marquart}, T.
  2006, \aap, 455, 119

\bibitem[{{Rauch} {et~al.}(2008){Rauch}, {Haehnelt}, {Bunker}, {Becker},
  {Marleau}, {Graham}, {Cristiani}, {Jarvis}, {Lacey}, {Morris}, {Peroux},
  {R{\"o}ttgering}, \& {Theuns}}]{Rauch2008}
{Rauch}, M., {Haehnelt}, M., {Bunker}, A., {et~al.} 2008, \apj, 681, 856

\bibitem[{{Rhoads} {et~al.}(2014){Rhoads}, {Malhotra}, {Richardson},
  {Finkelstein}, {Fynbo}, {McLinden}, \& {Tilvi}}]{Rhoads2014}
{Rhoads}, J.~E., {Malhotra}, S., {Richardson}, M.~L.~A., {et~al.} 2014, \apj,
  780, 20

\bibitem[{{Rivera-Thorsen} {et~al.}(2015){Rivera-Thorsen}, {Hayes},
  {{\"O}stlin}, {Duval}, {Orlitov{\'a}}, {Verhamme}, {Mas-Hesse}, {Schaerer},
  {Cannon}, {Ot{\'{\i}}-Floranes}, {Sandberg}, {Guaita}, {Adamo}, {Atek},
  {Herenz}, {Kunth}, {Laursen}, \& {Melinder}}]{Rivera-Thorsen2015}
{Rivera-Thorsen}, T.~E., {Hayes}, M., {{\"O}stlin}, G., {et~al.} 2015, \apj,
  805, 14

\bibitem[{{Robertson}(2013)}]{Robertson2013}
{Robertson}, J.~G. 2013, \pasa, 30, 48

\bibitem[{{Roth} {et~al.}(2010){Roth}, {Fechner}, {Wolter}, {Sandin}, {Kelz},
  {Bauer}, {Popow}, {Monreal-Ibero}, {Kehrig}, \& {Streicher}}]{Roth2010}
{Roth}, M.~M., {Fechner}, T., {Wolter}, D., {et~al.} 2010, in Society of
  Photo-Optical Instrumentation Engineers (SPIE) Conference Series, Vol. 7742,
  High Energy, Optical, and Infrared Detectors for Astronomy IV, 774209

\bibitem[{{Roth} {et~al.}(2005){Roth}, {Kelz}, {Fechner}, {Hahn}, {Bauer},
  {Becker}, {B{\"o}hm}, {Christensen}, {Dionies}, {Paschke}, {Popow}, {Wolter},
  {Schmoll}, {Laux}, \& {Altmann}}]{Roth2005}
{Roth}, M.~M., {Kelz}, A., {Fechner}, T., {et~al.} 2005, \pasp, 117, 620

\bibitem[{{S{\'a}nchez} {et~al.}(2007{\natexlab{a}}){S{\'a}nchez}, {Aceituno},
  {Thiele}, {P{\'e}rez-Ram{\'{\i}}rez}, \& {Alves}}]{Sanchez2007a}
{S{\'a}nchez}, S.~F., {Aceituno}, J., {Thiele}, U., {P{\'e}rez-Ram{\'{\i}}rez},
  D., \& {Alves}, J. 2007{\natexlab{a}}, \pasp, 119, 1186

\bibitem[{{S{\'a}nchez} {et~al.}(2007{\natexlab{b}}){S{\'a}nchez}, {Cardiel},
  {Verheijen}, {Mart{\'{\i}}n-Gord{\'o}n}, {Vilchez}, \& {Alves}}]{Sanchez2007}
{S{\'a}nchez}, S.~F., {Cardiel}, N., {Verheijen}, M.~A.~W., {et~al.}
  2007{\natexlab{b}}, \aap, 465, 207

\bibitem[{{S{\'a}nchez} {et~al.}(2012){S{\'a}nchez}, {Kennicutt}, {Gil de Paz},
  {van de Ven}, {V{\'{\i}}lchez}, {Wisotzki}, {Walcher}, {Mast}, {Aguerri},
  {Albiol-P{\'e}rez}, {Alonso-Herrero}, {Alves}, {Bakos}, {Bart{\'a}kov{\'a}},
  {Bland-Hawthorn}, {Boselli}, {Bomans}, {Castillo-Morales}, {Cortijo-Ferrero},
  {de Lorenzo-C{\'a}ceres}, {Del Olmo}, {Dettmar}, {D{\'{\i}}az}, {Ellis},
  {Falc{\'o}n-Barroso}, {Flores}, {Gallazzi}, {Garc{\'{\i}}a-Lorenzo},
  {Gonz{\'a}lez Delgado}, {Gruel}, {Haines}, {Hao}, {Husemann},
  {Igl{\'e}sias-P{\'a}ramo}, {Jahnke}, {Johnson}, {Jungwiert}, {Kalinova},
  {Kehrig}, {Kupko}, {L{\'o}pez-S{\'a}nchez}, {Lyubenova}, {Marino},
  {M{\'a}rmol-Queralt{\'o}}, {M{\'a}rquez}, {Masegosa}, {Meidt},
  {Mendez-Abreu}, {Monreal-Ibero}, {Montijo}, {Mour{\~a}o}, {Palacios-Navarro},
  {Papaderos}, {Pasquali}, {Peletier}, {P{\'e}rez}, {P{\'e}rez}, {Quirrenbach},
  {Rela{\~n}o}, {Rosales-Ortega}, {Roth}, {Ruiz-Lara},
  {S{\'a}nchez-Bl{\'a}zquez}, {Sengupta}, {Singh}, {Stanishev}, {Trager},
  {Vazdekis}, {Viironen}, {Wild}, {Zibetti}, \& {Ziegler}}]{Sanchez2012}
{S{\'a}nchez}, S.~F., {Kennicutt}, R.~C., {Gil de Paz}, A., {et~al.} 2012,
  \aap, 538, A8

\bibitem[{{Sandberg} {et~al.}(2015){Sandberg}, {Guaita}, {{\"O}stlin}, {Hayes},
  \& {Kiaeerad}}]{Sandberg2015}
{Sandberg}, A., {Guaita}, L., {{\"O}stlin}, G., {Hayes}, M., \& {Kiaeerad}, F.
  2015, \aap, 580, A91

\bibitem[{{Sandberg} {et~al.}(2013){Sandberg}, {{\"O}stlin}, {Hayes}, {Fathi},
  {Schaerer}, {Mas-Hesse}, \& {Rivera-Thorsen}}]{Sandberg2013}
{Sandberg}, A., {{\"O}stlin}, G., {Hayes}, M., {et~al.} 2013, \aap, 552, A95

\bibitem[{{Sandin} {et~al.}(2010){Sandin}, {Becker}, {Roth}, {Gerssen},
  {Monreal-Ibero}, {B{\"o}hm}, \& {Weilbacher}}]{Sandin2010}
{Sandin}, C., {Becker}, T., {Roth}, M.~M., {et~al.} 2010, \aap, 515, A35

\bibitem[{{Sandin} {et~al.}(2011){Sandin}, {Weilbacher}, {Streicher},
  {Walcher}, \& {Roth}}]{Sandin2011}
{Sandin}, C., {Weilbacher}, P., {Streicher}, O., {Walcher}, C.~J., \& {Roth},
  M.~M. 2011, The Messenger, 144, 13

\bibitem[{{Sandin} {et~al.}(2012){Sandin}, Weilbacher, Tabataba-Vakili, Kamann,
  \& Streicher}]{Sandin2012}
{Sandin}, C., Weilbacher, P., Tabataba-Vakili, F., Kamann, S., \& Streicher, O.
  2012, in Proceedings of SPIE, Vol. 8451, Software and Cyberinfrastructure for
  Astronomy II, ed. C.~G. Radziwill N.~M., Society of Photo-Optical
  Instrumentation Engineers (SPIE), 84510F

\bibitem[{{Schaerer} {et~al.}(2011){Schaerer}, {Hayes}, {Verhamme}, \&
  {Teyssier}}]{Schaerer2011a}
{Schaerer}, D., {Hayes}, M., {Verhamme}, A., \& {Teyssier}, R. 2011, \aap, 531,
  A12

\bibitem[{{Schaerer} \& {Verhamme}(2008)}]{Schaerer2008}
{Schaerer}, D. \& {Verhamme}, A. 2008, \aap, 480, 369

\bibitem[{{Shapley} {et~al.}(2003){Shapley}, {Steidel}, {Pettini}, \&
  {Adelberger}}]{Shapley2003}
{Shapley}, A.~E., {Steidel}, C.~C., {Pettini}, M., \& {Adelberger}, K.~L. 2003,
  \apj, 588, 65

\bibitem[{{Shimasaku} {et~al.}(2006){Shimasaku}, {Kashikawa}, {Doi}, {Ly},
  {Malkan}, {Matsuda}, {Ouchi}, {Hayashino}, {Iye}, {Motohara}, {Murayama},
  {Nagao}, {Ohta}, {Okamura}, {Sasaki}, {Shioya}, \&
  {Taniguchi}}]{Shimasaku2006}
{Shimasaku}, K., {Kashikawa}, N., {Doi}, M., {et~al.} 2006, \pasj, 58, 313

\bibitem[{{Shioya} {et~al.}(2009){Shioya}, {Taniguchi}, {Sasaki}, {Nagao},
  {Murayama}, {Saito}, {Ideue}, {Nakajima}, {Matsuoka}, {Trump}, {Scoville},
  {Sanders}, {Mobasher}, {Aussel}, {Capak}, {Kartaltepe}, {Koekemoer},
  {Carilli}, {Ellis}, {Garilli}, {Giavalisco}, {Kitzbichler}, {Impey},
  {LeFevre}, {Schinnerer}, \& {Smolcic}}]{Shioya2009}
{Shioya}, Y., {Taniguchi}, Y., {Sasaki}, S.~S., {et~al.} 2009, \apj, 696, 546

\bibitem[{{Sobral} {et~al.}(2015){Sobral}, {Matthee}, {Darvish}, {Schaerer},
  {Mobasher}, {R{\"o}ttgering}, {Santos}, \& {Hemmati}}]{Sobral2015}
{Sobral}, D., {Matthee}, J., {Darvish}, B., {et~al.} 2015, \apj, 808, 139

\bibitem[{{Song} {et~al.}(2014){Song}, {Finkelstein}, {Gebhardt}, {Hill},
  {Drory}, {Ashby}, {Blanc}, {Bridge}, {Chonis}, {Ciardullo}, {Fabricius},
  {Fazio}, {Gawiser}, {Gronwall}, {Hagen}, {Huang}, {Jogee}, {Livermore},
  {Salmon}, {Schneider}, {Willner}, \& {Zeimann}}]{Song2014}
{Song}, M., {Finkelstein}, S.~L., {Gebhardt}, K., {et~al.} 2014, \apj, 791, 3

\bibitem[{{Taniguchi} {et~al.}(2003){Taniguchi}, {Shioya}, {Ajiki}, {Fujita},
  {Nagao}, \& {Murayama}}]{Taniguchi2003}
{Taniguchi}, Y., {Shioya}, Y., {Ajiki}, M., {et~al.} 2003, Journal of Korean
  Astronomical Society, 36, 123

\bibitem[{{Tapken} {et~al.}(2006){Tapken}, {Appenzeller}, {Gabasch}, {Heidt},
  {Hopp}, {Bender}, {Mehlert}, {Noll}, {Seitz}, \& {Seifert}}]{Tapken2006}
{Tapken}, C., {Appenzeller}, I., {Gabasch}, A., {et~al.} 2006, \aap, 455, 145

\bibitem[{{Tapken} {et~al.}(2004){Tapken}, {Appenzeller}, {Mehlert}, {Noll}, \&
  {Richling}}]{Tapken2004}
{Tapken}, C., {Appenzeller}, I., {Mehlert}, D., {Noll}, S., \& {Richling}, S.
  2004, \aap, 416, L1

\bibitem[{{Tapken} {et~al.}(2007){Tapken}, {Appenzeller}, {Noll}, {Richling},
  {Heidt}, {Meink{\"o}hn}, \& {Mehlert}}]{Tapken2007}
{Tapken}, C., {Appenzeller}, I., {Noll}, S., {et~al.} 2007, \aap, 467, 63

\bibitem[{{Terlevich} \& {Melnick}(1981)}]{Terlevich1981}
{Terlevich}, R. \& {Melnick}, J. 1981, \mnras, 195, 839

\bibitem[{{Teyssier} {et~al.}(2010){Teyssier}, {Chapon}, \&
  {Bournaud}}]{Teyssier2010}
{Teyssier}, R., {Chapon}, D., \& {Bournaud}, F. 2010, \apjl, 720, L149

\bibitem[{Turner(2010)}]{Turner2010}
Turner, J.~E. 2010, Canary Islands Winter School of Astrophysics, Vol. XVII, 3D
  Spectroscopy in Astronomy, ed. E.~Mediavilla, S.~Arribas, M.~Roth,
  J.~Cepa-Nogue, \& F.~Sanchez (Cambridge University Press), 87--125

\bibitem[{{van Dokkum}(2001)}]{vanDokkum2001}
{van Dokkum}, P.~G. 2001, \pasp, 113, 1420

\bibitem[{{Verhamme} {et~al.}(2012){Verhamme}, {Dubois}, {Blaizot}, {Garel},
  {Bacon}, {Devriendt}, {Guiderdoni}, \& {Slyz}}]{Verhamme2012}
{Verhamme}, A., {Dubois}, Y., {Blaizot}, J., {et~al.} 2012, \aap, 546, A111

\bibitem[{{Verhamme} {et~al.}(2008){Verhamme}, {Schaerer}, {Atek}, \&
  {Tapken}}]{Verhamme2008}
{Verhamme}, A., {Schaerer}, D., {Atek}, H., \& {Tapken}, C. 2008, \aap, 491, 89

\bibitem[{{Verhamme} {et~al.}(2006){Verhamme}, {Schaerer}, \&
  {Maselli}}]{Verhamme2006}
{Verhamme}, A., {Schaerer}, D., \& {Maselli}, A. 2006, \aap, 460, 397

\bibitem[{{Wall}(1996)}]{Wall1996}
{Wall}, J.~V. 1996, \qjras, 37, 519

\bibitem[{{Wisnioski} {et~al.}(2015){Wisnioski}, {F{\"o}rster Schreiber},
  {Wuyts}, {Wuyts}, {Bandara}, {Wilman}, {Genzel}, {Bender}, {Davies},
  {Fossati}, {Lang}, {Mendel}, {Beifiori}, {Brammer}, {Chan}, {Fabricius},
  {Fudamoto}, {Kulkarni}, {Kurk}, {Lutz}, {Nelson}, {Momcheva}, {Rosario},
  {Saglia}, {Seitz}, {Tacconi}, \& {van Dokkum}}]{Wisnioski2015}
{Wisnioski}, E., {F{\"o}rster Schreiber}, N.~M., {Wuyts}, S., {et~al.} 2015,
  \apj, 799, 209

\bibitem[{Wisotzki {et~al.}(2015)Wisotzki, Bacon, Blaizot, Brinchmann, Herenz,
  Schaye, Bouché, Cantalupo, Contini, Carollo, Caruana, Courbot, Emsellem,
  Kamann, Kerutt, Leclercq, Lilly, Patrício, Sandin, Steinmetz, Straka,
  Urrutia, Verhamme, Weilbacher, \& Wendt}]{Wisotzki2015}
Wisotzki, L., Bacon, R., Blaizot, J., {et~al.} 2015 [\eprint{1509.05143}]

\bibitem[{{Wold} {et~al.}(2014){Wold}, {Barger}, \& {Cowie}}]{Wold2014}
{Wold}, I.~G.~B., {Barger}, A.~J., \& {Cowie}, L.~L. 2014, \apj, 783, 119

\bibitem[{{Yamada} {et~al.}(2012){Yamada}, {Matsuda}, {Kousai}, {Hayashino},
  {Morimoto}, \& {Umemura}}]{Yamada2012}
{Yamada}, T., {Matsuda}, Y., {Kousai}, K., {et~al.} 2012, \apj, 751, 29

\bibitem[{{Yang} {et~al.}(2015){Yang}, {Malhotra}, {Gronke}, {Rhoads},
  {Jaskot}, {Zheng}, \& {Dijkstra}}]{Yang2015}
{Yang}, H., {Malhotra}, S., {Gronke}, M., {et~al.} 2015, ArXiv e-prints
  [\eprint[arXiv]{1506.02885}]

\bibitem[{{Zheng} \& {Wallace}(2014)}]{Zheng2014a}
{Zheng}, Z. \& {Wallace}, J. 2014, \apj, 794, 116

\bibitem[{{Zitrin} {et~al.}(2015){Zitrin}, {Labb{\'e}}, {Belli}, {Bouwens},
  {Ellis}, {Roberts-Borsani}, {Stark}, {Oesch}, \& {Smit}}]{Zitrin2015}
{Zitrin}, A., {Labb{\'e}}, I., {Belli}, S., {et~al.} 2015, \apjl, 810, L12

\end{thebibliography}

% maybe someone finds this comment on ArXiV and tweets it.... :-)

\end{document}